%% file: main__5___1___1_.tex
\documentclass[aps,preprintnumbers,superscriptaddress,longbibliography,nofootinbib]{revtex4-1}

  \usepackage{pgfplots}
  \pgfplotsset{compat=1.16}

  \definecolor{cqgrey}{HTML}{8A8A8A}
  \definecolor{cqgreyb}{HTML}{9A9A9A}
  \definecolor{cqred}{HTML}{A8452C}
  \definecolor{cqorange}{HTML}{B8763A}
  \definecolor{cqblue}{HTML}{4F8BA3}

  \pgfplotsset{
    cqaxis/.style={
      width=0.88\linewidth, height=0.56\linewidth,
      axis lines=left,
      axis line style={cqgrey, line width=0.4pt},
      tick style={cqgrey, line width=0.4pt},
      every tick label/.append style={cqgrey, font=\footnotesize},
      label style={cqgrey, font=\small},
      title style={cqgrey, font=\small},
      xlabel={$x=(\alpha-3\beta)/\alpha$},
      tick align=outside,
      clip=false,
    }
  }
\definecolor{cqgreen}{HTML}{4F7A52}
\usepackage{amsmath, mathrsfs, amssymb,amsfonts,amsthm,graphicx, epsf, dcolumn, yfonts}
\usepackage[bottom]{footmisc}
\usepackage[normalem]{ulem}
\usepackage{xcolor}
\usepackage{amssymb}
\usepackage{slashed}
\usepackage{setspace}
\usepackage{cancel}
\usepackage{wasysym}
\usepackage{tikz-feynman}
\usepackage{float}
\usepackage{siunitx}
\usepackage[utf8]{inputenc}
\usepackage{todonotes}
\usepackage{tikz}
\usepackage{tikz-cd}
\usepackage{array}
\usepackage{tabularray}
\usepackage{tabularx}
\usepackage{physics}
\usepackage{hyperref}
\definecolor{subduedgreen}{RGB}{0,215,0} 
\definecolor{subduedblue}{RGB}{0,0,255} 
\pdfoutput=1
\input{definitions}

\newcommand{\MS}[1]{#1}

\definecolor{corrflag}{RGB}{180,80,0}
\definecolor{proseflag}{RGB}{0,110,60}
\definecolor{physflag}{RGB}{170,0,140}
\definecolor{kbflag}{RGB}{0,90,200}
\newcommand{\pdf}{\rho}      
\newcommand{\density}{m}     

\def\X{X}
\def\T{\bf T}

\def\z{{q,\dot{q}}}

\def\g{{g}}

\def\T00{{\bf T_{NN}}}

\def\0mom{{\bar{\Gamma}^{\rho\sigma}(\z)}}
\def\1mom{{\Gamma^{\rho\sigma}_1(\z)}}
\def\2mom{{\Gamma^{\rho\sigma}_2(\z)}}

\def\d2{\frac{D_2}{\delta t}}
\begin{document}

\title
{Stochastic degrees of freedom in postquantum classical gravity}

\author{Jonathan Oppenheim, Muhammad Sajjad}
\affiliation{Department of Physics and Astronomy, University College London, Gower Street, London WC1E 6BT, United Kingdom}

\begin{abstract}
We study fluctuations of the metric in the postquantum theory of classical gravity, a covariant theory which couples a classical spacetime with quantum matter fields. Mathematical consistency requires spacetime to evolve stochastically. Starting from the classical-quantum path integral, we linearize around Minkowski space and perform a scalar-vector-tensor decomposition, identifying the stochastic degrees of freedom: a classical spin-2 field and spin-0 scalar, both diffusing around their respective wave equations. These are related to the degrees of freedom found in quadratic gravity, but are here interpreted as stochastic contributions to spacetime. We show that the action is positive semi-definite (PSD) on all dynamical modes, which is a necessary condition for the theory to consistently treat spacetime as classical.  There is also a non-dynamical vector field and the Newtonian potential which is the non-dynamical scalar field. We compute the two-point function and power spectral density corresponding to fluctuations of the Newtonian potential, and compare it to the excess noise found in LISA Pathfinder. This sets a bound on one combination of the two dimensionless coupling constants of the theory, while bounds on the stochastic gravitational wave energy density in an FLRW background constrain another combination. We derive the effective action for matter distributions and find that the decoherence rate probed by table-top experiments depends on fluctuations in both the Newtonian potential $\Phi$ and the curvature perturbation $\psi$, so that such experiments bound a combination of the two.
Finally, we show consistency between different formulations of the pure gravity theory, the Onsager-Machlup form of the action, the Martin-Siggia-Rose form, and that given by stochastic differential equations.
\end{abstract}

\maketitle

\section{Introduction}
In general relativity, matter fields propagate on a dynamical background geometry, with the matter fields causing spacetime to curve via Einstein's equation. Spacetime thus serves in a dual role: both as a background for defining causal structures for quantum matter fields and as a dynamical variable influenced by those same fields. Whereas quantum field theory in curved space is well understood, it is not clear how to suitably modify general relativity so as to incorporate the back-reaction of matter fields which are now quantum in nature. Various proposals have been put forth on the strength of different underlying philosophical considerations. The dominant approach is quantum gravity, including string theory, loop quantum gravity and asymptotic safety. There are also a number of pragmatic approaches,
such as the mean field gravity approach (sometimes called ``semi-classical gravity''), which involves sourcing the geometry from the expectation value of the energy momentum tensor. This approach has a domain of applicability, but cannot be considered as a fundamental theory, since it suffers from various pathologies including superluminal signaling~\cite{gisin1989stochastic} and a breakdown in the statistical interpretation of the density matrix~\cite{page1981indirect}.
While the dominant view is that gravity is fundamentally quantum, the special status that spacetime holds within quantum field theory obliges us to consider the hypothesis that we should treat it as fundamentally classical.

While there has been considerable debate about whether there are consistent ways of coupling classical and quantum systems, a number of examples have been demonstrated since the mid-1990s, including via a master-equation~\cite{blanchard1993interaction,diosi1995quantum} and via a measurement and feedback approach~\cite{diosi1998coupling,kafri2014classical,tilloy2016sourcing}. In~\cite{oppenheim2023postquantum,oppenheim2022two} the most general form of consistent classical-quantum dynamics was derived. The resulting dynamics preserves positivity and normalization of probabilities and maps the state of the classical subsystem to a classical state and the state of the quantum subsystem to a quantum one. At longer distance scales, the back-reaction onto the classical degrees of freedom acts as if the quantum system is being weakly measured, resulting in decoherence, while at shorter distance scales, the dynamics can be highly non-Markovian with the classical degrees of freedom evolving stochastically. The two parameters (governing diffusion and decoherence) are thus related to each other via an inequality termed the `diffusion-decoherence' trade-off~\cite{diosi1995quantum,oppenheim2023gravitationally}. That any consistent classical-quantum coupling must be fundamentally irreversible has also been established using general probabilistic theories~\cite{galley2023any}.

The measurement and feedback approach has been used to describe Newtonian gravity~\cite{kafri2014classical,tilloy2016sourcing,tilloy2017principle}, and a theory coupling quantum fields to general relativity has been presented via a master-equation approach~\cite{oppenheim2023postquantum}, a trajectories approach~\cite{layton2024healthier} with Layton and Weller-Davies, and a manifestly covariant classical-quantum path-integral with Weller-Davies~\cite{oppenheim2023covariant}. In~\cite{grudka2024renormalisation} the formal renormalizability of the gravitational path integral was demonstrated, which stands in contrast to the non-renormalizability of perturbative quantum gravity. This further motivates probing the quantum vs classical nature of spacetime via experiment.
The aforementioned decoherence-diffusion trade-off allows one to tightly constrain the space of feasible experimental bounds~\cite{oppenheim2023gravitationally,janse2024current}, allowing one to possibly rule out classical gravity if the trade-off is experimentally found to be violated. Proposed tests which would witness the quantum nature of gravity include gravitationally-mediated entanglement experiments~\cite{bose2017spin,marletto2017gravitationally,carney2019tabletop} and non-entanglement witnesses of quantum gravity~\cite{lami2024testing,kryhin2025distinguishable}.

One of the aims of the present work is to reformulate the theory in such a way as to constrain its parameters via observation. Constraints coming from tabletop interference experiments~\cite{gerlich2011quantum} and precision gravity measurements such as LISA Pathfinder~\cite{armano2024indepth,janse2024current} have been derived in the Newtonian limit of the theory~\cite{oppenheim2023gravitationally,layton2023weak,grudka2024renormalisation} via the decoherence-diffusion trade-off. Constraints from the production of stochastic gravitational waves have been estimated via dimensional analysis~\cite{penington2024}, and from using the stochastic Klein-Gordon equation, in work with Panella, as a model~\cite{oppenheim2025diffusion}. Stochastic violation of the Hamiltonian constraint has been shown to lead to the appearance of phantom cold dark matter, in work with Panella and Pontzen, in a Friedmann–Lemaître–Robertson–Walker model~\cite{oppenheim2024emergence}, which may give constraints from considering stochastic fluctuations as a candidate for dark matter.

Contemporaneous studies \cite{hirotani2026testing,fabiano2026minimal} have also constrained the parameters of post-quantum gravity for various models.
More generally however, understanding relativistic extensions of hybrid dynamics, particularly in the case of gravity has been a challenge. Much work has been done on relativistic toy models~\cite{oppenheim2023covariant,oppenheim2024diffeomorphism,diosi2022there, grudka2024renormalisation,carney2024classicalquantum}, but the full theory, or even its linearized versions, have additional subtleties.

A significant issue is one of interpretability. Historically, there have been two different approaches towards the study of gravity: the original, geometro-dynamical one, and the particle physicist one where, starting from a Lagrangian describing a massless spin-2 particle around a fixed spacetime background, one is able to recover general relativity. Thus, the fundamental object for particle physicists in the path integral formulation of quantum gravity (with the Einstein-Hilbert action) is the spin-2 particle, with its two degrees of freedom. However, this is not necessarily the case in alternative theories of gravity; for example all the degrees of freedom of the spin-2 particle are excited in massive gravity, whereas in quadratic gravity one encounters a spin-0 particle as well as an additional ghost
particle. What, then, are the degrees of freedom for classical-quantum gravity, and how do they relate to the Newtonian potential? The first objective of the present paper is to understand what the degrees of freedom are.

Secondly, and not unrelatedly, we address the main obstacle to writing a classical gravitational path integral with the interpretation of a transition probability amplitude: the fact that the gravitational action we consider is unbounded from below. Were the unstable mode to represent a physical degree of freedom, it would leave us in the unenviable position of perturbing around a saddle-point, and would result in an indefinite two-point function. Since the two-point function has the interpretation of a covariance matrix in a classical stochastic theory, it must be positive semi-definite (PSD). This suggests a tension in the theory that must be resolved. The problem of unboundedness is not without precedent: in Euclidean Einstein-Hilbert gravity, the conformal mode has a wrong-sign kinetic term, rendering the action unbounded from below and the naive path integral divergent~\cite{gibbons1978path}. Various contour and measure prescriptions have been proposed; in particular, Mazur and Mottola argued that the conformal mode is tied to the constraint structure of the theory and does not propagate, so that the physically relevant spin-2 part of the action is positive definite~\cite{MAZUR}. Whether this fully resolves the conformal factor problem remains an active question~\cite{marolf2022gravitational}.

Here we show that if we perturb around flat spacetime, there exists a gauge-invariant, positive semi-definite sector that describes a massless, dynamical, classical spin-2 field undergoing diffusion. Additionally, we obtain a new dynamical, stochastic spin-0 field. We then address the issue of the unbounded sector, which corresponds to an off-shell vector mode, and show that it does not correspond to degrees of freedom, but to the momentum constraint. 

Thirdly, we use our understanding of the different stochastic modes of the path integral, to place bounds on the parameters of the theory from observational constraints. We relate the spin-0 and the spin-2 fields to the Newtonian potential, and compute the two-point function for the Newtonian field \eqref{eq:newtontwo} as well as the power spectral density of the acceleration, Eq.~\eqref{eq:spectral}. This is then used to place an upper bound on the diffusion constants, 
 using results from LISA Pathfinder~\cite{Armano_2024}. This improves upon results from \cite{grudka2024renormalisation} through an analysis of how the different coupling constants and scalar fields contribute. Another upper bound is obtained in Eq. \eqref{eq:tensorbound} by computing the energy-density of stochastic gravitational waves. We then obtain a lower bound from the minimum rate of decoherence (Section~\ref{effective}), and show that the theory is consistent with current observational constraints. We note that the decoherence rate of Section~\ref{effective} is non-Markovian due to coupling with the curvature scalar field $\psi$.

The paper is structured as follows: in Section \ref{CQPI} we briefly review the classical quantum path integral. In Section \ref{Linearized} we give the two-point function in terms of Barnes-Rivers projectors onto the spin-0 and spin-2 modes, and in Section \ref{SVT} in terms of a Scalar-Vector-Tensor (SVT) decomposition of the metric. We show that the dynamical degrees of freedom have positive semi-definite two-point functions. In particular, the action is PSD for the tensor mode corresponding to gravitational waves, and for the scalar sector. This is the condition we require in order to consistently interpret the action as giving transition probabilities for classical stochastic fields. We distinguish these dynamical modes of the theory from those found in the Einstein-Hilbert action and in Quadratic Gravity \cite{stelle1977renormalization,salvio2018quadratic,alvarez-gaume2016aspects,hell2023degreesfreedomr2gravity} in Subsection \ref{DOF}. In Subsection \ref{norm} we show that the only part of the action which is not PSD corresponds to a constraint equation for the vector mode. \MS{As a result, the path integral for the vector mode does not give transition probabilities in phase space, and we argue for a modification of this sector such that the resulting action gives a normalizable and well-behaved path integral.}

In Section~\ref{Exp} we constrain the parameter space of the theory using gravitational observations, as well as table-top decoherence experiments. Our bounds from gravitational waves are consistent with prior toy models as well as the study of
 Hirotani and Matsumura~\cite{hirotani2026testing}, which appeared while this draft was being prepared. We are additionally able to distinguish bounds which apply to the propagating modes, from those which place bounds on the constraints.
 Hirotani and Matsumura also introduce a noise model whose vacuum sector gives a theory related to that found in stochastic gravity~\cite{hu2008stochastic,moffat1997stochastic} and thus can describe a theory which is designed to be fundamentally quantum, but has a domain of validity which admits a consistent and non-local classical limit.

 Inspired by their work, we demonstrate our local path integral may be produced by a noise covariance tensor that satisfies the Bianchi identity. We also
 resolve a discrepancy between the MSR and Onsager-Machlup formalisms of the theory, and between these two and the stochastic differential equation (SDE) formalism.
This is done in the Appendix. In Subsection~\ref{sec:OMJD}, we discuss the Martin Siggia Rose (MSR) form of the theory. In subsections \ref{GPIM}, \ref{discretization},\ref{Jacobian} we review results on the gravitational path integral, as well as on classical stochastic path integrals in general, while in Appendices~\ref{Section:Poles} and \ref{Details} we review calculational details for Subsection~\ref{Spectral}. Finally, Appendix~\ref{Barnes-Rivers} of the appendix defines the Barnes-Rivers projectors that we heavily use.

\subsubsection{Conventions}
We use the mainly plus $(-,+,+,+)$ metric, and natural units, $c=\hbar=1$. Expressions of the form $\frac{1}{\Box},\frac{1}{\triangle}$ denote the inverses of these operators, given in the full and spatial Fourier domains by $-\frac{1}{p^2}$ and $-\frac{1}{k^2}$ respectively.

\section{The Classical Quantum Path Integral}
\label{CQPI}

A classical-quantum system may be described by a quantum density matrix living in a Hilbert space, conditioned on a classical system being at a point in classical phase space. The state of the classical system may be represented by its probability distribution, $p(z,\dot{z},t)$, where we obtain $p(z,\dot z,t)$ by tracing over the Hilbert space of the classical quantum (CQ) state, $\hat{\varrho}(z,\dot{z},t)$. The classical degrees of freedom can also be taken to live in phase space. The dynamics must, by consistency, preserve the phase space and map a CQ state to another, ensuring that classical degrees of freedom are mapped to classical degrees of freedom, and the quantum subsystem to a quantum subsystem. Since the CQ state is positive and normalized, the dynamics must be completely positive and trace-preserving (CPTP). In~\cite{oppenheim2023path} a path integral formulation for such dynamics was introduced, the key features of which we briefly summarize below.

The classical-quantum (CQ) state of a system is given by
\begin{align}
    \hat{\varrho}(z,\dot z,t):=\int d\phi^+ d\phi^- \varrho(\phi^+,\phi^-,z,\dot z,t)|\phi^+\rangle\langle\phi^-|
\end{align}
with the components, $\varrho(\phi^+,\phi^-,z,\dot z,t):=\langle\phi^+|\hat{\varrho}(z,\dot z,t)|\phi^-\rangle$, where $\phi^+,\phi^-$ represent the ket and bra fields, respectively, and can depend on $z,\dot z,t$. The evolution of these components is governed by a path integral:
\begin{align}
 \varrho(\phi^{+},\phi^{-},z_f,\dot{z_f},t_f)= \frac{1}{\mathcal{N}} \int D\phi^{+}D\phi^{-}Dz e^{-\mathcal{I}_{CQ}}\varrho(\phi^{+},\phi^{-},z_i,\dot{z_i},t_i)
\end{align}
where $\frac{1}{\mathcal{N}}$ is a normalization constant and $\mathcal{I}_{CQ}$ is of the following form:
\begin{align}
 \mathcal{I}_{CQ}=-i\mathcal{I}_Q(\phi^+,\phi^-)+\mathcal{I}_{FV}+\mathcal{I}_{Diff}
\end{align}
Here $i\mathcal{I}_Q(\phi^+,\phi^-)$ governs the unitary dynamics of the bra and ket fields, $\mathcal{I}_{FV}$ are Feynman-Vernon, or Schwinger-Keldysh terms which cause decoherence, and $\mathcal{I}_{Diff}$ is the diffusive part of the action which in general incorporates quantum back-reaction on a classical field. Actions containing the first two terms, $i\mathcal{I}_Q(\phi^+,\phi^-)+\mathcal{I}_{FV}$ are used in a formalism known as the Schwinger-Keldysh or Feynman-Vernon formalism~\cite{feynman1963theory}, and are widely used in the study of open quantum systems. As a simple example, we may consider the following action \cite{grudka2024renormalisation},
\begin{align}
    \mathcal{I}_{CQ}=\int^{t_f}_{t_i} dt-i((\dot\phi^+)^2-(\dot\phi^-)^2)+\frac{D_0}{8}((\phi^+)^2-(\phi^-)^2)^2+\frac{1}{2D_2}(\ddot{z}-\frac{D_1}{2}((\phi^+)^2+(\phi^-)^2))^2
\end{align}
Here, $\phi$ and $z$ can also be taken to be fields, in which case $dt\rightarrow d^4x$. The imaginary part corresponds to $i\mathcal{I}_Q(\phi^+,\phi^-)$ while the term $\frac{D_0}{8}((\phi^+)^2-(\phi^-)^2)^2$ is the Feynman-Vernon term which acts to decohere the field with a strength governed by $D_0$. The last term is a classical quantum version of the Onsager-Machlup (OM) functional, a probability density functional used to determine transition probabilities for stochastic processes \cite{onsager1953fluctuations,Chaichian2018PathI} and has the form of the square of an equation of motion. As $D_2$ decreases, deviations from the equations of motion are suppressed. These deviations are stochastic kicks to the acceleration $\ddot{z}$. If $D_2$ is not positive (or PSD in the case of a matrix of equations), then the action is unbounded, and paths which have large deviations from the equations of motion dominate. Finally, $D_1$ controls the strength of the back-reaction.

This generalizes to the field theoretic case, where the primary objects are the classical and quantum fields~\cite{oppenheim2023covariant}. In the gravitational case, we have the following action:
 \begin{equation}
 \label{eq:PQG-action}
\begin{split}
     \mathcal{I}_{CQ}[ g_{\mu \nu}, T^{\mu\nu+}, T^{\mu\nu-}]& = \int d^4x \bigg[ - i\sqrt{-g}\big(\mathcal{L}_{Q}[T^{\mu\nu+}] - \mathcal{L}_{Q}[T^{\mu\nu-}]\big) +\frac{\sqrt{-g}}{8}( T^{\mu \nu + } - T^{\mu \nu - }) D_{0,\mu \nu \rho \sigma}(T^{\rho \sigma+ } - T^{\rho \sigma- }) \\
    & + \frac{\sqrt{-g}}{2}\bigg( G^{\mu \nu} +\Lambda g^{\mu\nu} - 8\pi G_N\frac{T^{ \mu \nu +} + T^{ \mu \nu -}}{2}\bigg)D_{2, \mu \nu \rho \sigma}^{-1}\bigg( G^{\rho \sigma} +\Lambda g^{\rho\sigma} - 8\pi G_N\frac{T^{ \rho \sigma +} + T^{ \rho \sigma -} }{2}\bigg)\bigg].
    \end{split}
\end{equation}

This action governs the time evolution of the system through
\begin{align}
 \varrho(T^{+},T^{-},g,t_f)= \frac{1}{\mathcal{N}}\int D\varphi^{+}D\varphi^{-}D'g e^{-\mathcal{I}_{CQ}}\varrho(T^{+},T^{-},g(z_i,\dot{z_i},t_i))
\end{align}
where the stress-energy tensor $T^+,T^-$ is a function of $\varphi^+,\varphi^-$, and $G$ is the Einstein tensor for spacetime metric $g$, and the gravitational path integral is over configurations, up to diffeomorphisms (see Appendix~\ref{GPIM}). Let us break down the action into its constituent parts. The imaginary part of the action corresponds to the unitary action for the stress energy tensor, in component form, $T^{\mu\nu+},T^{\mu\nu-}$, while the term $( T^{\mu \nu + } - T^{\mu \nu - }) D_{0,\mu \nu \rho \sigma}(T^{\rho \sigma+ } - T^{\rho \sigma- })$ is a Feynman-Vernon term which acts to decohere the density matrix by causing off-diagonal terms to exponentially decay. The last term represents a classical-quantum version of the Onsager-Machlup (OM) functional. The classical OM functional is of the form: equations of motion (EOMs) squared, which in this case are Einstein's equations. The classical-quantum version has the dynamics of the classical system $G^{\mu\nu}$ sourced by the average $\bar T^{\mu\nu}:=\frac{1}{2}(T^{\mu\nu+}+T^{\mu\nu-})$ of the bra and ket fields. The classical field thus diffuses around its equations of motion due to a stochastic force, with the covariance of the stochastic force given by the diffusion matrix $\mathcal{D}_2$, with components $D_{2}^{\rho\sigma,\mu\nu}$. One could instead write a more general, possibly non-local diffusion matrix, $\mathcal{D}_2'(x,y)$. In this work, however, we will stick to a local diffusion matrix of the form $\mathcal{D}_2(x,x'):=\mathcal{D}_2\delta^4(x-x')$ for reasons of general covariance.

Additionally, we can characterize the strength of the classical-quantum interaction in the OM functional by the matrix $\mathcal{D}_1=8\pi G_N$. If we now set $\mathcal{D}_2=\mathcal{D}_1\mathcal{D}_0^{-1}\mathcal{D}_1$, as we will do moving forward, no cross terms of the form $T^{+}T^{-}$ survive, and quantum states retain their purity, conditioned on the classical trajectory. The bra fields evolve independently of the ket fields. This is an example of saturating the inequality, which we call the diffusion-decoherence trade-off, $\mathcal{D}_0\geq 64\pi^2G_N^2\mathcal{D}_2^{-1}$~\cite{oppenheim2023covariant}. As long as we are to have a completely positive dynamics, this trade-off must be preserved~\cite{oppenheim2023gravitationally}.

Our primary object will be the inverse diffusion matrix, which will determine our action. Now we would like $\mathcal{D}_{2}^{-1}(x,x')$ to have the following properties:
\begin{enumerate}
    \item Positivity: The matrix must be positive semi-definite (PSD) so that it is a bona-fide covariance matrix. Covariance matrices must be PSD, i.e., have non-negative eigenvalues, which follows from the fact that the variance is the sum of squares and must be non-negative. Another way to see this is that the matrix must be PSD in order to suppress deviations from the equations of motion in the OM action.
    \item General covariance: This implies that the diffusion matrix does not introduce any structure additional to the background metric and is only delta-correlated in time and space, i.e., through $\delta^4(x-x')$. This implies two conditions: (a) that the noise kernel is ultra-local, so that the noise contains no spatial or temporal derivatives in its covariance matrix. and (b) that it determines the geometry by determining the metric up to gauge transformations.
\end{enumerate}

One of the proposals of \cite{oppenheim2023covariant} was to take it to be proportional to the Generalized DeWitt metric~\cite{dewitt1967quantum}
\begin{align}
\label{eq:dewitt}
\mathcal{D}_2^{-1 \rho\sigma,\mu\nu}=(\alpha\frac{g^{\rho\mu }g^{\sigma\nu}+g^{\rho\nu}g^{\sigma\mu}}{2}-\beta g^{\mu\nu}g^{\rho\sigma})\delta^{4}(x-x')
\end{align}
which is indefinite for non-zero $\alpha$, and so, on the face of it, does not appear to satisfy condition (1). This may be contrasted with two proposals: (i) that $\alpha$ is set to zero, which gives an infinite diffusion rate for the spin-2 sector; and (ii) that $C^2$, the square of the Weyl-tensor, is set to evolve deterministically, which then leads, for example, to an infinite decoherence rate for tensorial matter. Since both of the alternative proposals lead to infinite rates of diffusion or decoherence, we will continue with the DeWitt metric and show that it is indeed positive on all dynamical diffusive modes.

In {Appendix~\ref{sec:OMJD}}, we will consider the stochastic differential equation form of Einstein's equations, and will introduce another diffusion matrix for which we may relax the second condition, but which gives us the same action as below.

\section{Spin-Projector Representation}
\label{Linearized}
\label{SpinProjector}
 Here, we explicitly compute the two-point function around the Minkowski background for our choice of diffusion matrix and show that it consists of spin-0 and spin-2 parts.
 The gravitational two-point function must be positive semi-definite if it is to indeed be a genuine covariance matrix, and here we find that the spin-0 part is PSD, but the spin-2 part is only PSD for non-tachyonic modes. The consequences of this indefiniteness will be discussed in Section~\ref{DOF}, where we argue that it is benign.

The purely gravitational `action' is given by
\begin{equation}
    \label{eq:LGR}\mathcal{S}^{(g)}=\frac{1}{2}\int d^4xd^4y\sqrt {-g} G_{\mu\nu}(x)(\mathcal{D}_2^{-1})^{\mu\nu,\rho\sigma}(x,y)G_{\rho\sigma}(y)=\frac{1}{2}\int d^4x\sqrt{-g} (\alpha R_{\mu\nu}R^{\mu\nu}-\beta R^2).
\end{equation}
This action is rather similar in form to that of quadratic gravity~\cite{stelle1978classical} \cite{salvio2018quadratic}
\begin{align}
\label{eq:QuadGrav}
    \mathcal{S}_{QG}=\frac{1}{2}\int d^4x\sqrt{-g}( \alpha R^{\mu\nu}R_{\mu\nu}-\beta R^2+\gamma R^{\mu\nu\rho\sigma} R_{\mu\nu\rho\sigma}-\frac{M^2}{2}R)
;\end{align} where $R^{\mu\nu\rho\sigma} $
is the Riemann tensor. Since one combination of the terms above gives us the topological Gauss-Bonnet term, which is a total derivative, \eqref{eq:QuadGrav}
is equivalent to \eqref{eq:LGR} up to the $R$ term. However, the presence of that term, which corresponds to the Einstein-Hilbert term in quadratic gravity and to the cosmological constant in classical-quantum gravity, spoils the PSD-ness of the spin-0 sector and will thus be omitted.

We also note that $C^2:=C_{\mu\nu\rho\sigma}C^{\mu\nu\rho\sigma}$, the Weyl-squared term, may be written as $\frac{1}{2}C^2=\frac{1}{2}R^{\mu\nu\rho\sigma}R_{\mu\nu\rho\sigma}-R_{\mu\nu}R^{\mu\nu}+\frac{1}{6}R^2$, or up to the Gauss-Bonnet term as $\frac{1}{2}C^2=R_{\mu\nu}R^{\mu\nu}-\frac{1}{3}R^2$. Note then that taking the limit $\alpha\rightarrow3\beta$ in \eqref{eq:LGR}
gives an action similar in form to that of pure Weyl-squared gravity.

Expanding \eqref{eq:LGR} around Minkowski space to second order in $h$, we obtain the following Lagrangian density:
 \begin{equation}
\begin{split}
\label{eq:Mink}
   2\mathcal{L}^{(g)}&= \frac{\alpha}{4}(\partial_{\lambda}\partial_\mu h^{\lambda}_{\nu}+\partial_{\lambda}\partial_\nu h^{\lambda}_{\mu}-\partial_\mu \partial_\nu h-\Box h_{\mu\nu})(\partial^{\sigma}\partial^\mu h_{\sigma}^{\nu}+\partial^{\sigma}\partial^\nu h_{\sigma}^{\mu}-\partial^\mu \partial^\nu h-\Box h^{\mu\nu})-\beta(\partial_{\mu}\partial_{\nu}h^{\mu\nu}-\Box h)^2\\&=h^{\mu\nu}[\frac{\alpha}{2}(\frac{1}{2}I_{\mu\nu,\rho\sigma}\Box^2+\frac{1}{2}\eta_{\mu\nu}\eta_{\rho\sigma}\Box^2-\eta_{\nu\beta}\Box\partial_{\rho}\partial_{\mu}-\frac{1}{2}\Box\eta_{\mu\nu}\partial_{\rho}\partial_{\sigma}-\frac{1}{2}\Box\eta_{\rho\sigma}\partial_{\mu}\partial_{\nu}+\partial_{\rho}\partial_{\sigma}\partial_{\mu}\partial_{\nu})\\&-\beta(\eta_{\rho\sigma}\eta_{\mu\nu}\Box^2-\eta_{\mu\nu}\Box\partial_\rho\partial_\sigma-\eta_{\rho\sigma}\Box\partial_{\mu}\partial_\nu+\partial_{\mu}\partial_{\nu}\partial_{\rho}\partial_\sigma)]h^{\rho\sigma}\\&=h^{\mu\nu}H_{\mu\nu,\rho\sigma}h^{\rho\sigma}
\end{split}
\end{equation}
It is convenient to write $\mathcal{H}$ in terms of the Barnes-Rivers spin projectors (Appendix~\ref{Barnes-Rivers}):
\begin{align}
   \label{eq:H}H_{\mu\nu,\rho\sigma}=\Box^2(\frac{\alpha}{4}(P^{(2)}+4P^{(0-s)})-3\beta P^{(0-s)})_{\mu\nu,\rho\sigma}=\Box^2(\frac{\alpha}{4}P^{(2)}+(\alpha-3\beta)P^{(0-s)})_{\mu\nu,\rho\sigma}
\end{align}
The three dimensional version of the spin-2 projector, $P^{ij,kl}$ \eqref{eq:projector3D} is a familiar sight in classical gravity where it is used to project out the transverse-traceless part of the spatial part of the metric perturbation, $h_{ij}$, to give us the two transverse, traceless modes corresponding to gravitational waves. In four dimensions, one obtains instead five transverse traceless modes, which one may write in terms of their polarization tensors \cite{Spin2}. After gauge-fixing and inverting \eqref{eq:H}, one has, up to gauge-fixing terms, the following two-point function:
\begin{align}
    \label{eq:Twopointfunction}
 \langle h_{\rho\sigma }h_{\mu\nu}\rangle=\frac{4}{\alpha\Box^2}(P^{(2)}+\frac{P^{(0-s)}}{4(1-3\frac{\beta}{\alpha})})_{\rho\sigma,\mu\nu}
\end{align}
One can easily check that the spin-0 part is positive definite for $\alpha>3\beta$\footnote{$\alpha\geq0$ throughout.}. Explicitly,
\begin{equation}
    \frac{1}{(\alpha-3\beta)\Box^2}S^{\mu\nu} P^{(0-s)}_{\mu\nu,\rho\sigma}S^{\rho\sigma} = \frac{1}{3(\alpha-3\beta)}\,(\frac{S^{\mu\nu}\theta_{\mu\nu}}{\Box})^2.
\end{equation}
Note however, that for the full trace sector to be positive definite (which we refer to as positivity on all ten components), we have the condition, $\alpha>4\beta$. The more stringent condition is however superfluous as long as diffeomorphism invariance is preserved: regardless, we explore this as a more strict alternative.

For the spin-2 part to be PSD, we require
\begin{align}
 S_{\rho\sigma}P^{\rho\sigma,\mu\nu(2)}S_{\mu\nu}\geq 0
 \label{eq: PSD}
\end{align}
for all $S_{\rho\sigma}$.
Acting on a tensor $S^{\mu\nu}$ with the projection operator takes us to the (symmetric) transverse traceless subspace, i.e., $\textbf{P}S=S'$,
\begin{align}
 S_{\rho\sigma}P^{\rho\sigma,\mu\nu}S_{\mu\nu}=S'_{\rho\sigma}S'^{\rho\sigma}
 \label{eq: PSD2}
\end{align}
\begin{align}
\label{eq: transverse}
    p_\rho S'^{\rho\sigma}=0, \quad \eta^{\rho\sigma}S'_{\rho\sigma}=0.
\end{align}
One can easily show for any 2 tensor that satisfies the above relation for all $p^2\leq0$
that the quadratic invariant,
\begin{align}
 S_{\mu\nu}'S'^{\mu \nu} \geq 0
\end{align}
To see this, first, note that for any tensor $\textbf{S}$ in Minkowski space,
\begin{align}
    S'^{\mu\nu}S'_{\mu\nu}=(S'^{00})^2-2\sum_i(S'^{0i})^2+\sum_{ij}(S'^{ij})^2.
\end{align}
Thus, it is the $S^{'0i}$ term which could potentially spoil PSD-ness. By~\eqref{eq: transverse} and the fact that $p^2\leq0$, there exists a coordinate system in which $p=(1,0,0,0)$, and for that co-ordinate system, $S'^{0\mu}=0$, and so $S'^{\mu\nu}S'_{\mu\nu}\geq0$ in any coordinate system.
For a null vector $p$, the Barnes-Rivers projectors seem to be undefined. However, as long as the source tensor $S^{\mu\nu}$ is conserved, i.e., $p_\mu S^{\mu\nu}=0$, the projector is in fact well-defined. Again, we can work in a coordinate system where $p=(1,0,0,\pm 1)$. Then, we see that
\begin{align}
    \begin{split}
 S'^{0i}&=\pm S'^{3i}\\ S'^{00}&=\pm S'^{30}\\ (S'^{00})^2+2\sum_i(-(S'^{0i})^2+(S'^{3i})^2)-(S^{30})^2+\sum_{i,j\neq3}(S'^{ij})^2&= \sum_{i,j\neq3}(S'^{ij})^2\ge0 .
    \end{split}
    \end{align}
On the other hand, if we wish to remove this restriction from the source tensor, the projection operator will have to be suitably altered to ensure transversality:
\begin{equation}
\label{eq:thetanull}
    \theta^{\mu \nu}=\eta^{\mu\nu}-\frac{p^{\mu}n^{\nu}+p^{\nu}n^{\mu}}{p^{\alpha}n_\alpha},
    \qquad
    \omega^{\mu \nu}=\frac{p^{\mu}n^{\nu}+p^{\nu}n^{\mu}}{p^{\alpha}n_\alpha}
\end{equation}
where $n^2=p^2=0$ and $p^{\alpha}n_\alpha\neq0$. These satisfy
$\theta+\omega=\eta$, $\theta^2=\theta$, $\omega^2=\omega$, $\theta\omega=0$ and
$p_\mu\theta^{\mu\nu}=n_\mu\theta^{\mu\nu}=0$ and give \begin{equation} 
\label{eq:spin2null}
    P^{(2)}_{\mu\nu,\rho\sigma}=\tfrac{1}{2}\bigl(\theta_{\mu\rho}\theta_{\nu\sigma}
    +\theta_{\mu\sigma}\theta_{\nu\rho}\bigr)-\tfrac{1}{2}\,\theta_{\mu\nu}\theta_{\rho\sigma},
   \qquad
    P^{(0-s)}_{\mu\nu,\rho\sigma}=\tfrac{1}{2}\,\theta_{\mu\nu}\theta_{\rho\sigma} .
\end{equation}

One can also check that this method does not work for spacelike $p$. Setting $p=(0,0,0,1)$ in some coordinate system, we have $p_\rho S'^{\rho\sigma}=0$, i.e. $S'^{3\sigma}=0$. Thus, now the best we can say is that $S'^{30}=0$. More generally, one can only use a space-like vector to draw relations between the various components of $S'^{0i}$, and so can say nothing about the relative size of the troublesome term, $\sum_i(S'^{0i})^2$, which may be arbitrarily large. Thus, in this case, $S'^{\mu\nu}S'_{\mu\nu} $ is indefinite.

Thus, for linearized gravity, the spin-2 sector has a positive semi-definite correlation function only for timelike and null Fourier modes, unlike the spin-0 sector, which is positive semi-definite for all modes. However, in what follows, we will show using a scalar-vector-tensor decomposition that the actual degrees of freedom (which we will define in due course) have two-point functions that are PSD.

\section{The Scalar-Vector-Tensor Decomposition}
\label{SVT}
In this section, our aim will be to identify the true degrees of freedom. To this end we will perform a standard scalar-vector-tensor decomposition of the metric, and consequently of the action, and show that the scalar and tensor sectors are PSD. In addition, we will show that the indefinite vector sector, in our framework, does not represent a degree of freedom.

To perform the scalar-vector-tensor decomposition around Minkowski space, we write the perturbed metric as
\begin{equation}
   g_{\mu\nu}^{(1)}= -(1-2\phi)dt^2 + 2(B_{,i}+S_i)dtdx^{i}+((1+2\psi)\delta_{ij}+2E_{,ij}+F_{i,j}+F_{j,i}+h_{ij}^{TT})dx^{i}dx^{j}
\end{equation}
where the vector modes, $S_i,F_{i}$, are transverse, while the tensor modes, $h_{ij}^{TT}$, are both transverse and traceless.
For infinitesimal coordinate changes, $x^{\mu}\rightarrow x^{\mu}+\xi^{\mu}$, one has the following transformation of modes:
\begin{align}
\phi&\rightarrow \phi-\dot\xi_0 & B&\rightarrow B-\xi_0-\dot \zeta
      & S_i&\rightarrow S_i-\dot{\xi}_i^{T} &h^{TT}_{ij}&\rightarrow h^{TT}_{ij}\\ \psi&\rightarrow\psi & E&\rightarrow E-\zeta & F_{i}&\rightarrow F_{i}-\xi_i^{T}
\end{align}
where $\xi_\mu=(\xi_0,\xi_i^{T}+\zeta_{,i})$. We can then combine the different modes to form the following gauge-invariant modes:
\begin{align}
\begin{split}
    \Phi&=\phi+\ddot{E}-\dot{B} \\V_{i}&=S_{i}-\dot{F_i}
    \end{split}
\end{align}
where
\begin{align}
S_{i,i}&=F_{i,i}=h^{TT}_{ii}=h^{TT}_{ij,i}=0
\end{align}
The pure gravity action, to quadratic order, is then given by
\begin{align}
\mathcal{L}^{(g)}=\mathcal{L}_{tensor}+\mathcal{L}_{vector} +\mathcal{L}_{scalar}
\end{align}
with:
\begin{equation}
    \label{eq:tensor}
\mathcal{L}_{tensor}=\frac{\alpha}{8} h^{ij{TT}}\Box^2h_{ij}^{TT}
\end{equation}
\begin{equation}
\label{eq:vector}
    \mathcal{L}_{vector}=-\frac{\alpha}{4}V_i\Box \Delta V^{i}
\end{equation}
\begin{align}
\begin{split}
\label{eq:scalar}
    \mathcal{L}_{scalar}&=\frac{1}{2}\Big((12\alpha-36\beta)(\partial^2_0\psi)^2+(2\alpha-4\beta)(\triangle\Phi)^2+(6\alpha-16\beta)(\triangle\psi)^2+ (-16\alpha+48\beta)(\triangle\psi\partial_0^2\psi)\\&+(8\alpha-24\beta)(\triangle\Phi\partial_0^2\psi)+(-4\alpha+16\beta)\triangle\Phi\triangle\psi\Big).
\end{split}
\end{align}
Taking the Newtonian gauge amounts to setting $\Phi=\phi, V_i=S_i$.

Clearly, the tensor sector is positive semi-definite, while the vector one is not, since the $\Box$ operator is 
negative on spacelike modes. 
To check for the scalar sector, we pass to Fourier space (the finite-time case is similar and does not alter the result) ($\partial_0^2\to -\omega^2$, $\triangle\to -k^2$, $p^2\equiv -\omega^2+k^2$) and write \eqref{eq:scalar} as a quadratic form in the column vector $(\psi,\Phi)^T$:
\begin{align}
\label{eq:scalarkernel}
\mathcal{L}_{scalar} = \frac{1}{2}\begin{pmatrix}\psi \\ \Phi\end{pmatrix}^T K \begin{pmatrix}\psi \\ \Phi\end{pmatrix}, \qquad K = \begin{bmatrix} K_{\psi\psi} & K_{\psi\Phi} \\ K_{\Phi\psi} & K_{\Phi\Phi}\end{bmatrix}
\end{align}
where, reading off the coefficients from \eqref{eq:scalar},
\begin{align}
\begin{split}
K_{\psi\psi} &= (12\alpha-36\beta)\omega^4+(-16\alpha+48\beta)k^2\omega^2+(6\alpha-16\beta)k^4 \\
K_{\Phi\Phi} &= (2\alpha-4\beta)k^4 \\
K_{\psi\Phi} = K_{\Phi\psi} &= (4\alpha-12\beta)k^2\omega^2+(-2\alpha+8\beta)k^4
\end{split}
\end{align}
The matrix $\mathcal{M}=K^{-1}$ is then obtained by the standard $2\times2$ inversion,
with $\det K = 8\alpha(\alpha-3\beta)k^4 p^4$:
\begin{align}
\label{eq:twopointscalar}
  \mathcal{M}= \frac{1}{8\alpha(\alpha-3\beta)k^4p^4} \begin{bmatrix}
    (2\alpha-4\beta)k^4 & -k^4(-2\alpha+8\beta)-k^2\omega^2(4\alpha-12\beta)\\ -k^4(-2\alpha+8\beta)-k^2\omega^2(4\alpha-12\beta)&k^4(6\alpha-16\beta)+k^2\omega^2(-16\alpha+48\beta)+\omega^4(12\alpha-36\beta)
\end{bmatrix}
\end{align}
As long as $\mathcal{M}$, and by extension its inverse, is PSD for all Fourier modes, the scalar action is PSD. We can easily check if this matrix is PSD: if the determinant and a given diagonal element are positive for this matrix, then this matrix is PSD in Fourier space. The determinant of this matrix is given by
\begin{align}
\begin{split}
  \mathrm{Det}[\mathcal{M}]&=\frac{1}{8k^4p^4\alpha(\alpha-3\beta)} \\ M_{00}&={(2\alpha-4\beta)k^4} \mathrm{Det}[\mathcal{M}]
\end{split}
\end{align}
Then, the condition for positive definiteness is
\begin{align}
	\label{eq:alphabeta}
\begin{split}
     3\beta &< \alpha
     \end{split}
\end{align}
where $\alpha$ is a positive constant, and $\beta$ may take negative values.

Note that at second order, the action contains no time derivatives of $\Phi$, and so $\Phi$ is non-dynamical.
In order to obtain the effective scalar action, $\mathcal{L}_{eff}$, which determines the dynamics of the scalar system, we vary the action with respect to $\Phi$ to obtain
\begin{align}
\label{eq:constraint}
    2(2\alpha-4\beta)\triangle^2\Phi+(8\alpha-24\beta)\triangle\partial_0^2\psi+(-4\alpha+16\beta)\triangle^2\psi=0
\end{align}
which is analogous to the \textit{Hamiltonian constraint}\cite{DiracHamiltonianGR,arnowitt1959dynamical}. \MS{Here we note that this equation of constraint is only satisfied on expectation: $\Phi$ \emph{diffuses} around its equation of constraint.}

In the Einstein-Hilbert case, the constraint is simply $\triangle\psi=0$. Here, $\Phi$, which was only a Lagrange multiplier in the Einstein-Hilbert action, is constrained instead of the now dynamical field $\psi$). The solution of $\Phi$ may be inserted back into the action to obtain
\begin{align}
\label{eq:MPP}\mathcal{L}_{eff}=\frac{2\alpha(\alpha-3\beta)}{\alpha-2\beta}\psi\Box^2\psi.
\end{align}
The scalar action for $\psi$ then has the same form as the action for the tensor modes and is clearly PSD whenever 
the condition \eqref{eq:alphabeta} holds. This procedure is equivalent to integrating out $\Phi$. 

We thus see that the scalar and tensor sectors are PSD, while the vector sector is indefinite. In addition, we can now move towards identifying the true degrees of freedom, which we will see correspond to $\psi$ and $h^{TT}_{ij}$.

\subsection{The Degrees of Freedom}
\label{DOF}
Before discussing gravitational degrees of freedom in our stochastic path integral, it is of benefit to recall the method of identifying degrees of freedom in quantum (Einstein-Hilbert and quadratic) gravity, so as to better illustrate procedural differences in the two cases and their physical nature. In the case of Einstein-Hilbert gravity, we have the action (up to constants)
\begin{align}
    \label{eq:Einstein-Hilbert}
    \mathcal{S}=\int d^4x\sqrt{-g}R=\int d^4x \mathcal{L}^{{EH}}_{scalar}+\mathcal{L}^{{EH}}_{vector}+\mathcal{L}^{{EH}}_{tensor}
\end{align}
where in particular,
\begin{align}
\mathcal{L}^{{EH}}_{scalar}=(2\Phi\triangle\psi+3\ddot{\psi}\psi-\psi\triangle\psi)
\end{align}
\begin{align}
    \mathcal{L}^{{EH}}_{vector}=-\frac{1}{2}V_i\triangle V^i
\end{align}
\begin{align}
    \mathcal{L}^{{EH}}_{tensor}=\frac{1}{4}h_{ij}^{TT}\Box h^{ijTT}
\end{align}
We then vary the scalar action with respect to $\Phi$ and substitute the constraint, $\triangle\psi=0$, back in the action. From this we have that the entire scalar sector is constrained. A similar analysis of the vector and tensor sectors confirms that the only degrees of freedom are those of the tensorial sector. Thus, the procedure is to identify the constrained variable, and use the constraint to eliminate spurious modes. Similarly, in quadratic gravity, the constrained modes (which do not have time-derivatives acting on them) are integrated out to identify the actual degrees of freedom.

The procedure for our classical stochastic theory is slightly different. Here the `action',~\eqref{eq:LGR} is \textit{not} the actual Lagrangian of the gravitational sector; rather, it is the Onsager-Machlup (OM) action, representing diffusion around the gravitational equations of motion. Thus, identifying a degree of freedom is no longer equivalent to integrating out constrained modes and then counting the number of $\partial_0^2$ operators that survive in the effective OM action; instead, one must perform an initial data count to correctly identify the degrees. An equation of motion (EOM) obtained from the OM action, implies the diffusion of a degree of freedom around its EOM, while a constraint equation implies that the constrained mode diffuses around its constraint equation, and is not an independent degree of freedom. This is the case for $\Phi$ since the action contains no time-derivatives of $\Phi$ at second order.

One can see this by referring to the probability density function (pdf). The pdf, $\pdf(q_f,p_f|q_i,p_i)$ allows one to compute conditional transition probabilities, given a particular equation of motion. That is, suppose one has a second order equation of motion. Then, one must specify four pieces of boundary data for the conditioned pdf. This, however, does not mean that one now has additional degrees of freedom; instead, one is simply calculating a probability for reaching a particular final configuration. One can do this since the motion is non-deterministic and so, setting initial conditions does not determine the final ones.

The situation becomes clearer when one instead considers the unconditioned pdf. Now, one must specify two pieces of information at the initial time slice, as in the deterministic case, while integrating over all possible final conditions. Thus, a variable represents a true degree of freedom if and only if one can non-trivially assign two initial conditions for the unconditioned pdf, or equivalently four initial and final conditions for the conditioned pdf. 

Let us now apply our analysis to $\Phi$. Starting from the conditioned pdf of the scalar sector, $ \pdf(\Phi(T),\dot{\Phi}(T),0,0;\allowbreak\psi(T),\dot{\psi}(T),0,0;k)$ one has
\begin{align}
    \pdf(\Phi(T),\dot{\Phi}(T),0,0;\psi(T),\dot{\psi}(T),0,0;k)=\int D\Phi D\psi\exp{-\int dt \mathcal{L}(|\vec{k}|,t)_{scalar}}.
\end{align}
where we have worked in spatial Fourier space.

One can then `diagonalize' the density
\begin{align}
\label{eq:Diagonal}
   \mathcal{L}_{scalar}(|\vec{k}|,t)=(\alpha-2\beta)k^4\Phi'^2+\frac{2\alpha(\alpha-3\beta)}{\alpha-2\beta}((\partial_0^2+k^2)\psi)^2
\end{align}
using the following field re-definition: $\Phi'=\Phi-\frac{2(\alpha-3\beta)\partial_0^2+(\alpha-4\beta)k^2}{(\alpha-2\beta)k^2}\psi$. Since $\Phi'=\Phi-\langle\Phi\rangle_\psi$ is the part of $\Phi$ not determined by $\psi$, $\langle\Phi'\psi\rangle=0$ exactly and the density factorises allowing us to treat $\Phi'$ and $\psi$ as independent even though $\Phi'$ contains $\psi$. The \emph{OM} action is stationary for $\Phi'=0$, i.e., in the pure $\Phi$ direction. Unlike GR, $\Phi$ is not a Lagrange multiplier here since $K_{\Phi\Phi}\neq0$; it becomes one only at $\alpha=2\beta$ (which is however ruled out by positivity). With this field redefinition, the pdf changes to
\begin{align}
    \pdf=\int JD\Phi'D\psi\exp{-\int dt \mathcal{L}_{scalar}(\Phi',\psi)},
\end{align} where $J$ is the Jacobian of the field transformation (and being independent of $\Phi',\psi$ may be absorbed in the normalization). We may then split the action in a semi-classical, $\mathcal{L}(\Phi'_{cl})$, and a fluctuation, $\mathcal{L}(\tilde{\Phi}')$, part, where $\tilde{\Phi}'$, by definition, vanishes at the boundaries. It is easy to see that the semi-classical action of $\Phi'$ is zero. Endpoints of $\Phi'$ may be chosen arbitrarily but have no relation to the field in the interior, as the action for $\Phi'$ simply factorizes.

Meanwhile, the semi-classical action for $\psi$ admits an initial and final value formulation. Integrating over all final conditions then gives us the unconstrained Green's function, $\mathcal{G}$ for the operator $(\partial_0^2+k^2)^2$ subject to the boundary conditions, $\mathcal{G}(0)=\dot{\mathcal{G}}(0)=\hat{L}\mathcal{G}(t_f)=\dot{\hat{L}}\mathcal{G}(t_f)=0$ where $\hat{L}=(\partial_t^2+k^2)$. That is, we set the initial conditions but leave the final conditions to be free, subject to the fact that the stochastic force and its derivatives do not act at the final boundary. One can always set this condition by taking the final boundary to $t_f\rightarrow\infty$, since $\mathcal{G}$ is independent of $t_f$ for $t,s<t_f$. Note that $\mathcal{G}$ is in fact identical to the convolution of retarded and advanced Green's functions for the operator, $\hat{L}$. In momentum space, one thus finds, after making the change of variables, $t-s\rightarrow\tau,\frac{t+s}{2}\rightarrow T$
\begin{align}
    \mathcal{G}(\tau,T;k)=\frac{1}{4k^3}\left((2T-|\tau|)k\cos(k|\tau|)-\sin(2kT)+\sin(k|\tau|)\right).
\end{align}

One can then read off the inverse matrix in terms of the original variables $\psi,\Phi$. This is:
\begin{align}
\label{eq:twopointscalar2}
  \mathcal{M}= \begin{bmatrix}
    \frac{(2\alpha-4\beta)}{8\alpha(\alpha-3\beta)}\mathcal{G} & \frac{2(\alpha-3\beta)\,\partial_{0}^{2}+(\alpha-4\beta)\,k^{2}}
  {4\alpha(\alpha-3\beta)\,k^{2}}\mathcal{G}\\\ \frac{2(\alpha-3\beta)\,\partial_{0}^{2}+(\alpha-4\beta)\,k^{2}}
  {4\alpha(\alpha-3\beta)\,k^{2}}\mathcal{G}&\left(\frac{(\alpha-4\beta)+2(\alpha-3\beta)\,\partial_{0}^{2}/k^{2}}{\alpha-2\beta}\right)^{\!2}
  \frac{(\alpha-2\beta)}{4\alpha(\alpha-3\beta)}\,\mathcal{G}
  \;+\;\frac{1}{k^{4}\,(2\alpha-4\beta)}
\end{bmatrix},
\end{align}
where we have suppressed the $t,s$ arguments. Note that this is the matrix~\eqref{eq:twopointscalar} adapted to the finite-time case, and hence a function of the time and the spatial momentum.

From~\eqref{eq:MPP} or \eqref{eq:Diagonal} we thus see that we have a single degree of freedom in the scalar sector, which diffuses around the wave equation. This is the stochastic Klein-Gordon equation~\cite{oppenheim2025diffusion,balan2010stochastic},
\begin{align}
\label{SDE:scalar}
  \Box\psi=\xi
\end{align}
where $\xi$ is a mean-zero random noise process, with covariance given by
\begin{align}
\langle\xi\xi\rangle\propto\frac{\alpha-2\beta}{2\alpha(\alpha-3\beta)}
\end{align}
and similarly for the two tensor modes:
\begin{align}
\label{SDE:tensor}
    \Box h_{ij}^{TT}=\xi_{ij}^{'TT}
\end{align}
where again $\xi_{ij}^{'TT}$ is another mean-zero random noise process, with covariance given by
\begin{align}
\langle\xi^{ij'TT}\xi^{'kl TT}\rangle\propto\frac{1}{\alpha}P^{ij,kl}
\end{align}
where \begin{align}
\begin{split}
      \label{eq:projector3D} P^{ij,kl}=&\frac{1}{2}(P^{ik}P^{jl}+P^{il}P^{jk})-\frac{1}{2}P^{ij}P^{kl}\\P^{ij}=&\delta^{ij}-\frac{k^{i}k^{j}}{k^2}
\end{split}
\end{align}

On the other hand, we will see that the vectorial density does not correspond to a degree of freedom, but to the \textit{momentum constraint}~\cite{arnowitt2008republication}. The vectorial action~\eqref{eq:vector} may be written in spatial frequency space as
\begin{align}
\mathcal{L}=\frac{\alpha}{4}(({k\dot{V}})^2-k^2(kV)^2)
\end{align}
One may recognize this as the action for a Brownian particle in a quadratic potential, written in terms of the scaled variable $u:=kV_i$, in which $\mathcal{L}=\frac{\alpha}{4}(\dot{u}^2-k^2u^2)$. The potential term in this case is inverted, and in fact satisfies the following differential equation for the Feynman-Kac weight $w(u_i(t),t;u_i(0))$:
\begin{align}
    \partial_tw=\frac{1}{\alpha}\partial^2_{u}w+\frac{\alpha k^2}{4} u^2w
\end{align}
where $w$ satisfies the initial condition $w(u_i,0)=\delta(u(0)-u_i)$. Note, however, that this indicates the presence of a probability source instead of a probability sink in the un-inverted case.
Thus, even for convergent $|\vec{k}|$, the vector sector results in the probability for large spacelike $V$ growing without bound in the sense that $w$ grows with time. Consequently, to ensure that probability is conserved, one would have to normalize this weight for \textit{all} convergent values of $k$. However, the critical time, at which the normalized weight ceases to exist, and can consequently no longer have a pdf interpretation is at $t_{c}=\frac{\pi}{2k}$. This follows from the solution of the above equation,
\begin{align}
    w(u,t;u_i)=\sqrt{\frac{\alpha k}{4\pi\sin kt}}\,\exp\Big[-\frac{\alpha k}{4\sin kt}\big((u_i^2+u^2)\cos kt-2u_iu\big)\Big],
\end{align}
in which the coefficient of $u^2$ in the exponent, $-\frac{\alpha k}{4}\cot kt$, changes sign at $kt=\pi/2$; the variance of the normalized weight, $\frac{2}{\alpha k}\tan kt$, diverges there. If the total time of evolution, $t\rightarrow\infty$, 
the vector sector can have a pdf interpretation only if the potential term is zero.

On the other hand, the fact that it has fewer time derivatives indicates that it corresponds to a constraint rather than a dynamical degree of freedom. The reason is that one may only assign an initial and final field value in the conditioned 'pdf' or equivalently, only an initial field value to $V_i$ in the unconditioned 'pdf'. One cannot set both an initial field value as well as an initial momentum without over-determining the Green's function of $V_i$. The initial data count therefore gives us three degrees of freedom: two in the tensor, and one in the scalar sector.

One may once more compare this with the degrees of freedom in classical Einstein-Hilbert gravity. There we only have two degrees of freedom, whereas we now have diffusion around an additional scalar degree of freedom $\psi$, while we expect that the vector potential must be fixed in some way. One can see why instead of diffusion around only the tensorial degrees of freedom, we now have an additional, diffusive degree of freedom by looking at the tensorial two-point function,
\begin{align}
\langle h^{TTij}h^{TTkl}\rangle\propto P^{ij,kl}/\Box^2=\lim_{m\rightarrow0}\frac{P^{ij,kl}}{m^2}(\frac{1}{\Box-m^2}-\frac{1}{\Box}).
\end{align}
We can always write this or the corresponding action as that of two helicity-2 `modes', one massless and the other massive, in the $m\rightarrow0$ limit: this is, however, only a partial-fraction representation of the double pole, and nothing in the theory has a mass. The essential point is that the double
  pole changes which helicities propagate. Off shell, the spin-2 projector $P^{(2)}$ contains helicity-0 and helicity-$\pm1$ components as well as
  helicity-$\pm2$ ones. With a single pole $1/p^2$ these components, when coupled to a conserved source, give only contact terms, and the massless
  graviton carries helicity $\pm2$ alone. With the double pole $1/p^4$ the factor of $p^2$ no longer cancels the pole: the helicity-$\pm1$ components
  acquire a simple pole, and the helicity-0 components, which here combine with the spin-0 term with a coefficient different from $-\tfrac12$, no longer
  cancel. Hence the additional time derivatives in the scalar and vector sectors, e.g.\ the vector action $\propto V_i\triangle\Box V^i$ in place of
  $V_i\triangle V^i$ in general relativity.

\subsection{Non-dynamical modes: $\Phi$ and $V_i$}
\label{norm}

The OM action determines the status of each SVT field through the number of its time derivatives: four ($\psi$, $h^{TT}_{ij}$) correspond to dynamical stochastic differential equations, zero ($\Phi$) to an algebraic constraint, and two ($V_i$) to a first-order constraint, i.e., initial data on phase space. In this section, we explore the consequences of this by showing that the path integral for fields with two or zero time derivatives in the action factorizes over time slices (in phase and configuration space, respectively), failing to enforce temporal continuity. As a result, $\Phi$ and $V_i$ are specified by two constraints. However, this does not mean that $\Phi$ is discontinuous in configuration space: the apparent discontinuity is an artifact of the truncation of the action at quadratic order in the potentials.

While in quantum gravity, the unstable mode is associated with the Hamiltonian constraint, here, it is the vector mode whose action ~\eqref{eq:vector} is not PSD. In particular, it would appear that we need to restrict the path integral for the field $V^i$ to non-tachyonic modes, since the action converges only for $\omega^2\geq k^2$. This restriction is however insufficient: zero weight on $k^2 \geq \omega^2$ implies infinite diffusion in that region, and a divergent gaussian integral.

The first thing to note is that the action for $V_i$ has only second time-derivatives and corresponds to a pure constraint equation. While a second time derivative corresponds to dynamical degrees of freedom for a quantum theory, it does not for an action of OM form, which is an equation of motion squared. The action Eq.~\eqref{eq:MPP} for $\psi$ contains a term $\int d^4x(\Box\psi)^2$, with four time derivatives, whereas Eq.~\eqref{eq:vector} for the vector mode has only two. The fact that $\psi$ becomes dynamical is the result of two conditions, the first being that the Hamiltonian constraint is no longer required to vanish, and the second being that the dynamical part of the action has second time derivatives of $\psi$ as can be seen from Eq.~\eqref{eq:scalar}. This is not the case for the action of $V^i$ which represents equations of constraints in phase space. We have already seen that $\Phi$ is non-dynamical.

To understand the case of $\Phi$ better, it is worthwhile to consider the continuity of the action, given that no $\dot{\Phi}^2$ term appears at leading order. We will find that continuity in $\Phi$ is enforced at higher order. Consider a stochastic equation of constraint in which the configuration of a mode is tied directly to the noise. For such a mode $q_c$, constrained in configuration space, the naive transition probability no longer enforces the continuity condition   \begin{equation}
  \label{eq:continuity}
\lim_{\epsilon\rightarrow0}\big[q_{c}(t+\epsilon)-q_{c}(t)\big]=0 ,
  \end{equation}
  understood as a limit in probability. This holds even when the action is PSD. A simple example suffices to illustrate our argument. Starting with a linear equation of constraint
\begin{align}
\begin{split}
    q_{c}(t)&=\xi
\end{split}
\end{align}
we have the following propagator, $K(q_t,t_f|q_i,t_i)$ for $\pdf(q)$, the classical probability density of the system:
\begin{align}
\begin{split}
    \label{eq:ToyPI}
K(q_t,t_f|q_i,t_i)&=\int^{\infty}_{-\infty} Dq\exp{-\int dt q^2}\\&=\int^{\infty}_{-\infty} \prod_{k=1}^{N-1} dq_k\exp{-\sum_k\epsilon  q_k^2}\\&=\prod_{k=1}^{N-1} \int^{\infty}_{-\infty}dq_k\exp{-\epsilon q_k^2}
\end{split}
\end{align}
where $q_i$ and $q_f$ indicate the initial and final positions, and $q_k$ is the value of $q$ at each discretized time step $k$. The path integral factorizes over $q_k$ and thus $q_{k+1}$ is uncorrelated with $q_{k}$, even in the limit $\epsilon\rightarrow0$. Continuous paths appear in the path integral with measure zero. This is hardly surprising: $q$ is directly tied to white noise, which is only defined in a distributional sense, and is delta correlated in time, properties that $q_c$ inherits. To enforce continuity in $q_c$ without introducing additional dynamics, one could, for example, add a delta function to the measure, $\delta(q_{c,k+1}-q_{kc,})$ at every time slice, resulting in deterministic evolution, which is equivalent to setting $T$ in $\langle\xi(t) \xi(t')\rangle=T\delta(t,t')$ to zero, i.e., a zero-temperature heat bath.

Contrast this with the dynamical case, where the action, $S_{dyn}$, is of the form $\mathcal{S}_{dyn}=\frac{1}{2D}\sum_k\epsilon (\frac{q_{k+1}-q_k}{\epsilon}-F(q_k))^2$, where $D$ is the diffusion constant. Due to the time-derivative, the action couples adjacent time slices rather than factorizing over single slices, and in the short time limit
$\epsilon\rightarrow0$ the transition weight tends to $\delta(q_{k+1}-q_{k})$, so that $q_{k+1}-q_{k}=O(\sqrt{D\epsilon})\rightarrow0$ as required.

As a second example, we take a stochastic form of Newton's equation in vacuum, such as that appearing in other stochastic theories of gravity~\cite{hu2008stochastic,moffat1997stochastic,tilloy2016sourcing}
\begin{align}
\begin{split}
\triangle\phi(t)&=\xi(t)\\\langle\xi(t)\xi(t')\rangle&=2D_N\delta(t-t')
\end{split}
\end{align}
with a PSD diffusion matrix, $D_N$. The probability distribution at the spatial slice $\Sigma_{t_f}$ at final time, given the distribution at the initial slice, $\Sigma_{t_i}$, then reads \begin{align}
\begin{split}
    \pdf(\phi(\Sigma_{t_f}))=&\int D\phi\exp{-\frac{1}{4D_N}\int d^4x (\triangle\phi)^2} \pdf(\phi(\Sigma_{t_i})).
\end{split}
\end{align}
It is easy to see that spatially connected fields contribute to the path integral with measure $1$. However, the path integral factorizes in time, and so the field is discontinuous in time.

Let us now consider $\Phi$. Prima facie, it seems that $\Phi$ may only be defined distributionally in time: the $K_{\Phi\Phi}$ has no time derivatives acting on it and $\Phi=\Phi'+\langle\Phi\rangle_{\psi}$ and both $\Phi'$ as well as $\partial_0^2\psi$ are directly tied to white noise. At any instant, the variance in $\Phi$, $\langle\Phi(t)\Phi(t)\rangle\sim\delta(0)$ and therefore would need smearing with a smooth function. However, at 4th order in the potentials, one has terms of the form $S(\dot\Phi)^2$, where $S$ is a function of other potentials. This term has the correct sign, and thus, is sufficient to render $\Phi$ continuous, \emph{conditional} on $S$, notwithstanding the fact that it occurs at fourth order. Therefore, the apparent discontinuity of $\Phi$ is an artifact of the linearized theory. This however, has no bearing on the $\omega$ independent experimental results.

\subsubsection{The Vector Sector}
From the above discussion, it is clear that a) the vector sector is non-dynamical and b) it is a normalizable weight only for $t\leq t_c=\frac{\pi}{2k}$ in the unconditioned case. For an evolution time, $t\rightarrow\infty$, the cut-off momentum would be unacceptably low. In the vacuum case, one could possibly impose the momentum constraint, $\triangle V_i=0$, in effect limiting the noise to a subset of the spin-2 projector, $P^{(2)}$. However, such a restriction would seemingly inherit the non-covariance of the helicity decomposition (or equivalently SVT). 

Further, non-normalizability is not the only issue with the vector sector. The real part of the vectorial sector, with matter added in, is given by
\begin{equation}
\mathcal{L}_{\rm vector}=\frac{\alpha}{4}\bigg[\Big(\dot V_{i,j}
+16\pi G_N\triangle^{-1}\dot{\bar q}_{i,j}\Big)^{2}
-\Big(\triangle V_i+16\pi G_N\bar q_i\Big)^{2}\bigg]
+\frac{\alpha}{4}\bigg[\Big(8\pi G_N\triangle^{-1}\Delta\dot q_{i,j}\Big)^{2}
-\Big(8\pi G_N\Delta q_i\Big)^{2}\bigg].
\end{equation}
where $q_i$ comes from the vectorial sector of the energy-momentum tensor.
The second term, in the second bracket indicates that re-coherence is favored for $q_i$: off diagonal elements grow over time. Therefore, even if the gravitational vector sector is somehow fixed, the evolution would no longer be positive. To fix this, one can restrict the path integral to exclude tachyonic $V$, at the expense of having a non-Markovian noise kernel. This would be done by either fixing $V_i$ for those modes, or setting it free: either choice appears to be problematic for non-relativistic $q_i$. One can also construct a PSD noise-kernel by introducing a preferred frame, or additional field. Current lorentz-symmetry violation (LSV) searches need not be sensitive to those violations. In Appendix \ref{VectorSection}, for example, we discuss a PSD diffusion matrix, to which LSV searches are insensitive. Regardless, at this juncture it is not fully clear to us how the vectorial sector is to be handled in the path integral formulation.

\section{Experimental Implications}
\label{Exp}
Our purpose in this section will be to obtain upper and lower bounds on both diffusion constants $\alpha$ and $\beta$ in order to constrain the theory. We first compare the two-point functions for the scalar and tensor sectors and then obtain bounds for the diffusion coefficients of the two sectors from LISA Pathfinder and LIGO. The two sectors have different diffusion constants, but the scalar one effectively upper bounds the tensorial one. Finally, for a point of comparison we reproduce the effective decoherence rate
which we use to set a lower bound on the diffusion constant(s). In comparison to previous results \cite{oppenheim2023gravitationally,janse2024current,grudka2024renormalisation,oppenheim2025diffusion,hirotani2026testing}, our analysis allows us to distinguish the role of both constants $\alpha$ and $\beta$ in determining observational constraints. Our results here should be considered preliminary for the following reasons; since they rely on the linearized action, and since we do not know at what scale perturbation theory breaks down..

\subsection{Comparison between the Scalar and the Tensor Sectors}
\label{Comp}
How is the purely radiative scalar diffusion related to the tensor diffusion, and can each be used to bound the other, given the additional parameter $\beta$ in the scalar sector?

The correlation matrix for the scalar sector is given (in the $\psi,\Phi$ basis) by
\begin{align}
\label{eq:twopointscalar21}
  \mathcal{M}_{2pf}= \begin{bmatrix}
    \frac{(2\alpha-4\beta)}{8\alpha(\alpha-3\beta)}\mathcal{G} & \frac{2(\alpha-3\beta)\,\partial_{0}^{2}+(\alpha-4\beta)\,k^{2}}
  {4\alpha(\alpha-3\beta)\,k^{2}}\mathcal{G}\\\ \frac{2(\alpha-3\beta)\,\partial_{0}^{2}+(\alpha-4\beta)\,k^{2}}
  {4\alpha(\alpha-3\beta)\,k^{2}}\mathcal{G}&\left(\frac{(\alpha-4\beta)+2(\alpha-3\beta)\,\partial_{0}^{2}/k^{2}}{\alpha-2\beta}\right)^{\!2}
  \frac{(\alpha-2\beta)}{4\alpha(\alpha-3\beta)}\,\mathcal{G}+\frac{1}{k^4(2\alpha-4\beta)}
\end{bmatrix}.
\end{align}

One can decompose the $\langle\Phi\Phi\rangle$ component into frequency-dependent and frequency-independent parts: $\langle\Phi\Phi\rangle=\frac{3}{2\alpha k^4}+(1/\alpha p^4)(\frac{\omega^2}{k^2}+\frac{(-3\alpha+10\beta)}{4(\alpha-3\beta)})$ i.e., one part that is correlated across time, while the other is only delta-correlated in time. The frequency-independent part, $\frac{3}{2\alpha k^4}$ is the two-point function that we would expect from diffusion around a constraint equation of the form $\nabla^2\Phi=\xi$, $\langle\xi(x)\xi(x')\rangle\propto \frac{3}{2\alpha}\delta^3(x-x')$ and is in fact obtained as the sum of delta correlated two-point function of $\tilde{\Phi}$ in \eqref{eq:Diagonal} and the time-independent component obtained from acting with the operator in $\left(\frac{(\alpha-4\beta)+2(\alpha-3\beta)\,\partial_{0}^{2}/k^{2}}{\alpha-2\beta}\right)^{\!2}$ on $\mathcal{G}$ in \eqref{eq:twopointscalar2}. For $\beta=(3/10)\alpha$ the purely radiative part ($\frac{(-3\alpha+10\beta)}{4\alpha(\alpha-3\beta)p^4}$) disappears, but a frequency-dependent part still survives.

 From~\eqref{eq:scalar} one notes that in the limit $(\alpha-3\beta)\rightarrow0$, the kernel $K$ becomes degenerate, $\det K\rightarrow0$, so that the two-point function is singular. One can also see this directly from the two-point function~\eqref{eq:twopointscalar} by expanding $\mathcal{M}_{2pf}$ in $\alpha-3\beta$,
\begin{align}
    \lim_{(\alpha-3\beta)\rightarrow0^+}\mathcal{M}_{2pf}=\frac{1}{12(\alpha-3\beta) p^4}\begin{bmatrix}
    1 & -1\\ -1&1
\end{bmatrix}+O((\alpha-3\beta)^0)
\end{align}
The other limit is where $\beta \rightarrow-\infty$:
\begin{align}
    \lim_{\beta\rightarrow-\infty}\mathcal{M}_{2pf}=\frac{1}{6\alpha k^4p^4}\begin{bmatrix}
    k^4 & 2k^4-3k^2\omega^2\\ 2k^4-3k^2\omega^2&(2k^2-3\omega^2)^2
\end{bmatrix}
\end{align}
Clearly, for most of its allowed parameter space, the amount of radiative scalar diffusion is directly proportional to that of tensorial diffusion, before diverging as $(\alpha-3\beta)\rightarrow0$. Thus, the scalar bound constrains the tensor one as long as only transverse positivity is required (under full positivity the  stronger tensor bound also constrains the scalar). Experiments based on the amount of tensorial diffusion, on the other hand, may only be used to set an upper bound on the value of the diffusion constant, $\alpha^{-1}$, which is relevant only for those frequency branches in the scalar sector that are independent of $(\alpha-3\beta)$. In conclusion, it is to experiments based on tensorial diffusion that we must turn to obtain the most stringent bounds on noise.

\begin{figure}[htbp]
\centering
\begin{tikzpicture}
\begin{axis}[cqaxis, xmode=log,
  xmin=0.05, xmax=10, ymin=2.40, ymax=6.40,
  xtick={0.05,0.1,0.25,0.5,1,2,4,10},
  xticklabels={\ensuremath{0.05},\ensuremath{0.1},\ensuremath{\tfrac14},
               \ensuremath{\tfrac12},\ensuremath{1},\ensuremath{2},
               \ensuremath{4},\ensuremath{10}},
  ytick={2.6667,3,4,5,6},
  yticklabels={\ensuremath{\tfrac83},\ensuremath{3},\ensuremath{4},
               \ensuremath{5},\ensuremath{6}},
  ylabel={$D_{2,\,scalar}/D_{2,\,tensor}$},
  title={Low-frequency scalar vs. tensor diffusion in parameter space},
]
  \fill[cqred,  opacity=0.10] (axis cs:0.05,2.40)   rectangle (axis cs:0.25,6.40);
  \fill[cqblue, opacity=0.09] (axis cs:0.25,2.6667) rectangle (axis cs:10,4);

  \draw[cqgreyb, dotted, line width=0.5pt]
        (axis cs:0.05,2.6667) -- (axis cs:10,2.6667);
  \draw[cqblue,  dotted, line width=0.5pt]
        (axis cs:0.05,4)      -- (axis cs:10,4);
  \draw[cqred,   dashed, line width=0.6pt]
        (axis cs:0.25,2.40)   -- (axis cs:0.25,6.40);

  \addplot[cqblue, line width=1.6pt, domain=-1.0492:1, samples=300]
        ({10^x}, {(1+8*10^x)/(3*10^x)});
  \addplot[cqblue, only marks, mark=*, mark size=1.6pt]
        coordinates {(0.25,4)};

  \node[cqblue,  font=\scriptsize] at (axis cs:0.068,5.55)
        {$\sim\tfrac{1}{3x}\to\infty$};
  \node[cqred,   font=\tiny, rotate=90] at (axis cs:0.085,3.30)
        {forbidden if $\beta\leq\alpha/4$};
  \node[cqgreyb, font=\tiny, rotate=90] at (axis cs:0.155,3.30)
        {allowed if only $\alpha\geq3\beta$};
  \node[cqred,   font=\scriptsize, anchor=west] at (axis cs:0.265,6.25)
        {$\alpha=4\beta$};
  \node[cqblue,  font=\small, anchor=west] at (axis cs:0.62,5.35)
        {$\dfrac{D_{2,\,scalar}}{D_{2,\,tensor}}=\dfrac{1+8x}{3x}$};
  \node[cqblue,  font=\scriptsize] at (axis cs:3.5,3.55)
        {trapped in $[\tfrac83,4]$};
  \node[cqred,   font=\tiny, anchor=west] at (axis cs:0.052,2.50)
        {$\leftarrow\;\alpha=3\beta$};
  \node[cqblue,  font=\tiny, anchor=east] at (axis cs:9.6,2.50)
        {$\alpha\to0$ or $\beta\to-\infty\;\rightarrow$};
\end{axis}
\end{tikzpicture}
\caption{The effective scalar and tensor diffusion constants differ only through the
ratio $x=(\alpha-3\beta)/\alpha$, as
$D_{2,scalar}/D_{2,tensor}=(1+8x)/(3x)$, which is monotonically decreasing.  The
plotted axis is the whole of the region left open by positivity on the transverse
six: $\alpha=3\beta$ at the left end, $\alpha\to0$ or $\beta\to-\infty$ at the right.
Imposing positivity on all ten components deletes the shaded strip $x<\tfrac14$ and so
traps the ratio in  $(\tfrac83,4]$ --- the $4$ saturated at the boundary $\beta=\alpha/4$,
the $8/3$ approached as $\beta\to-\infty$ --- whereupon the LIGO bound
$D_{2,tensor}=\alpha^{-1}\lessapprox 3.3\times10^{-67}$ carries over to the scalar sector as
$D_{2,scalar}\lessapprox4\times10^{-66}$.  On $\alpha\geq3\beta$ alone the ratio runs
away as $1/3x$ and no gravitational-wave measurement constrains the scalar sector
at all.}
\label{fig:tensor-scalar}
\end{figure}

One may also use the low-frequency/non-relativistic regime, where the effective diffusion constant, $D_{2, scalar}:=\frac{(3\alpha-8\beta)}{\alpha(\alpha-3\beta)}$ (see the following section). Then, one can plot the ratio of $\frac{D_{2,scalar}}{D_{2,tensor}}$ over parameter space, when $D_{2,tensor}\equiv\alpha^{-1}$. In the $\alpha \rightarrow3\beta$ limit, the ratio behaves as in the scalar diffusion case, while in the ten-component positivity case, the ratio remains bounded.

Alternatively, one may consider the situation in terms of diffusion around the trace, and the traceless parts of the Einstein tensor. From $\mathcal{M}$, one notes that as $\alpha\rightarrow3\beta$ the dynamical part of $\langle\Phi\Phi\rangle$ dominates, while for large, negative values of $\beta$, the non-dynamical part is of increased importance, but the dynamical part does not disappear. Note from the action $\int d^4x (\alpha G^{\mu\nu}G_{\mu\nu}-\beta G^2)$ that the latter is the regime in which the trace part of the Einstein equations is highly constrained. One can also compare our results with the regime where diffusion takes place around the trace, $G$, and the action has the form, $-\int d^4x\beta G^2$. From the positivity condition, only negative values of $\beta$ are permissible. In this case, the correlation matrix does not exist, as the matrix of the quadratic action, $\mathcal{M}^{-1}$ is singular, indicating a breakdown in the theory, as we would expect from taking the $\alpha \rightarrow0$ limit in $\mathcal{M}$. We can thus identify two regimes in which the theory is singular: when $\alpha\rightarrow0$ and when $\alpha\rightarrow3\beta^+$. In the former regime, the tensorial action will also be singular, and the theory as a whole breaks down. The latter limit corresponds to the action $\int d^4x C^2$ where $C$ is the Weyl tensor; the scalar action now has an additional symmetry (conformal symmetry) which renders it degenerate.

\subsection{Spectral Acceleration}
\label{Spectral}
We can now use the two-point function for the scalar sector to find the variance in acceleration for a non-relativistic particle. We may then bound the diffusion coefficient using precision gravity measurements from Lisa Pathfinder. The dominant contribution to the variance in acceleration, $\langle a_ia^i\rangle$, in the non-relativistic limit, is then given by the spectral acceleration of $\langle\Phi\Phi\rangle$, i.e., $\langle\partial_i\Phi(x)\partial^i\Phi(y)\rangle$, where we note that in the absence of matter, $\langle a_i \rangle=0$ for the linearized action.
We first calculate the partial Fourier transform of the two-point function (in space only) and then the spectral density. We have
\begin{align}
\langle\Phi\Phi\rangle=\frac{1}{(2\pi)^3}\left(\frac{(\alpha-4\beta)+2(\alpha-3\beta)\,\partial_{0}^{2}/k^{2}}{\alpha-2\beta}\right)^{\!2}
  \frac{(\alpha-2\beta)}{4\alpha(\alpha-3\beta)}\,\mathcal{G}(t,s;|k|)+\frac{1}{(2\pi)^3}\frac{1}{(2\alpha-4\beta)k^4}
\end{align}
where $\langle\Phi\Phi\rangle$ is symmetric in $t,s$ and where $\mathcal{G}$ is, for the reassignment, $t-s\rightarrow\tau,t+s\rightarrow2T$ \begin{align}
    \mathcal{G}(\tau,T;|k|)=\frac{1}{4|k|^3}\left((2T-|\tau|)|k|\cos(|k||\tau|)-\sin(2|k|T)+\sin(|k||\tau|)\right).
\end{align}
Note that $\mathcal{G}$ is equivalent to the convolution of retarded and advanced Green's functions for the operator $\Box$ (see Appendix~\ref{Section:Poles} and \cite{oppenheim2024anomalous,oppenheim2025diffusion}).

Simplifying and computing the Fourier transform, we obtain (see Appendix~\ref{Alternate} for details)

\begin{align}
\begin{split}
\label{eq:newtontwo}
    \langle \Phi \Phi\rangle(\omega,T,r)&=\frac{(\alpha-2\beta)}{4\alpha(\alpha-3\beta)}(\frac{2T-r}{8\pi r}(\frac{\sin(\omega r)}{\omega}-\Theta(r-2T)\pi\delta(\omega)))\\&-\frac{1}{2\alpha}\frac{1}{4\pi r}(2\pi r\delta(\omega)-\frac{4\sin{^2(\frac{\omega r}{2})}}{\omega^2})-\frac{3}{2\alpha}\frac{r}{8\pi}
\end{split}
\end{align}
and the spectral acceleration:
\begin{align}
\label{eq:spectral}
\begin{split}
  -\nabla^2 \langle \Phi \Phi\rangle(\omega,T;r)&=\frac{\alpha-2\beta}{4\alpha(\alpha-3\beta)}(\frac{\omega(2T-r)\sin{(\omega r)}}{8\pi r}+\frac{\cos{(\omega r)}}{4\pi r}-\frac{\delta(\omega)\delta(r-2T)}{16T})\\&-\frac{1}{2\alpha}\frac{\cos{(\omega r) }}{2\pi r}+\frac{3}{2\alpha}\frac{1}{4\pi r}
\end{split}
\end{align}

 One can integrate this over test masses with uniform density $\density(x)=\frac{M}{V}$, to compute the covariance of the force exerted by the Newtonian potential:
\begin{align}
\boldsymbol{E}(F_iF^i)(\omega)= \int_0^{R_2} d^3x\int_0^{R_1}d^3y\density(\vec{x})\density(\vec{y})\boldsymbol{E} (\partial_i \Phi(\vec{x}) \partial^i\Phi(\vec{y}))
\end{align}
Matters may be simplified by taking the low-frequency limit of $-\nabla^2\langle \Phi\Phi\rangle$ before integrating it over a test mass:
\begin{align}
    \label{eq:acceleration_potential}\lim_{\omega\rightarrow0}-\nabla^2\langle \Phi \Phi\rangle=\frac{1}{64(\alpha-3\beta)\alpha\pi}(\frac{4(3\alpha-8\beta)}{|x-y|}+ \omega^2(4(\alpha-4\beta)|x-y|+(2\alpha-4\beta)(x^0+y^0))+O(\omega^4))
\end{align}
Then, we obtain,
\begin{align}
    \boldsymbol{E}(F_iF^i)(\omega)=\frac{1}{64(\alpha-3\beta)\alpha\pi}(\frac{24(3\alpha-8\beta)M^2}{5R}+ \omega^2(4(\alpha-4\beta)\frac{36M^2R}{35}+2(2\alpha-4\beta)TM^2)+O(\omega^4))
\end{align}
where we assume spherical geometry and $T=:\frac{x^0+y^0}{2}$.

The trace $\langle\partial_i\Phi\,\partial^i\Phi\rangle=-\nabla^2\langle\Phi\Phi\rangle$ is however, not what an accelerometer measures along a fixed axis, so one needs the full tensor. 
Using $\partial/\partial y^j=-\partial/\partial r^j$, we have $\langle\partial_i\Phi(\vec x)\,\partial_j\Phi(\vec y)\rangle=-\partial_i\partial_j\langle\Phi\Phi\rangle$, i.e.
\begin{align}
\label{eq:transverse11}
\langle\partial_i\Phi(\vec x)\,\partial_j\Phi(\vec y)\rangle=\frac{1}{16\pi\alpha(\alpha-3\beta)}\left[\frac{3\alpha-8\beta}{2}\,\frac{\delta_{ij}-\hat r_i\hat r_j}{r}
+\omega^2\left(\frac{\alpha-4\beta}{4}\,r\,\big(\delta_{ij}+\hat r_i\hat r_j\big)+\frac{(\alpha-2\beta)\,T}{3}\,\delta_{ij}\right)\right]+O(\omega^4),
\end{align}
whose trace reproduces Eq.~\eqref{eq:acceleration_potential}. At leading order the correlator is transverse to $\vec r$: its component along the separation vanishes. The $T$-dependent part is independent of $\vec x$ and $\vec y$.

One can compare this result with the noise results obtained during the LISA Pathfinder (LPF) mission. The purpose of the mission was to show that a test-mass could be placed in near perfect free-fall, with the noise-levels below a certain threshold. For that purpose, the acceleration of a test mass, called TM1, was measured relative to a second test mass whose centre lies $L_0=37.6\,$cm away, along the sensitive $x$-axis joining their centres of mass, and its one-sided power spectral density computed~\cite{armano2024indepth}. Three branches were noted in the power spectral density (PSD): a low-frequency branch with an approximate $\omega^{-2}$ behavior, a frequency independent branch above 1 mHz, and a rising branch above 10 mHz. For LISA Pathfinder (LPF) the measured quantity is therefore the differential acceleration along that axis, $\langle\Delta g_x\Delta g_x\rangle (\omega)=\langle g_{1x} g_{1x}\rangle+\langle g_{2x} g_{2x}\rangle-2\langle g_{1x} g_{2x}\rangle$. The test masses are Au--Pt cubes of side $a=46\,$mm and mass $M=1.928\,$kg. Since this is an order of magnitude calculation, the spherical assumption is sufficient. We obtain for the variance in acceleration,
\MS{
\begin{align}
\langle\Delta g_x\Delta g_x\rangle(\omega)=2\,\big\langle\langle\partial_x\Phi\,\partial_x\Phi\rangle\big\rangle_{11}-2\,\big\langle\langle\partial_x\Phi\,\partial_x\Phi\rangle\big\rangle_{12} .
\end{align}
i.e.,
\begin{align}
\langle\Delta g_x\Delta g_x\rangle(\omega)&=\frac{1}{16\pi\alpha(\alpha-3\beta)}\bigg\{(3\alpha-8\beta)\left[\Big\langle\frac{1-\hat r_x^2}{r}\Big\rangle_{11}-\Big\langle\frac{1-\hat r_x^2}{r}\Big\rangle_{12}\right]
\\&+\frac{\omega^2(\alpha-4\beta)}{2}\Big[\big\langle r(1+\hat r_x^2)\big\rangle_{11}-\big\langle r(1+\hat r_x^2)\big\rangle_{12}\Big]\bigg\}+O(\omega^4).
\end{align}
Note that the average total time elapsed therefore does not appear in the differential measurement, as the term corresponding to it is position independent and thus cancels,
although it would dominate the noise of a single test mass.

Dropping the remaining $\omega^2$ term, we obtain
\begin{align}
\langle\Delta g_x\Delta g_x\rangle(\omega)=\frac{3\alpha-8\beta}{16\pi\alpha(\alpha-3\beta)}\left[\Big\langle\frac{1-\hat r_x^2}{r}\Big\rangle_{11}-\Big\langle\frac{1-\hat r_x^2}{r}\Big\rangle_{12}\right]+O(\omega^2).
\end{align}
For a sphere or a cube, we compute the self-interaction term using $\langle\hat r_x^2/r\rangle_{11}=\frac13\langle 1/r\rangle_{11}$, so $\langle(1-\hat r_x^2)/r\rangle_{11}=\frac23\langle1/r\rangle_{11}$. The cross term, on the other hand is obtained by using the fact that $(1-\hat r_x^2)/r=\partial_x^2\,r$, so $\langle(1-\hat r_x^2)/r\rangle_{12}=\partial^2_{L_x}\langle|\vec L+\vec u-\vec v|\rangle$ at $\vec L=L_0\hat x$, where $\vec u,\vec v$ range over the two bodies. Because $r$ is biharmonic away from the origin, averaging it over a uniform body shifts it by $\tfrac12\langle u_x^2\rangle\nabla^2 r=\langle u_x^2\rangle/r$, with $\langle u_x^2\rangle=R^2/5$. Hence $\langle|\vec L+\vec u-\vec v|\rangle=L+2R^2/5L$ for spheres (exact for $L>2R$). With $\partial_{L_x}^2L=(1-\hat L_x^2)/L=0$ and $\partial_{L_x}^2L^{-1}=(3\hat L_x^2-1)/L^3=2/L_0^3$,
\begin{align}
\Big\langle\frac{1-\hat r_x^2}{r}\Big\rangle_{12}=
\dfrac{4R^2}{5L_0^3}.
\end{align} Hence
\begin{align}
\label{eq:Dgxgeom}
\langle\Delta g_x\Delta g_x\rangle (\omega)=\frac{3\alpha-8\beta}{16\pi\alpha(\alpha-3\beta)}\left[\frac{2}{3}\Big\langle\frac{1}{r}\Big\rangle_{11}-\Big\langle\frac{1-\hat r_x^2}{r}\Big\rangle_{12}\right]+O(\omega^2)
=
\dfrac{3\alpha-8\beta}{20\pi\alpha(\alpha-3\beta)}\,\dfrac{1}{R}\left(1-\dfrac{R^3}{L_0^3}\right)
\end{align}
up to $O(\omega^2)$.}
Notably, the average total time elapsed does not appear in this expression, irrespective of the geometry considered. Reintroducing factors of $c$, we have
\begin{align}
\langle\Delta g_x\Delta g_x\rangle (\omega)=
\dfrac{(3\alpha-8\beta)\,c^3}{20\pi\alpha(\alpha-3\beta)}\,\dfrac{1}{R}\left(1-\dfrac{R^3}{L_0^3}\right)
 +\,O(\omega^2).
\end{align}
This is the two-sided spectral density $\int d\tau\, e^{-i\omega\tau}\langle\Delta g_x(\tau)\Delta g_x(0)\rangle$. LPF reports the one-sided PSD per Hz, which is twice the two-sided density at $\omega=2\pi f$, so
\begin{align}
\label{eq:LPFonesided}
S^{(1)}_{\Delta g}(f)=
\dfrac{c^3}{10\pi R}\left(1-\dfrac{R^3}{L_0^3}\right)D_{2,scalar},
\qquad D_{2,scalar}:=\frac{3\alpha-8\beta}{\alpha(\alpha-3\beta)} .
\end{align}

Since the prediction is white at leading order, the constraint comes from the minimum of the measured spectrum: the flat branch, $S^{1/2}_{\Delta g}=(1.74\pm0.05)\,\text{fm}\,\text{s}^{-2}\,\text{Hz}^{-1/2}$ above $2\,$mHz~\cite{armano2018beyond}, i.e.\ $S^{(1)}_{\Delta g}=3.0\times10^{-30}\,\text{m}^2\,\text{s}^{-4}\,\text{Hz}^{-1}$. Modeling the test masses as spheres gives $D_{2,scalar}\le 10\pi R\,S^{(1)}_{\Delta g}/[c^3(1-R^3/L_0^3)]$, i.e.\ $1.0\times10^{-55}$ for a sphere of the same volume ($R=28.5\,$mm):
\begin{align}
    \boxed{D_{2,\text{scalar}}\leq10^{-55}.}
\end{align}

\subsection{Waves in FLRW}
\label{Tensor}
We now upper bound the tensorial diffusion constant using bounds from LIGO. One may recall that around Minkowski space the two-point functions for the scalar and tensor sectors are non-stationary (even in the wider sense, as they do not depend solely on $(t_2-t_1)$), and indeed diverge if we integrate from $t=-\infty$. However, the addition of a finite amount of friction is sufficient to make the two-point functions stationary. We may thus consider the MSR-type path integral (see Appendix \ref{sec:OMJD}) for the tensorial sector around an FLRW background. Since the tensorial (or transverse traceless) components of the auxiliary field, $\tilde{h}$, are gauge-invariant, we can identify the correct auxiliary variable without trouble.
The tensorial action is then given by
\begin{equation}
    S_{tensor}=\int d^4x\sqrt{-g}(\tilde{h}_{ij}^{TT}\Box'{h}_{ij}^{TT}-\frac{2}{\alpha} (\tilde{h}_{ij}^{TT})^2)
\end{equation}
where
\begin{align}
\Box'=\partial_t^2+2\eta\partial_t-\triangle+2K
\end{align}
where $\eta=aH$ is the conformal Hubble rate, $\eta\simeq H_0$ today ($a=1$), $K$ is the scalar three-dimensional curvature, and $H_0=67.9\,\text{km}\,\text{s}^{-1}\,\text{Mpc}^{-1}=2.20\times 10^{-18}\,\text{s}^{-1}$. Assuming spatial flatness, we set $K=0$. The two-point function, in Fourier space is then given by
\begin{align}
\label{HTT}
    \langle h^{TT}_{ij}(p)\,h^{TT}_{kl}(p')\rangle=(2\pi)^4\delta^4(p+p')\,\frac{4\alpha^{-1} P_{ij,kl}}{(k^2-\omega^2+2i\omega\eta)(k^2-\omega^2-2i\omega\eta)}
\end{align}
where $P^{ij,kl}$ is given by~\eqref{eq:projector3D}.
\MS{  It is simple enough to obtain the one-sided energy density per frequency:
  \begin{align}
  \rho_{GW}(\omega)=\frac{c^4}{8\pi G_N}\,\frac{1}{8\pi}\,\lim_{\vec x'\rightarrow \vec x}\Big(\frac{\omega^2}{c^2}-\nabla^2\Big)\langle
  h^{TT}_{ij}h^{TT\,ij}\rangle(\omega,\vec x-\vec x')
  \end{align}
  where $x^0=ct$, the factor $\frac{1}{8}$ makes $\frac{c^4}{8\pi G_N}\frac18\langle\partial_0h\,\partial_0h+\partial_kh\,\partial_kh\rangle$ equal to the
  Isaacson energy density on shell, and $\frac{1}{\pi}$ converts the two-sided spectral density (per $d\omega/2\pi$) into a one-sided density per $d\omega$.
  The spatial Fourier transform is given by
  \begin{align}
   \langle h^{TT}_{ij}h^{TT\,ij}\rangle(\omega,\vec{x}-\vec{x'}) \approx \frac{\alpha^{-1}\,e^{-{\eta}|x|}}{\pi \eta\omega
  |x|}\sin\Big({|x|}\big(\omega+\frac{\eta^2}{2\omega}\big)\Big)
  \end{align}
  The energy density per frequency in units of $\text{kg}\,\text{m}^{-1}\,\text{s}^{-1}$ is given by
  \begin{align}
      \mathcal{\rho}_{GW}(x)(\omega)\approx\frac{\alpha^{-1}c^2e^{-\frac{\eta}{c}|x|}}{16\pi G_N(2\pi)^2\eta\frac{\omega}{c} |x|}\Big(2\omega^2\sin\Big(\frac{|x|}{c}\big(\omega+\frac{\eta^2}{2\omega}\big)\Big)+2\eta\omega\cos\Big(\frac{|x|}{c}\big(\omega+\frac{\eta^2}{2\omega}\big)\Big)\Big)
  \end{align}
  where we have restored factors of $8\pi G_N,c$.} Taking the coincident limit, $|x-x'|\rightarrow 0$, we have
\begin{align}
\label{eq:EnergyDensity}
\mathcal{\rho}_{GW}(0)(\omega)\approx\frac{\alpha^{-1}c^2}{8\pi G_N(2\pi)^2}\Big(\frac{\omega^2}{\eta}+c\frac{1}{|x|}\Big)+O(\eta)
\end{align}
  The second term is a coincident-point (ultraviolet) divergence from the gradient energy of off-shell modes $k\gg\omega$ which is neither part of the radiative energy, nor accessible to gravitational wave measurement. The Isaacson energy
  density is in any case defined only after averaging over a region large compared to
  the wavelength and small compared to the background curvature radius: the second term is negligible against the first for any short-distance cutoff $|x|\gg c\eta/\omega^2\approx3\times10^{-14}\,$m, with the ratio of the two terms being $\sim 10^{-20}$ for gravitational wave detectors. We can therefore safely drop this term. Keeping only the leading-order terms in $\eta$, the energy density per logarithmic frequency is
  \begin{align}
  \label{eq:tensorintegrated}
  \frac{d\mathcal{\rho}_{GW}}{d\ln\omega}=\omega\,\mathcal{\rho}_{GW}(0)(\omega)\approx\frac{\alpha^{-1}c^2}{8\pi G_N(2\pi)^2}\,\frac{\omega^3}{\eta}.
  \end{align}

\MS{We can now use
\begin{align}
\label{eq:OmegaGW}
\Omega_{GW}(\omega)=\frac{1}{\rho_c c^2}\frac{d\rho_{GW}}{d\ln\omega}=\frac{\alpha^{-1}\,\omega^3}{12\pi^2H_0^2\,\eta},
\end{align}
where $\rho_c=3H_0^2/8\pi G_N=8.66\times10^{-27}\,\text{kg}\,\text{m}^{-3}$ is the critical density for $H_0=67.9\,\text{km}\,\text{s}^{-1}\,\text{Mpc}^{-1}$, the value adopted by~\cite{abac2025upper} to define $\Omega_{GW}$. The spectrum~\eqref{eq:OmegaGW} is a power law $\Omega_{GW}\propto f^{3}$, which corresponds to a flat strain power spectral density. For that spectral index, \cite[Table III]{abac2025upper} gives the 95\% upper limit $\Omega_{GW}(25\,\text{Hz})\leq1.0\times10^{-9}$ under a uniform amplitude prior ($3.2\times10^{-10}$ under the log-uniform prior used for that paper's headline results), from which we obtain}
\MS{\begin{align}
\label{eq:tensorbound}
\alpha^{-1}:=D_{2,tensor}\le12\pi^2\,\Omega_{GW}(f)\left(\frac{H_0}{2\pi f}\right)^{3}\lessapprox3.3\times10^{-67}
\end{align}
($1.0\times10^{-67}$ with the log-uniform prior)}
that is, 11.5 orders of magnitude below $D_{2,scalar}$ obtained from the noise threshold in LISA Pathfinder. A consistent estimate, $D\lessapprox 10^{-68}$ (see also \cite{penington2024}), was obtained in~\cite{oppenheim2025diffusion} using the stochastic Klein-Gordon equation as a toy model for the tensor sector, and a similar bound appears in~\cite{hirotani2026testing}.

Coherence experiments on molecular superpositions lower-bound the dimensionless diffusion coefficient at $D_2\gtrapprox 10^{-71}$~\cite{oppenheim2025diffusion} via the decoherence-diffusion trade-off. This leaves a viable window of roughly four and a half orders of magnitude, $10^{-71}\lesssim D_{2, \rm tensor} \lesssim 10^{-67}$, although further theoretical work on decoherence bounds could narrow this window.

  \begin{figure}[htbp]
    \centering
    \begin{tikzpicture}
    \begin{axis}[
      scale only axis, width=0.6\linewidth, height=0.423\linewidth,
      xmin=-72.6, xmax=-49, ymin=46, ymax=75.5,
      axis lines=left,
      axis line style={cqgrey, line width=0.4pt},
      tick style={cqgrey, line width=0.4pt},
      every tick label/.append style={cqgrey, font=\scriptsize},
      label style={cqgrey, font=\footnotesize},
      tick align=outside, clip=false,
      xtick={-72,-68,-64,-60,-56,-52},
      xticklabels={\ensuremath{10^{-72}},\ensuremath{10^{-68}},\ensuremath{10^{-64}},
                   \ensuremath{10^{-60}},\ensuremath{10^{-56}},\ensuremath{10^{-52}}},
      ytick={48,56,64,72},
      yticklabels={\ensuremath{10^{48}},\ensuremath{10^{56}},\ensuremath{10^{64}},\ensuremath{10^{72}}},
      xlabel={channel diffusion constant $\mathcal{D}=D_2$},
      ylabel={channel decoherence $\mathcal{C}$ $/\,(4\pi G_N)^{2}$},
      xlabel style={yshift=2pt}, ylabel style={yshift=-4pt},
    ]
      \fill[cqred, opacity=0.085]
            (axis cs:-72.6,46) -- (axis cs:-49,46) --
            (axis cs:-49,48.6990) -- (axis cs:-72.6,72.2990) -- cycle;
      \fill[cqgreen, opacity=0.45]
            (axis cs:-72.6,72.2990) -- (axis cs:-49,48.6990) --
            (axis cs:-49,49.1427) -- (axis cs:-72.6,72.7427) -- cycle;
      \fill[cqgreyb, opacity=0.13]
            (axis cs:-71,71.1427) -- (axis cs:-54.985,55.1273) --
            (axis cs:-54.985,75.5) -- (axis cs:-71,75.5) -- cycle;
      \fill[cqblue, opacity=0.16]
            (axis cs:-71,71.1427) -- (axis cs:-65.885,66.0280) --
            (axis cs:-65.885,75.5) -- (axis cs:-71,75.5) -- cycle;
      \fill[cqgreen, opacity=0.14]
            (axis cs:-71,70.6990) -- (axis cs:-66.487,66.1863) --
            (axis cs:-66.487,75.5) -- (axis cs:-71,75.5) -- cycle;

      \draw[cqred,   line width=0.6pt]
            (axis cs:-72.6,72.7427) -- (axis cs:-49,49.1427);
      \draw[cqgreen, line width=0.5pt, dash pattern=on 4pt off 1.5pt]
            (axis cs:-72.6,72.2990) -- (axis cs:-49,48.6990);

      \draw[cqblue,   line width=0.6pt] (axis cs:-71,46)      -- (axis cs:-71,75.5);
      \draw[cqgreen,  line width=0.6pt] (axis cs:-66.487,46)  -- (axis cs:-66.487,75.5);
      \draw[cqorange, line width=0.6pt] (axis cs:-65.885,46)  -- (axis cs:-65.885,75.5);
      \draw[cqgreyb,  line width=0.5pt] (axis cs:-54.985,46)  -- (axis cs:-54.985,75.5);
      \draw[cqgreen,  line width=0.3pt, dashed]
            (axis cs:-66.487,66.1863) -- (axis cs:-49,66.1863);

      \node[circle, fill=cqgreen,  inner sep=1.1pt] at (axis cs:-66.487,66.1863) {};
      \node[circle, fill=cqorange, inner sep=1.1pt] at (axis cs:-65.885,66.0280) {};
      \node[circle, fill=cqgreyb,  inner sep=1.0pt] at (axis cs:-54.985,55.1273) {};

      \node[cqblue,   font=\tiny, rotate=90, anchor=south, inner sep=1.5pt]
            at (axis cs:-71,55) {matter-wave coherence: $\mathcal{D}\gtrsim10^{-71}$};
      \node[cqgreen,  font=\tiny, rotate=90, anchor=south, inner sep=1.5pt]
            at (axis cs:-66.487,55) {LVK, tensor: $\mathcal{D}\lessapprox3.3\times10^{-67}$};
      \node[cqorange, font=\tiny, rotate=90, anchor=north, inner sep=1.5pt]
            at (axis cs:-65.885,55) {LVK $+\,\beta\leq\alpha/4$, scalar: $\mathcal{D}\lessapprox1.3\times10^{-66}$};
      \node[cqgreyb,  font=\tiny, rotate=90, anchor=south, inner sep=1.5pt]
            at (axis cs:-54.985,66.5) {LISA Pathfinder, scalar};

      \node[cqgreen, font=\tiny, rotate=-29.4, anchor=north west, inner sep=1pt]
            at (axis cs:-72.4,69.3) {tensor floor $\mathcal{D}\mathcal{C}=\tfrac12$};
      \node[cqred,   font=\tiny, rotate=-29.4, anchor=south west, inner sep=1pt]
            at (axis cs:-63.4,65.3) {scalar floor $\mathcal{D}\mathcal{C}=\tfrac{25}{18}$};
      \node[cqred, font=\tiny, anchor=east] at (axis cs:-49.3,47.3)
            {forbidden in both channels};

      \draw[cqgrey, line width=0.3pt] (axis cs:-63.7,69.4) -- (axis cs:-66.1,66.4);
      \node[cqgrey, font=\tiny, anchor=west, align=left, inner sep=1pt] at (axis cs:-63.7,69.4)
            {$\mathcal{C}_{tensor}\gtrsim1.5\times10^{66}$\\$\mathcal{C}_{scalar}\gtrsim1.1\times10^{66}$};
      \node[cqgreyb, font=\tiny, anchor=east, inner sep=1.5pt] at (axis cs:-55.4,52.6)
            {$\mathcal{C}_{scalar}\gtrsim1.3\times10^{55}$};

      \draw[cqgreen, <->, line width=0.4pt] (axis cs:-71,74.9) -- (axis cs:-66.487,74.9)
            node[midway, above, font=\tiny, align=center, inner sep=1.5pt] {scalar--tensor\\4.5 orders};
      \draw[cqblue,  <->, line width=0.4pt] (axis cs:-71,72.9) -- (axis cs:-65.885,72.9)
            node[midway, below, font=\tiny, align=center, inner sep=1.5pt] {scalar\\5.1 orders};
      \draw[cqgreyb, <->, line width=0.4pt] (axis cs:-65.885,72.9) -- (axis cs:-54.985,72.9)
            node[midway, below, font=\tiny, align=center, inner sep=1.5pt] {transverse positivity only\\10.9 orders more};
    \end{axis}
    \end{tikzpicture}
    \caption{The decoherence--diffusion trade-off in the $(\mathcal{D},\mathcal{C})$ plane, with $\mathcal{C}$ in units of $(4\pi G_N)^2$. At saturation of
  the trade-off the tensor channel lies on its floor (dashed green line); the scalar floor (red line) lies a factor $25/9$ higher, so the strip between the
  two lines is allowed for the tensor channel but not the scalar one, and the region below both is forbidden. Vertical lines mark the diffusion bounds: the
  matter-wave coherence lower bound (blue), the LVK upper bound on the tensor channel (green), its extension to the scalar channel under full positivity
  (orange), and the LISA Pathfinder bound, which alone constrains the scalar channel under transverse positivity (grey). Where each ceiling meets its
  channel's floor it gives a lower bound on the decoherence. The shaded bands and arrows mark the viable windows: $4.5$ orders of magnitude common to the
  scalar and tensor channels, $5.1$ orders for the scalar channel under full positivity, and a further $10.9$ orders on transverse positivity alone.}
    \label{fig:floors}
  \end{figure}

If one instead assumes the spectrum holds throughout and uses the integrated energy density $\rho_{GW}\propto D_2\Lambda^3 T$ (as in~\cite{oppenheim2025diffusion}), the resulting bound requires the UV cutoff to satisfy $\ell\gtrsim 10^{2}\,\mathrm{m}$, which is far above the millimeter scale to which gravity has been probed. With scale-independent couplings, this would rule out the theory. However, asymptotic freedom~\cite{grudka2024renormalisation} causes the coupling to run, and matter coupling plus non-linear effects and cosmological considerations may further suppress the integrated energy density; a definitive assessment requires a full treatment on an expanding background.

\subsection{The effective decoherence rate}
\label{effective}

Given the dependence of the diffusion and decoherence terms on two parameters, $\alpha,\beta$, does the diffusion-decoherence trade-off hold for static potentials if we integrate out $\psi$? More generally, in decoherence experiments, the measurement is of the decoherence rates, and not of the gravitational fields. Thus, to obtain the correct decoherence rate, we must integrate out all mediating gravitational fields. Such a calculation is beyond the scope of this work, but we include the analysis of \cite{oppenheim2026non} to better understand the role of the various coupling constants.

As an example, let us take \eqref{eq:PQG-action}, with $T^{\mu\nu}=\density(x)U^{\mu}U^{\nu}$, i.e., pressureless dust with energy density $\density(x)$. The trace, $T$, is given by $-\density(x)$, and $T_{\mu\nu}T^{\mu\nu}=\density(x)^2$. Working in the rest-frame of this distribution, in units of $8\pi G_N$, we obtain the following Lagrangian density:
 \begin{align}
  \begin{split}
  \mathcal{S}_{dust}&=S^{(g)}+\int d^4x\frac{(\alpha-\beta)}{8}(4\bar{\density}^2+(\Delta\density)^2)\\
  &\quad+{\frac{1}{2}}((4\alpha-8\beta)\bar{\density}\triangle\psi+12\beta\partial_0^2\psi\bar{\density}+4\beta\bar{\density}\triangle\Phi)\\
  &\quad-i\Delta\density(1-\phi+\nabla^2E+3\psi)
  \end{split}
  \end{align}
where $\mathcal{S}^{(g)}$, given by \eqref{eq:LGR}, is the pure gravitational action and $\bar{\density}=:(\density^{+}+\density^{-})/2$, and $\Delta\density=:\density^{+}-\density^{-}$. Since the vector and tensor fields do not interact with the dust distribution in its rest frame, we will ignore them henceforth and focus only on the scalar sector. We then have the following Lagrangian density:
  \begin{align}
  \label{eq:DustAction}
      \begin{split}
      \mathcal{L}_{dust}&=\frac{1}{2}(\alpha((2\nabla^2\psi+\bar{\density})^2+2(\nabla^2(\psi-\Phi))^2+12(\partial^2_0\psi)^2-16\partial_0^2\psi\nabla^2\psi
  +8\partial_0^2\psi\nabla^2\Phi)-\beta(2\nabla^2(2\psi-\Phi)-6\partial_0^2\psi+\bar{\density})^2)\\&+\frac{\alpha-\beta}{8}(\Delta\density)^2-i\Delta\density(1-\phi+\nabla^2E+3\psi)
  \end{split}
  \end{align}
In the case of slowly changing $\psi,\Phi$, $\partial_t \ll \partial_i$, the terms with time-derivatives drop out, allowing us to integrate out $\psi$ without adding non-local in time terms to the action. Restoring the factors of $8\pi G_N$, We obtain the following, effective action, which is non-local in space:
  \begin{align}
  \begin{split}
  \mathcal{S}_{dust,stat,eff}[\Phi]&=\int d^4x\Big[(\frac{\alpha(\alpha-3\beta)}{2(3\alpha-8\beta)}(2\triangle\Phi+8\pi
  G_N\bar{\density})^2+(\frac{64\pi^2G_N^2(\alpha-\beta)}{8})(\Delta{\density})^2
  \\&-i\Delta\density[1-\phi+\nabla^2E-\frac{3}{(3\alpha-8\beta)}((-\alpha+4\beta)\Phi)])(x)\\&+\int
  d^3y(\frac{9}{4(3\alpha-8\beta)\nabla^4}(\Delta{\density})^2 -i\Delta\density\frac{3(8\pi
  G_N)}{(3\alpha-8\beta)}(\frac{(-\alpha+2\beta)\bar{\density}}{\nabla^2}))(x,y)\Big]
  \end{split}
  \end{align}
where $-\frac{1}{\nabla^2}$ refers to the spatial Fourier transform of $\frac{1}{k^2}$, and where we have explicitly written out the arguments of the terms involved.

This has an obvious interpretation in the Newtonian gauge. The Newtonian potential $\Phi$ is sourced by the average of the bra and ket dust fields but diffuses around the solution to Poisson's equation, with an effective diffusion constant, $D_{diff,eff}=\frac{(3\alpha-8\beta)}{\alpha(\alpha-3\beta)}$. This is the effective diffusion constant, $D_{2,scalar}$.
Compare this with the approach of~\cite{layton2023weak}. Therein, one assumed for the scalar sector that only the $G_{00}$ component of Einstein's equations was diffusive. That assumption would have allowed us to set $\psi=\Phi$, and we would have consequently obtained an effective diffusion constant, $D'_{diff,eff}=\frac{1}{(\alpha-\beta)}$.

In our case, the rate of decoherence, $D_{dec,eff}=8\pi^2G_N^2(\alpha-\beta)+\frac{9}{4k^4(3\alpha-8\beta)}$, where the second term represents the secondary decoherence, does not set an upper bound on the rate of diffusion, $D_{diff,eff}$ which diverges as $\beta \rightarrow \alpha/3$, even for a finite decoherence rate. Similarly, $\lim_{\beta\rightarrow-\infty} D_{diff,eff}=\frac{8}{3\alpha}$, where clearly the decoherence rate, $D_{dec,eff}$ diverges.

By integrating out $\psi$, which mediates interactions between the bra and the ket fields, we also obtain terms of the form $\density^{+}\density^{-}$. A pure initial state will now be sent to a mixed one, even when conditioning on a particular classical configuration  of $\Phi$.
This is a direct consequence of the fact that the mass-field interaction term is of the form $\bar{T}^{\mu \nu}G_{\mu\nu}$, instead of $\bar{T}^{\mu\nu}h_{\mu\nu}$. Physically, since we have integrated out a mediating field, we would expect an initially pure state to decohere. \MS{The mixed term is
\begin{align}
\label{eq:mixed}
-\density^{+}(x)\left[\frac{32\pi^{2}G_N^{2}(\alpha-2\beta)^{2}}{3\alpha-8\beta}+\frac{9}{2(3\alpha-8\beta)\nabla^{4}}\right]\density^{-}(x),
\end{align}
and the evolution remains completely positive. In the Markovian limit the map on matrix elements is a Schur multiplier, $\varrho\mapsto e^{-\mathcal I_{CQ}}\varrho$, which is completely positive if and only if the kernel is positive semi-definite. In Fourier modes the kernel is
\begin{align}
\frac{1}{3\alpha-8\beta}\Big[32\pi^{2}G_N^{2}(\alpha-2\beta)^{2}+\frac{9}{2k^{4}}\Big]\geq0 \iff 3\alpha-8\beta>0,
\end{align}
which is implied by the positivity conditions already assumed, since $3\alpha-8\beta=\tfrac{8}{3}(\alpha-3\beta)+\tfrac{1}{3}\alpha$. Complete positivity therefore imposes no new constraint. The second term of the kernel is the secondary decoherence, which is unopposed; the first is the residual imbalance between primary decoherence and the renormalised back-reaction, and vanishes only at $\beta=\alpha/2$, excluded by $\alpha>3\beta$.}
Additionally, the unitary part now has terms of the form $(\density^{+})^2,(\density^{-})^2$; however, note that there are no $\density^{+}\density^{-}$ mixed terms in the unitary part.

Finally, we take the Newtonian gauge, setting $\Phi=\phi$ and thus $\nabla^2E=0$, and integrate out $\phi$. This gives the action
  \begin{align}
  \begin{split}
  \mathcal{S}_{dust,static}[\bar{\density},\Delta\density]&=\int d^4x[((\Delta\density)^2\frac{(64\pi^2G_N^2)(\alpha-\beta)}{8}-i\Delta\density)(x)\\&+\int
  d^3y(\frac{(3\alpha-\beta)}{4\alpha(\alpha-3\beta)\nabla^4}(\Delta\density)^2+i\Delta\density\frac{8\pi G_N\bar{\density}}{\nabla^2})(x,y)]
      \end{split}
  \end{align}
After integrating out all the interacting classical fields, the real part consists solely of a decoherence term, which, by our initial positivity condition $\alpha\geq3\beta$, is always larger than the original decoherence term (which had the coefficient $8\pi^2 G_N^2(\alpha-\beta)$). We have thus made the theory only \textit{more} constrained than before. One may also note that the limit $\alpha\rightarrow3\beta^+$ leads to instantaneous decoherence.

\subsubsection{Non-Markovian Effects}
If we drop the assumption, $\partial_t \ll \partial_i$, and integrate out $\psi$, we obtain, in the Newtonian gauge, the following non-local action:
 \begin{align}
  \begin{split}
  \mathcal{S}_{dust,effective}[\Delta\density,\bar{\density},\Phi]&=\int d^4x \Bigg( 64\pi^2
  G_N^2\bar{\density}(\frac{\alpha-\beta}{2})\bar{\density}+64\pi^2 G_N^2\Delta\density(\frac{\alpha-\beta}{8})\Delta\density-i\Delta\density(1-\Phi)+16\pi
  G_N\beta\bar{\density}\nabla^2\Phi\\&+\int d^4y\Big(\Phi\frac{2\alpha(\alpha-3\beta)\Box^2\nabla^4}{S}\Phi+\Delta\density(\frac{9}{4S})\Delta\density-3i\Delta\density\frac{\nabla^2((\alpha-4\beta)\nabla^2-2(\alpha-3\beta)\partial_0^2)}{S}\Phi\\&+16\pi
  G_N((\alpha-2\beta)\nabla^2+3\beta\partial_0^2)\bar{\density}\frac{\nabla^2((\alpha-4\beta)\nabla^2-2(\alpha-3\beta)\partial_0^2)}{S}\Phi\\&
  -\frac{64\pi^2 G_N^2(((\alpha-2\beta)\nabla^2+3\beta\partial_0^2)\bar{\density})^2}{S}+24\pi G_N
  i\Delta\density\frac{(\alpha-2\beta)\nabla^2+3\beta\partial_0^2}{S}\bar{\density}\Big)\Bigg)\\
  S&:=6(\alpha-3\beta)\partial_0^4-8(\alpha-3\beta)\nabla^2\partial_0^2+(3\alpha-8\beta)\nabla^4.
  \end{split}
  \end{align}

Integrating out $\Phi$, we obtain the following non-local action:
   \begin{align}
    \begin{split}
        \mathcal{S}_{dust,effective}[\Delta\density,\bar{\density}]&=\int d^4x\frac{64\pi^2
    G_N^2(\alpha-\beta)}{8}((\Delta\density)^2+4\bar{\density}^2)-i\Delta\density \\&+\int d^4y\Delta\density\frac{((3\alpha-\beta)\nabla^4+2(\alpha-3\beta)
  \nabla^2\partial_0^2+3(\alpha-3\beta)\partial_0^4)}{4\alpha(\alpha-3\beta)(\nabla^2-\partial_0^2)^2\nabla^4}\Delta\density\\&-64\pi^2
    G_N^2\bar{\density}\frac{((\alpha-3\beta)(\alpha-\beta) \nabla^4+2\beta(\alpha-3\beta)\nabla^2\partial_0^2+3\beta^2\partial_0^4)}{2(\alpha-3\beta)(\nabla^2-\partial_0^2)^2}\bar{\density}\\&+i\Delta\density\frac{8\pi
    G_N((\alpha-3\beta)\nabla^2+(\alpha-\beta)\partial_0^2)}{(\alpha-3\beta)(\nabla^2-\partial_0^2)^2}\bar{\density}
    \end{split}
    \end{align}
In momentum-space, one thus obtains the non-Markovian decoherence coefficient,
$
\mathcal{D}_{\mathrm{decoh}}(\omega,k) = 8\pi^2 G_N^2(\alpha-\beta)+\frac{(3\alpha-\beta)k^4+2(\alpha-3\beta)k^2\omega^2+3(\alpha-3\beta)\omega^4}{4\alpha k^4(\alpha-3\beta)(\omega^2-k^2)^2}$, which, like the corresponding static rate obtained once both $\psi$ and $\Phi$ have been integrated out,
$\mathcal{D}_{\mathrm{decoh,stat}}(k)=8\pi^2 G_N^2(\alpha-\beta)+\frac{(3\alpha-\beta)}{4\alpha(\alpha-3\beta)k^4}$,to be distinguished from $D_{dec,eff}$ above, which follows from integrating out $\psi$ alone, is larger than the initial minimum decoherence rate, $8\pi^2G_N^2(\alpha-\beta)$ and therefore leads to a tighter bound on the parameters, $\alpha,\beta$.

\MS{
  \subsection*{Summary of bounds}
  Write $\mathcal{D}$ for a channel's dimensionless diffusion constant $D_2$, and $\mathcal{C}$ for its decoherence coefficient --- the coefficient $D_0/8$
  of $(\Delta\density)^2$ in the action, as in $D_{dec,eff}$ above --- in units of $(4\pi G_N)^2$. Saturation of the decoherence--diffusion trade-off,
  $D_0=(8\pi G_N)^2D_2^{-1}$, then reads $\mathcal{D}\,\mathcal{C}=\tfrac12$ in the tensor channel. The two channels are
  \begin{align}
  \text{tensor:}\quad &\mathcal{D}_{tensor}=\alpha^{-1},\qquad \mathcal{C}_{tensor}=\frac{\alpha}{2},\qquad
  \mathcal{D}_{tensor}\,\mathcal{C}_{tensor}=\frac12,\\
  \text{scalar:}\quad &\mathcal{D}_{scalar}=D_{2,scalar}=\frac{3\alpha-8\beta}{\alpha(\alpha-3\beta)},\qquad \mathcal{C}_{scalar}=\frac{\alpha-\beta}{2},
  \end{align}
  and, with $x=(\alpha-3\beta)/\alpha$,
  \begin{equation}
  \label{eq:floor}
  \mathcal{D}_{scalar}\,\mathcal{C}_{scalar}=f(x)=\frac{(1+8x)(2+x)}{18x}\;\geq\;\frac{25}{18},
  \end{equation}
  with the minimum at $x=\tfrac12$, i.e.\ $\beta=\alpha/6$. Two positivity conditions enter. \emph{Transverse} positivity, $\alpha>3\beta$ ($x>0$), is
  required by the transverse two-point functions; \emph{full} positivity on all ten components adds $\beta\leq\alpha/4$ ($x\geq\tfrac14$). The minimum
  $x=\tfrac12$ satisfies both. The channels are linked by $D_{2,scalar}/D_{2,tensor}=(1+8x)/(3x)$ (Fig.~\ref{fig:tensor-scalar}): under full positivity this
  ratio lies in $(\tfrac83,4]$, so the gravitational-wave bound also bounds the scalar channel, whereas on transverse positivity alone it is unbounded and
  only LISA Pathfinder constrains the scalar channel.

  Through~\eqref{eq:floor} and $\mathcal{D}_{tensor}\mathcal{C}_{tensor}=\tfrac12$, each diffusion ceiling becomes a decoherence floor
  (Figs.~\ref{fig:floors}):
  \begin{center}
  \begin{tabular}{lllll}
  \hline
  channel & positivity & diffusion bound & from & decoherence floor \\
  \hline
  tensor & none & $\alpha^{-1}\lessapprox3.3\times10^{-67}$ & LVK O1--O4a, Eq.~\eqref{eq:tensorbound} & $\mathcal{C}_{tensor}\gtrsim1.5\times10^{66}$ \\
  scalar & full & $D_{2,scalar}\lessapprox1.3\times10^{-66}$ & LVK, ratio $\leq4$ & $\mathcal{C}_{scalar}\gtrsim1.1\times10^{66}$ \\
  scalar & transverse & $D_{2,scalar}\lessapprox1.0\times10^{-55}$ & LISA Pathfinder, Eq.~\eqref{eq:newtontwo} &
  $\mathcal{C}_{scalar}\gtrsim1.3\times10^{55}$ \\
  scalar & --- & $D_{2,scalar}\gtrapprox10^{-71}$ & matter-wave coherence~\cite{oppenheim2025diffusion} & --- \\
  \hline
  \end{tabular}
  \end{center}
  The LVK entries use the 95\% upper limit for the $\Omega_{GW}\propto f^3$ spectrum predicted here under a uniform amplitude prior,
  $\Omega_{GW}(25\,\text{Hz})\leq1.0\times10^{-9}$; the log-uniform prior used for that paper's headline results would tighten them by a factor $3.1$, to
  $\alpha^{-1}\lessapprox1.0\times10^{-67}$ and $\mathcal{C}_{tensor}\gtrsim4.8\times10^{66}$. The LISA Pathfinder entry uses the measured one-sided noise
  floor $S_{\Delta g}=3.0\times10^{-30}\,\text{m}^2\,\text{s}^{-4}\,\text{Hz}^{-1}$ along the sensitive axis and the cubic test masses; modelling them as
  spheres changes it to $(1.0\times10^{-55})$.

  The scalar channel is therefore bounded on both sides. Under full positivity the window is $10^{-71}\lesssim D_{2,scalar}\lesssim1.3\times10^{-66}$, about
  $5.1$ orders of magnitude; on transverse positivity alone a further $10.9$ orders, up to $1.0\times10^{-55}$, remain open. $\mathcal{C}_{tensor}$ and
  $\mathcal{C}_{scalar}$ decohere different observables --- transverse-traceless anisotropic stress and mass density respectively --- so it is the scalar
  floor that matter-wave interferometry tests, and the tensor channel has at present only an upper bound on its diffusion. All of these bounds rest on the
  linearised, Markovian treatment of this section and should be regarded as preliminary.}

\section{Discussion}
\label{Discussion}
The path integral formulation of postquantum gravity is locally Lorentz invariant, since the covariance matrix (the DeWitt metric) only depends on the metric $g$. This creates a tension with the action being indefinite, since the metric's Lorentzian signature means that in $3+1$ dimensions the DeWitt metric has negative eigenvalues.
In quantum gravity the action is also indefinite, but due to the conformal mode.
Fortunately, we have shown here that the sectors that correspond to the gravitational degrees of freedom are positive semi-definite, in a gauge-invariant manner. The indefinite sector is non-dynamical: it corresponds to constraint equations rather than propagating modes.
 By performing a scalar-vector-tensor decomposition, we have been able to cleanly separate the Newtonian sector from the gravitational wave sector, and identify the physical degrees of freedom. In the scalar sector, the dynamical scalar $\psi$ is promoted to a stochastic degree of freedom, while the Newtonian potential $\Phi$ diffuses around an equation of constraint. The same pattern holds in the vector sector: $V_i$ satisfies a constraint equation, rather than an independent stochastic equation of motion. As a result, the indefiniteness of the gravitational action appears to be relatively benign: the sectors with positive-semi-definite two-point functions are precisely the dynamical ones, while the indefinite vector sector represents a constraint rather than a propagating mode.

This does not mean that $V_i$ is determined once and for all. $V_i$ must acquire stochastic contributions through the momentum constraint and has a dependence on $\partial\psi$. Indeed, this must be so both because the momentum constraint is sourced by quantum fields in superposition, and because covariance must be preserved. The vector sector is also the least constrained observationally: LISA Pathfinder bounds the Newtonian potential and the stochastic gravitational wave background bounds the tensor modes, but no experiment places significant bounds on noise in $V_i$. Because a gyroscope responds to the circulation $\oint h_{0i}dx^i$, frame-dragging experiments such as the proposed superfluid $^4$He Josephson gyrometers\cite{ellers2025quantum} are the natural instrument class for this sector, though their projected sensitivities remain far from the diffusion scales considered here.
A fuller understanding of how the constraints are imposed in the path integral at higher order, where the vector-scalar coupling becomes dynamically relevant, remains an open question.

In addition, whereas classical, deterministic general relativity has only two dynamical modes, we now have an additional, scalar mode. This is to be expected, as the conservation of the energy-momentum tensor no longer suffices to make four of the remaining six modes vanish or cancel on-shell \cite{Spin2}. 

Further, we discussed possible experimental implications of a stochastic model of gravity, and whether such a model is ruled out by current experimental bounds on noise and coherence times via the decoherence-diffusion trade-off. From the SVT decomposition, we are able to refine possible bounds from the LISA Pathfinder via the Newtonian potential, and from stochastic gravitational waves. We found that the theory is viable given current observational bounds, but care needs to be taken, given our poor understanding of a number of subtleties, such as the effects of nonlinearities, non-Markovianity, and scale-invariance. These calculations point the way to further theoretical improvements, which, together with more precise experiments, can be used to test the quantum vs classical nature of gravity.

Indeed, this work highlights a subtlety with practical consequences. In the Newtonian limit with pressureless dust~\cite{oppenheim2026non}, our results differ significantly from those of~\cite{layton2023weak}, as well as \cite{tilloy2016sourcing}. The origin of this difference is illuminating. The work of \cite{layton2023weak}, which considers the scalar sector of the metric, first reduces the gravitational phase space by solving the constraints classically, and then introduces stochasticity on the remaining degrees of freedom. Here, the stochastic dynamics acts on the full metric and the constraints emerge from the path integral. These two procedures do not commute, a tension that mirrors the long-standing debate in quantum gravity between Dirac quantization and reduced phase space quantization~\cite{isham1992canonical,kuchar1992time}. The present setting, being classical-stochastic rather than quantum, offers a considerably more tractable arena in which to study this inequivalence.



\noindent\textbf{ Acknowledgements:}
We would like to thank John Donoghue, Rhys Evans, Bob Holdom, Isaac Layton, Emanuele Panella, Geoffrey Penington, and Elizabeth Wilson for valuable discussions, as well as EPSRC (UKRI) for financial support. The calculations were performed by humans, and indeed were used to better understand the failure modes of LLMs, as described in \href{https://superposer.substack.com/p/teaching-an-ai-to-do-physics-by-wiping}{this post}.

\bibliography{cq_unified}

\newpage
\appendix

\section{The Onsager-Machlup and the Janssen–De Dominicis path integrals }
\label{sec:OMJD}

 In this section we motivate the stochastic gravitational path integral, and introduce an alternate formalism, the Janssen-De Dominicis
  (JD)~\cite{Janssen1976-hh,De_Dominicis1976-zy} or Martin--Siggia--Rose (MSR) formalism~\cite{martin1973statistical}. We note that unless the diffusion
  matrix is conserved, the gauge-fixed JD and OM formalisms give different two-point functions: a non-conserved diffusion matrix couples the transverse and
  longitudinal sectors (for the generalized DeWitt metric, whenever $\beta\neq0$), so the gauge-fixed longitudinal modes feed back into the physical
  propagator. We show, however, that the ultra-local generalized DeWitt metric~\eqref{eq:dewitt} yields the same OM action as a conserved diffusion matrix,
  even though it is not itself transverse: contracting it with the conserved Einstein tensor projects out its longitudinal parts. With the conserved
  diffusion matrix, the JD and OM two-point functions coincide. In addition, we discuss stochastic differential equation forms of the linearized Einstein equations and present a local, conserved diffusion matrix which does satisfy the Bianchi identity. We thus see that we can have a local stochastic theory which nonetheless generates a valid spacetime.

 A caveat before we start: many of the equations are formal, in the sense that they must be regularized before they can be used in the gravitational case. Our presentation mainly follows~\cite{CriticalDynamics}.
Consider a Langevin type equation of the form
\begin{align}
    \partial_t^2\Phi(x,t)=\triangle\Phi(x,t)+\xi
\end{align}
where the covariance of $\xi$, the Gaussian white noise, is given by
\begin{align}
    \langle \xi(x,t)\xi(x',t')\rangle=2D_2\delta(x-x')\delta(t-t')
\end{align}
where $D$ could possibly be a function of $t$ and $x$. The probability distribution from which the correlations can be obtained is then given by
\begin{align}
\label{eq:IntroA1}
\pdf[\xi]=\mathcal{N}\int D\xi \exp({-\int d^4x \xi D_2^{-1}\xi/4})
\end{align}
and so one obtains the OM path integral by a change of variables
\begin{align}
\label{eq:IntroA11}
     \pdf[\Phi]\propto\int J D\Phi \exp{-\frac{1}{4}\int d^4x (\partial_t^2\Phi-\triangle\Phi) D_2^{-1}(\partial_t^2\Phi-\triangle\Phi)}
\end{align}
where the Jacobian $J$ depends on the discretization used (see~\ref{Jacobian}).

An entirely equivalent formalism is the Janssen–De Dominicis (JD), or the Martin-Siggia-Rose (MSR) response functional one, in which the covariance matrix, $D_2$ does not appear in the denominator, and additionally, $\Phi$ appears linearly. In this formalism, we have
\begin{align}
     \pdf[\Phi]\propto\int D[\Phi] D[i\tilde{\Phi}] \exp{-\int d^4x \left[(\partial_t^2\Phi-\triangle\Phi)\tilde{\Phi}-\tilde{\Phi}D_2\tilde{\Phi}\right]}
\end{align}
One can verify their equivalence by integrating over the Gaussian response functional.

We now replicate the steps above for the gravitational case twice: firstly for a non-conserved diffusion matrix, and secondly for a conserved one.

\subsubsection{The Gravitational Case I: The DeWitt Metric}

This whole procedure is no longer as straightforward in the gravitational case. Suppose that we start from
\begin{align}
\label{eq:SD}
    G_{\mu \nu}^{(1)}&=\xi_{\mu \nu}+\xi'_{\mu\nu}\\
 \langle\xi_{\mu \nu}\xi_{\rho\sigma}\rangle&=\alpha^{-1}(I_{\mu\nu,\rho\sigma}-\frac{\beta}{4\beta-\alpha}\eta_{\mu\nu}\eta_{\rho\sigma})\\\mathcal{D}_{2,\mu \nu\rho\sigma,}&:= \langle\xi_{\mu \nu}\xi_{\rho\sigma}\rangle
\end{align}
where $G_{\mu \nu}^{(1)}$ is the linearized Einstein tensor and $I_{\mu\nu,\rho\sigma}:=\frac{1}{2}(\eta_{\rho\nu}\eta_{\sigma\mu}+\eta_{\rho\mu}\eta_{\sigma\nu})$ and we have suppressed $\delta^4(x-x')$. The first equation is a linearized version of that appearing in \cite{hu2008stochastic,moffat1997stochastic}, except that here the noise kernel is local. We here consider the vacuum case, since when matter is added, the dynamics takes on a different form in comparison to stochastic gravity.

There are two issues with this formulation. Firstly, $\xi_{\mu\nu}$ is not a conserved tensor. Writing $\gamma\equiv\beta/(4\beta-\alpha)$ for the coefficient appearing in~\eqref{eq:SD}, one has in momentum space $\partial^{\mu}\partial^{\rho}\langle\xi_{\mu \nu}\xi_{\rho\sigma}\rangle\propto\tfrac{1}{2}k^2\eta_{\nu\sigma}+(\tfrac{1}{2}-\gamma)k_{\nu}k_{\sigma}$, which is non-zero for every $\alpha,\beta$, the first term carrying no $\gamma$. Its longitudinal part does vanish, however, precisely when the positivity constraint for the scalar sector is saturated, $\alpha=3\beta$: there $\gamma=1$ and the expression reduces to the transverse $\tfrac{1}{2}k^2\theta_{\nu\sigma}$. An additional, non-conserved tensor must therefore be added for~\eqref{eq:SD} to be consistent. Secondly, the DeWitt metric can serve as a covariance matrix only in a formal sense given that it is not positive semi-definite; however, we have argued above in Section~\ref{DOF} that after contracting the inverse diffusion matrix with the Einstein tensor, only the PSD sector of the action obtained is dynamical and so physically relevant.

Following the procedure of \eqref{eq:IntroA1},\eqref{eq:IntroA11} we obtain the following, diffeomorphism invariant OM action
\begin{align}
\begin{split}
     \label{eq:OM}\pdf[h^{\mu\nu}]&\propto\int J Dh^{\mu\nu} \exp{-\mathcal{S}_{OM}},\\ \mathcal{S}_{OM}&=\frac{1}{2}\int d^4x \int d^4yG_{\mu\nu}(x)\mathcal{D}_2^{-1\mu\nu,\rho\sigma}(x,y)G_{\rho\sigma}(y)=\frac{1}{2}\int d^4x (\alpha R^{\mu\nu}R_{\mu\nu}-\beta R^2)
     \end{split}
\end{align}

Can we now claim that~\eqref{eq:OM} implies the following JD action with the auxiliary, symmetric two-tensor $\tilde{h}^{\rho\sigma}$, i.e.,
\begin{align}
     \label{eq:MSR}\pdf[h^{\mu\nu}]&\propto^{?}\int J D[h] D[i\tilde{h}] \exp{-\int d^4x \left[G_{\rho\sigma}\tilde{h}^{\rho\sigma}-\frac{1}{2}\tilde{h}^{\mu\nu}\mathcal{D}_{2\mu\nu,\rho\sigma}\tilde{h}^{\rho\sigma}\right]}
\end{align}in the sense that both actions give us the same two-point function?

The JD action thus has the following Lagrangian density:
\begin{align}
\label{eq:ACTIONJD} \mathcal{L}_{JD}=G_{\rho\sigma}\tilde{h}^{\rho\sigma}-\frac{1}{2\alpha}(\tilde{h}^{\mu\nu}\tilde{h}_{\mu\nu}-\gamma \tilde{h}^2)
\end{align}
Integrating out the auxiliary variable returns the original action,
\begin{align}
     \mathcal{L}_{OM}=\frac{1}{2}(\alpha R_{\mu\nu}R^{\mu\nu}-\beta R^2).
\end{align}

However, one does not recover the original propagator for any non-zero $\beta$. To see this, we can re-write the JD action in terms of spin-projection operators:
\begin{align}
\label{eq:JD}
    S_{JD}=\int d^4x h^{\mu\nu}\hat{O}_{\mu\nu,\rho\sigma}\tilde{h}^{\rho\sigma}-\frac{1}{2\alpha}\tilde{h}^{\mu\nu}(\frac{1}{2}(\eta_{\mu\rho}\eta_{\nu\sigma}+\eta_{\mu\sigma}\eta_{\nu\rho})-\gamma\eta_{\mu\nu}\eta_{\rho\sigma} )\tilde{h}^{\rho\sigma}
\end{align}
where
\begin{align}
      \hat{O}_{\mu\nu,\rho\sigma}=\Box(-\frac{1}{2}P^{(2)}+P^{(0-s)})_{\mu\nu,\rho\sigma}
\end{align} and the spin-projectors are given by~\eqref{eq:Spin-2} and \eqref{eq:spin-0}. Re-writing \eqref{eq:JD} in a matrix form,
  \begin{align}
  \begin{split}
  \label{eq:MJD}
  S_{JD}&=\frac{1}{2}\int\frac{d^4p}{(2\pi)^4}\begin{pmatrix}h(-p) \\ \tilde{h}(-p)\end{pmatrix}^T \mathcal{J}(p) \begin{pmatrix}h(p) \\
  \tilde{h}(p)\end{pmatrix},\\
  \mathcal{J}(p)&=
  \begin{bmatrix}
      \mathbf{0} & \hat{\mathbf{O}}(i\omega,k)\\
      \hat{\mathbf{O}}(-i\omega,k) & -\boldsymbol{\mathcal{D}}_2
  \end{bmatrix}.
  \end{split}
  \end{align}

This matrix is non-invertible, due to the non-invertibility of $\hat{O}$. Adopting the gauge $\partial_{\mu}h^{\mu\nu}=0$, we have the altered operator,
 \begin{align}
   \hat{O}'_{\mu\nu,\rho\sigma}=\Box(-\frac{1}{2}P^{(2)}+P^{(0-s)}-\frac{1}{4\Delta}(P^{(1)}+2P^{(0-w)}))_{\mu\nu,\rho\sigma}
\end{align}
where $\frac{1}{\Delta}$ governs the strength of gauge-fixing. Inverting $\mathcal{J}$ with $\hat{O}'$ in place of $\hat{O}$, we obtain for the propagator, $\langle h_{\rho\sigma }h_{\mu\nu}\rangle$
\begin{align}
   \lim_{\lambda\rightarrow0}\langle h_{\rho\sigma}h_{\mu\nu}\rangle_{JD}=\frac{1}{\Box^2}\left(\frac{4}{\alpha}P^{(2)}+\frac{\alpha-\beta}{\alpha(\alpha-4\beta)
  }P^{(0-s)}\right)_{\rho\sigma,\mu\nu}
\end{align}
where we have re-written $\gamma$ in terms of $\alpha,\beta$.
One notes that this is markedly different in its coefficients from the propagator obtained directly from the quadratic action, for any non-zero $\beta$.
\begin{align}
   \label{eq:2ppp}\lim_{\lambda'\rightarrow0} \langle h_{\rho\sigma }h_{\mu\nu}\rangle_{OM}=\frac{4}{\alpha\Box^2}(P^{(2)}+\frac{P^{(0-s)}}{4(1-\frac{3\beta}{\alpha})})_{\rho\sigma,\mu\nu}
\end{align} which one may verify by first integrating out $\tilde{h}_{\mu\nu}$,
\begin{align}
 \mathcal{L}(h^{\mu\nu})= \frac{1}{2}(\Box(-\frac{1}{2}P^{(2)}+P^{(0-s)}))_{\alpha\beta,\rho\sigma}h^{\rho\sigma}\mathcal{D}_2^{-1\alpha\beta,\kappa\gamma}(\Box(-\frac{1}{2}P^{(2)}+P^{(0-s)}))_{\kappa\gamma,\mu\nu}h^{\mu\nu}
\end{align}
and then inverting the resulting action for the propagator.

One can also see this using gauge-invariant variables. Allowing for integration by parts, the scalar, vector and tensor sectors decouple. The action for the three sectors is of the generic form
\begin{align}
    \mathcal{S}=\frac{1}{2}\int_{\omega,k}(h(-k,-\omega),\tilde{h}(-k,-\omega))\mathcal{J}'(h(k,\omega),\tilde{h}(k,\omega))
\end{align}
where $\mathcal{J}'$ is given by
\begin{align}
  \mathcal{J}'=
\begin{bmatrix}
    \bf0 &&\hat{\bf G}(i\omega,k)\\\hat{\bf G}(-i\omega,k) && -\mathcal{D}_{2,sector}
\end{bmatrix}
\end{align}
and $(\hat{G}h)_{\rho\sigma}$ denotes the scalar (respectively vector, tensor) sector of the linearized Einstein action, to which we have added an infinitesimal term of the form $i\omega$ for regularization, while $\mathcal{D}_{2,sector}$ is the coefficient matrix of the terms quadratic in the scalar (respectively vector, tensor) potentials.

However, it is no longer as simple to work with gauge-invariant potentials. The action in~\eqref{eq:OM} is gauge invariant, whereas in~\eqref{eq:MSR}, only the first term is invariant under the following transformation of the auxiliary variables, $\tilde{h}_{\mu\nu}\rightarrow\tilde{h}_{\mu\nu}+2\partial_{(\mu}\tilde{\xi}_{\nu)}$ (by the linearized Bianchi identity). Although the Einstein tensor can be written as a function of six variables, one has to contend with ten auxiliary variables corresponding to the ten potentials in a symmetric rank-2 tensor in four dimensions. Further, even if one amends the second term in~\eqref{eq:MSR} by hand\footnote{This is equivalent to adding the term $-\beta g^{\mu \nu}$ to the DeWitt metric and setting all scalar potentials apart from $\psi,\phi$ to zero.} to an action of the form, for example for the scalar sector\footnote{$G_{\rho\sigma}\tilde{h}^{\rho\sigma}=4(-\tilde{\Phi}\triangle\psi+\tilde{\psi}((\triangle-3\partial_0^2)\psi-\triangle\Phi))$},
\begin{align}
    \label{eq:MSRSCALAR}S[\Phi,\psi,\tilde{\Phi},\tilde{\psi}]=\int d^4x \left[G_{\rho\sigma}\tilde{h}^{\rho\sigma} (\Phi,\psi,\tilde{\Phi},\tilde{\psi})-(\tilde{\psi},\tilde{\Phi})^{T}\mathbf{A}(\alpha,\beta)(\tilde{\psi},\tilde{\Phi})\right]
\end{align}
one does not obtain~\eqref{eq:scalar} even with $\beta$ set to zero on integrating out the auxiliary variables, nor the correlation matrix~\eqref{eq:twopointscalar} upon inverting~\eqref{eq:MSRSCALAR}. Instead, the scalar JD action corresponding to~\eqref{eq:scalar} is given by an action of the form
\begin{align}
    S[\Phi,\psi,\tilde{\Phi},\tilde{\psi}]_{JD}=-\int d^4x \left[\Omega_1(a_1\partial_0^2\psi+\triangle(b_1\psi+c_1\Phi))\tilde{\psi}+\Omega_2(a_2\partial_0^2\psi+\triangle(b_2\psi+c_2\Phi))\tilde{\Phi}-(\alpha-\beta)(\tilde{\Phi}\tilde{\Phi}+\tilde{\psi}\tilde{\psi})\right]
\end{align}
where $\Omega_i,a_i,b_i,c_i$ are functions of $\alpha,\beta$ that may be obtained by diagonalizing the action in \eqref{eq:scalar} in the $\triangle\psi,\partial_0^2\psi,\Phi$ basis.\footnote{The equations of motion also clearly differ from the Einstein equations for the scalar sector.}

The vector sector proves to be much more instructive. Starting from~\eqref{eq:vector}, the JD action for this sector reads
\begin{align}
    \mathcal{L}_{vector}=V_i\triangle \tilde{V^i}-\frac{1}{\alpha}(-\tilde{S_i}\tilde{S^i}-\tilde{F_i}\triangle\tilde{F^i})
\end{align}
One can easily verify that inverting the action does not give us the two-point function we would obtain from inverting~\eqref{eq:vector}. Interestingly, if we `gauge-fix' here by hand by setting $\tilde{\dot{F_i}}=0$, the resulting action is non-dynamical. Let us recall why that is the case. The physically meaningful quantity is not the vectorial sector of $h_{\mu\nu}$, but rather the special combination, $V_i=S_i-\dot{F_i}$. The second term in~\eqref{eq:ACTIONJD} however, breaks diffeomorphism invariance of the action and thus makes meaningful all ten components of the auxiliary variable. This is of particular importance for the vectorial sector, for the original equation of motion (for $V_i$) is non-dynamical. For the action~\eqref{eq:ACTIONJD}, we then have two choices: either we preserve diffeomorphism invariance\footnote{without introducing additional fields as in the Stueckelberg trick.} and work with the invariant auxiliary variables, $V_i$, or we break it, as in~\eqref{eq:ACTIONJD}. Neither choice recovers the action in~\eqref{eq:vector}.

Finally, for the tensorial sector no complication arises, as $\tilde{h}_{\mu\nu}^{TT}$ is diffeomorphism invariant.

\subsubsection{The Gravitational Case II: A Conserved Diffusion Matrix}

In this section, we perform the calculations of the above section using a conserved diffusion matrix instead. This approach thus resolves one of the two issues with~\eqref{eq:SD}, namely its violation of the Bianchi identity, while also resolving the inconsistency between the propagators obtained from the OM and the JD actions.

Inspired by the work of Hirotani and Matsumura~\cite{hirotani2026testing}, we consider the alternative stochastic differential equation (SDE) formulation,
\begin{align}
\begin{split}
\label{SD:BR}
    G_{\mu \nu}^{(1)}&=\xi_{\mu \nu}\\\langle\xi_{\mu \nu}\xi_{\rho \sigma}\rangle&=\alpha^{-1}((1-3\frac{\beta}{\alpha})^{-1}P^{(0-s)}+P^{(2)})_{\mu\nu,\rho\sigma} \\ \langle\xi_{\mu \nu}\xi_{\rho \sigma}\rangle&=\mathcal{D}'_{2\rho\sigma,\mu\nu}
\end{split}
\end{align}
Although this matrix is also indefinite, one can verify that the diffusion matrix is indeed conserved by the transversality of the projectors used here. This matrix is, however, degenerate, and additional `gauge-fixing' terms must be added to invert it so as to give, for example, an inverse diffusion matrix of the form
\begin{align}
\label{eq:Diff2}
    (\langle\xi\xi\rangle)^{-1 \mu\nu,\rho\sigma}&=\alpha((1-3\frac{\beta}{\alpha})P^{(0-s)}+P^{(2)}+\frac{1}{\lambda}(P^{(0-w)}+P^{(1)}))^{\mu\nu,\rho\sigma}
\end{align}
where we can set the parameter $\lambda$ to be as small as we please. The key here is that for all the choices above, the inverse diffusion matrix is saturated by conserved tensors, and so the gauge-fixing terms all drop out. Thus, only the transverse part of the inverse diffusion matrix remains and so we obtain the same action as was obtained from the DeWitt metric:
\begin{align}
   S'_{G}= \frac{1}{2}\int d^4x \int d^4y G_{\mu\nu}(x) \mathcal{D}_2^{'-1\mu\nu,\rho\sigma}G_{\rho\sigma}(y)=\frac{1}{2}\int d^4x (\alpha R^{\mu\nu}R_{\mu\nu}-\beta R^2).
\end{align} Thus, in the OM formalism, both choices of diffusion matrix give us the same action and, consequently, the same two-point function.

Now, applying the JD formalism with $\mathcal{D}_2'$, we have the JD action, \begin{align}
\label{eq:JD2}
    S'_{JD}=\int d^4x h^{\mu\nu}\hat{O}_{\mu\nu,\rho\sigma}\tilde{h}^{\rho\sigma}-\frac{1}{2\alpha}\tilde{h}^{\mu\nu}((1-3\frac{\beta}{\alpha})^{-1}P^{(0-s)}+P^{(2)})_{\mu\nu,\rho\sigma}\tilde{h}^{\rho\sigma}\, ,
\end{align} giving us (up to gauge terms) the following two-point function,
\begin{align}
    \langle h^{\mu\nu}h^{\rho\sigma}\rangle=\frac{4}{\alpha\Box^2}(P^{(2)}+\frac{1}{4(1-\frac{3\beta}{\alpha})}P^{(0-s)})^{\mu\nu,\rho\sigma}
\end{align}
This matches perfectly with~\eqref{eq:2ppp}.

Note that in the JD approach, since we do not have to invert our diffusion matrix, there is no need to add `gauge-fixing' terms to it.

\subsubsection{The reason for the inconsistency and the role of Gauge-fixing}

The inconsistency between the propagators obtained from the OM action, and the JD action when using the diffusion matrix \eqref{eq:SD}, arises from the fact that the action \eqref{eq:JD} is not invariant under linearized diffeomorphisms, while \eqref{eq:Mink} is. This follows from the fact that the diffusion matrix, \eqref{eq:SD} is not a conserved tensor. One can indeed write $\mathcal{D}_2$ in terms of projection operators:
\begin{align}
\begin{split}
\label{eq:spindiff}
   \mathcal{D}_{2,\mu\nu,\rho\sigma}&=  \frac{1}{\alpha}(I_{\mu\nu,\rho\sigma}-\frac{\beta}{4\beta-\alpha}\eta_{\mu\nu}\eta_{\rho\sigma})\\&=D_2(P^{(0-s)}+P^{(0-w)}+P^{(1)}+P^{(2)}-\frac{\beta}{4\beta-\alpha}(3P^{(0-s)}\\&+\sqrt{3}(P^{(0-sw)}+P^{(0-ws)})+P^{(0-w)}))_{\mu\nu,\rho\sigma}
\end{split}
\end{align}
The longitudinal terms thus render the entire action non-gauge invariant. The OM action, on the other hand, remains gauge-invariant, due to the fact that it is saturated by the Einstein tensor, which is conserved. Saturating the inverse diffusion matrix with the Einstein tensor, and then inverting the resultant action is therefore \textit{not} equivalent to saturating the diffusion matrix with the inverted Einstein-tensor for non-conserved matrices. In order to have a diffeomorphism invariant JD path integral, and thus consistent two-point functions, one must use a conserved diffusion matrix.
On the other hand, as we have noted above, this issue simply does not arise in the OM formalism.

Further, one can discuss gauge-fixing in the context of diffusion. Suppose one has a transverse diffusion matrix, indicating that diffusion only occurs on the transverse subspace. Pseudo-inverting it and placing the inverse diffusion matrix in the action results in an action with a weight on only the transverse subspace. Prima facie, this would mean that the longitudinal subspace is free to diffuse, at zero cost (i.e., the longitudinal direction is a zero eigenvalue of the action). Gauge-fixing therefore ensures that the longitudinal direction is completely fixed, and does not diffuse \emph{at all}, and that diffusion only takes place in the transverse subspace. By switching to the diffeomorphism invariant modes in the SVT decomposition, we implicitly gauge-fix the longitudinal sector.

\subsection{The Gravitational Path Integral Measure}
\label{GPIM}

In this section, we review the path integral measure for the gravitational action, largely following the presentation in~\cite{MAZUR}.
  \MS{The metric $g$} may be considered as a point in \MS{the space of metrics equipped with the DeWitt supermetric}, and the measure constructed on the
  space of the deformations at that given point. The space of metrics at a point (before quotienting by diffeomorphisms) is
  \MS{$GL(D,\mathbb{R})/O(D-1,1)$}, which is not a vector space; however, we may instead consider the tangent space \MS{at} $g$ which is a vector space, and
  then use the exponential map to arrive at the full manifold of metrics (assuming that the space of metrics is contractible). Thus, for each topology, we
  would have
  $Dg=D\delta g$, where $\delta g$ spans the tangent space \MS{at} $g$.
  Now we would like to separate out the physical modes in $\delta g$ from the gauge ones. The key point here is that the probability measure is not to be on
  the space of metrics, but on the space of metrics modulo diffeomorphisms. We must thus factor out the volume
  of the diffeomorphism group, as in the quantum gravitational path integral. Further, we must be mindful of the fact that although the linearized
  gravitational action is of the form of a Wiener measure, it is unbounded from below and must be carefully regularized. As a reminder, the diffeomorphism
  $x\rightarrow x+\xi$ results in the following metric transformation:
  \begin{align}
      \delta_\xi g_{\mu\nu}=2\nabla_{(\mu}\xi_{\nu)}
  \end{align}
  which can be rewritten in the following form
  \begin{align}
      \delta_\xi g_{\mu\nu}=\frac{1}{2}\nabla_{\alpha}\xi^{\alpha}g_{\mu\nu}+(L\xi)_{\mu\nu}
  \end{align}
  where $(L\xi)_{\mu\nu}$ is the traceless part of $\delta_\xi g_{\mu\nu}$ or, equivalently, the operator $L$ maps the vector $\xi_{\mu}$ to the traceless
  subspace. It can further be shown that the kernel of the operator $L^\dagger$ is spanned by transverse, traceless metric perturbations, i.e.,
  $h_{\alpha}^{\alpha TT}=\nabla^{\mu}h_{\mu\nu}^{TT}=0$. \MS{Since a diffeomorphism shifts both the trace $h=g^{\mu\nu}h_{\mu\nu}$ and the traceless part
  of $h_{\mu\nu}$, we} decompose the metric perturbation as
  \begin{align}
      \MS{h_{\mu\nu}=\frac{1}{4}\sigma g_{\mu\nu}+h_{\mu\nu}^{TT}+2\nabla_{(\mu}\xi_{\nu)}
      =\frac{1}{4}\big(\sigma+2\nabla\cdot\xi\big)g_{\mu\nu}+h_{\mu\nu}^{TT}+(L\xi)_{\mu\nu},}
  \end{align}
  \MS{where $\sigma\equiv h-2\nabla\cdot\xi$. Under $\xi\rightarrow\xi+\zeta$ both $\sigma$ and $h^{TT}_{\mu\nu}$ are invariant, so the action depends only
  on $(\sigma,h^{TT})$, and the change of variables $(h,\xi)\rightarrow(\sigma,\xi)$ has unit Jacobian. The trace, transverse-traceless and $L\xi$ parts are
  mutually orthogonal with respect to the DeWitt supermetric
  $G^{\rho\sigma\mu\nu}=\frac{1}{2}(g^{\rho\mu}g^{\sigma\nu}+g^{\rho\nu}g^{\sigma\mu}-C\,g^{\rho\sigma}g^{\mu\nu})$ --- the trace part is orthogonal to the
  traceless ones pointwise, for any $C\equiv\frac{2\beta}{\alpha}$, and $h^{TT}$ is orthogonal to $L\xi$ after integrating by parts, since $L^\dagger h^{TT}=0$. The trace direction
  has $G$-eigenvalue $1-2C$, so the normalization of the measure requires $C<\tfrac12$ or $\beta<\frac{\alpha}{4}$.} Then one can define the functional measure with respect to this
  metric
  \begin{align}
      \MS{\int D\delta g\, e^{-\frac{1}{2}\langle \delta g,\delta g\rangle}=1,\qquad
      \langle \delta g,\delta g\rangle=\int_\mathcal{M} d^4x\sqrt{-g}\,\delta g_{\rho\sigma}\,G^{\rho\sigma\mu\nu}\,\delta g_{\mu\nu}.}
  \end{align}
Thus, starting from a path integral of the form
  \begin{align}
       \int D\delta g e^{-S[g]}
  \end{align}
  one has two equivalent choices. The first is the usual \MS{Faddeev--Popov} procedure: adding gauge-fixing terms and factoring out the volume of the
  diffeomorphism group from the measure, while in the second one can decompose $\delta g$ in the form given above and factor out the part corresponding to
  the range of $L$. Thus,
  \begin{align}
     \MS{\int D\delta g\, e^{-S[g]}=\int J_1\,D\sigma\, D\xi\, Dh^{TT} e^{-S[g]}}
  \end{align}
  \MS{The only non-trivial Jacobian is $J_1=\big[\mathrm{Det}'(L^\dagger L)\big]^{1/2}$, where the prime denotes that the zero modes of $L$ (conformal
  Killing vectors) are removed. In flat space, up to constant factors, $J_1=\mathrm{Det}'(\Box)^{1/2}_S\,\mathrm{Det}'(\Box)^{1/2}_V$, relative to the
  unsplit measure $D\xi$; if one instead splits $\xi_\mu=\xi^T_\mu+\partial_\mu\chi$, that split carries its own Jacobian $\mathrm{Det}'_S(-\Box)^{1/2}$,
  and the fourth-order scalar block then contributes a full power $\mathrm{Det}'_S(-\Box)$, consistently. Since the action does not depend on $\xi$, the
  $\xi$ integral yields the volume of the diffeomorphism group,}
  \begin{align}
      \MS{\int J_1\,D\sigma\, D\xi\, Dh^{TT} e^{-S[g]} =Vol(Diff(M))\,J_1\int D\sigma\, Dh^{TT} e^{-S[g]},}
  \end{align}
  \MS{and dividing by $Vol(Diff(M))$ gives the path integral over geometries, $J_1\int D\sigma\,Dh^{TT} e^{-S[g]}$. But} this is evidently the same as
  \begin{align}
      \MS{J_1\int D\sigma\, Dh^{TT} e^{-S[g]}=K_a\int DV\,D\,{}^3h^{TT}\,D\Phi\, D\psi\, e^{-S[g]}}
  \end{align}
  where \MS{$K_a$ is the product of $J_1$ with the Jacobian of the change of basis from $(\sigma,{}^4h^{TT})$} to the gauge invariant coordinates, $(\psi,
  \Phi, \MS{{}^3h^{TT}},V)$. What we have essentially done is to write the $5+1$ modes of $(\MS{{}^4h^{TT},\sigma})$ in a new, non-manifestly covariant
  basis. Further, \MS{$\mathrm{Det}'(\Box)^{1/2}_V=\mathrm{Det}'(\Box)^{3/2}$ is the
  determinant over the three transverse components of $\xi_\mu$, and $\mathrm{Det}'(\Box)^{1/2}_S$ that over its longitudinal component}. We can use this
  then to check \MS{whether} the non-manifestly covariant approach gives us the correct number of \MS{box operators, each of which would correspond to a
  dynamical mode in the usual, non-stochastic sense}: each factor of $\mathrm{Det}(\Box)^{1/2}$ transforms a mode into a constraint, and hence we have
  $2\times5+2-4=8$ \MS{box operators}. This is indeed the same number of box operators as we obtain in the non-manifestly covariant formulation\MS{ --- two
  for each of the two tensor modes, two for $\psi$, one for each of the two components of $V_i$, and none for $\Phi$ --- although, as argued in
  Section~\ref{DOF}, only the tensor modes and $\psi$ are dynamical, while $V_i$ and $\Phi$ are constraints}.

\subsection{Choice of discretization}
\label{discretization}
Here we show that the choice of discretization does not matter, mainly based on the presentation in~\cite{CriticalDynamics}. Consider the following Langevin equation
\begin{align}
    \frac{\phi_{i+1}-\phi_{i}}{\epsilon}=\lambda F_{i+1}+(1-\lambda)F_{i}+\xi_i
\end{align}
where $0\leq \lambda \leq1$ and $F_i$ is the force at $t_i$. Note that the choice of discretization is equivalent to setting the value of the Heaviside function at the origin, $\Theta(0)=\lambda$. The Jacobian of the change of variable from $\phi_{i+1}\rightarrow\xi_i$ is a lower triangular matrix with entries only on the main, and the lower sub-diagonal. Thus the functional determinant is given by
\begin{align}
    \frac{1}{\epsilon^f} \prod_i(1-\lambda\epsilon\frac{\partial F_i}{\partial \phi_{i}})\approx\frac{1}{\epsilon^f}\exp(-\lambda\sum_i\epsilon\frac{\partial F_i}{\partial \phi_{i}})
\end{align}
Taking the limit $\epsilon\rightarrow0$ and absorbing the factor $\frac{1}{\epsilon^f}$ in the functional measure, we obtain the term $-\lambda\int d^3x\int dt\frac{\partial F}{\partial \phi}$. This is exactly canceled by the contribution from closed response loops contained in (the non-linear contribution to) terms of the form $\int d^3x\int dt F\tilde{\phi} $, which necessarily involve a contraction of $\tilde{\phi}$ with the internal field $\phi$ in the JD formalism
\begin{align}
    \int d^3x\int dt \phi\tilde{\phi}\frac{\partial F}{\partial \phi}=\int d^3x\int dt \Theta(0)\frac{\partial F}{\partial \phi}
\end{align}
as long as we set $\lambda=\Theta(0)$.

\subsection{Second Order SDEs and Heat Baths}\label{Jacobian}
Since the stochastic path integral in either of its iterations (OM or JD) is formulated using stochastic differential equations that are first order in time, let us explicitly write down the second order equations as first order ones. It is convenient to do so, as working with second order SDEs requires one to compute a Jacobian that is non-lower triangular. Our exposition mainly follows~\cite{Chaichian2018PathI}. Working in spatial Fourier space we have
\begin{align}
\begin{split}
    \partial_t\varphi&=\pi_{\varphi}\\
    \partial_t\pi_{\varphi}+k^2\varphi&=\xi
    \end{split}
\end{align}
We can consider this as a coupled system of the variables, $\varphi,\pi_{\varphi}$, undergoing stochastic diffusion in baths of different temperatures, $T_1,T_2$.
\begin{align}
\begin{split}
    \label{eq:SDE11}\partial_t\varphi-\pi_{\varphi}&=T_1\xi\\
    \partial_t\pi_{\varphi}+k^2\varphi&=T_2\xi
    \end{split}
\end{align}
Compare this with the SDE for Brownian motion of two particles:\begin{align}
\begin{split}
    \label{eq:SDE111}\partial_t\vartheta_1&=T_1\xi \\ \partial_t\vartheta_2&=T_2\xi\end{split}
\end{align}
We can write \eqref{eq:SDE11}, \eqref{eq:SDE111} in matrix form:
\begin{align}
    \begin{split}
\partial_t\boldsymbol{\vartheta}&=\boldsymbol{\xi}\\\partial_t\boldsymbol{\varphi}+\boldsymbol{D}\boldsymbol{\varphi}&=\boldsymbol{\xi}
    \end{split}
\end{align}
where
\begin{align}
    \boldsymbol{D}=\begin{bmatrix}
        0&-1\\k^2&0
    \end{bmatrix}
\end{align}
we obtain the Jacobian,
\begin{align}J_N=
\begin{bmatrix}
    \boldsymbol{1}+\boldsymbol{D}\frac{\epsilon}{2}\ &.&.&.&0\\\boldsymbol{D}\frac{\epsilon}{2}&\boldsymbol{1}+\boldsymbol{D}\frac{\epsilon}{2}&.&.&0\\.&.&.&.&.\\.&.&.&.&.\\\boldsymbol{D}\frac{\epsilon}{2} &.&.&.&\boldsymbol{1}+\boldsymbol{D}\frac{\epsilon}{2}
\end{bmatrix}=\exp{\frac{1}{2}t\text{Tr}\boldsymbol{D}}=1
\end{align}
where we have used the fact that the coefficient matrix is lower triangular. The Jacobian is therefore simply unity in this case. We thus have the following probability distribution,
\begin{align}
   \pdf(\varphi_f,\pi_{\varphi_f}|\varphi_i,\pi_{\varphi_i}) =\lim_{T_1\rightarrow 0, T_2\rightarrow1}\int D\varphi D\pi_{\varphi}\exp{-\int dt d^3k (\frac{1}{T_1}( \partial_t\varphi-\pi_{\varphi})^2+\frac{1}{T_2}( \partial_t\pi_{\varphi}+k^2\varphi)^2)}
\end{align}
This gives us a delta function at every time slice, which we may use to integrate out $\pi_{\varphi}$ for intermediate times. In addition, one must integrate over all states, $\pi_{\varphi_i},\pi_{\varphi_f}$, if we want the probability distribution to be solely a function of $\varphi_i,\varphi_f$.

Alternatively, one may simply start with a second order action:
\begin{align}
    \pdf(\varphi_f,\pi_{\varphi_f}|\varphi_i,\pi_{\varphi_i}) =\int D\varphi \exp{-\int dt d^3k ( \partial_t^2 \varphi+k^2\varphi)^2}
\end{align}

and obtain the same results.

\subsection{Pole Prescription}
\label{Section:Poles}
Here we discuss the pole prescriptions for the two-point function, as justified in \cite{oppenheim2025diffusion}. The pole prescription for the infinite time two-point function is given by
\begin{align}
\label{eq:pole prescription}
    \frac{1}{p^4}=\frac{1}{(k^2-(\omega-i\epsilon)^2)(k^2-(\omega+i\epsilon)^2)}
\end{align}
We can obtain this directly from the action~\eqref{eq:JD}, where the limits of integration run from $-\infty$ to $+\infty$, and we can work in momentum space:
\begin{align}
      \hat{O}_{\mu\nu,\rho\sigma}=\Box_{\epsilon}(-\tfrac{1}{2}P^{(2)}+P^{(0-s)})_{\mu\nu,\rho\sigma},\qquad \Box_{\epsilon}:=-\big(k^2-(\omega+i\epsilon)^2\big)
\end{align} where we directly add to the (Fourier transform of the) box operator the term in $i\epsilon$ corresponding to the retarded prescription for its Green's function. Then, we can integrate out the auxiliary variable tensor field, $\tilde{h}^{\mu\nu}$ to obtain the following action,

\begin{align}
    \mathcal{S}_{G}=\frac{1}{2}\int d^4p\int d^4p'\,\Box_{\epsilon}(-\tfrac{1}{2}P^{(2)}+P^{(0-s)})_{\mu\nu,\kappa\lambda}h^{\kappa\lambda}(p)\, D_2^{-1}I^{\mu\nu,\mu'\nu'}(p,p')\,\Box_{\epsilon}(-\tfrac{1}{2}P^{(2)}+P^{(0-s)})_{\mu'\nu',\rho\sigma}h^{\rho\sigma}(p')
\end{align}where
\begin{align}
  I^{\mu\nu,\rho\sigma}(p,p')=\delta^4(p+p')(\frac{1}{2}(\eta^{\mu\rho}\eta^{\nu\sigma}+\eta^{\mu\sigma}\eta^{\nu\rho})-\frac{\beta}{\alpha}\eta^{\mu\nu}\eta^{\rho\sigma})
\end{align}Integrating over $p'$, we obtain
\begin{align}
\begin{split}
        \mathcal{S}_{G}&=\frac{1}{2}\int d^4p\,\Box_{\epsilon}(-\tfrac{1}{2}P^{(2)}+P^{(0-s)})_{\mu\nu,\kappa\lambda}h^{\kappa\lambda}(p)\, D_2^{-1}I^{\mu\nu,\mu'\nu'}\,\Box_{-\epsilon}(-\tfrac{1}{2}P^{(2)}+P^{(0-s)})_{\mu'\nu',\rho\sigma}h^{\rho\sigma}(-p)\\&=\frac{1}{2}\int d^4 p (\alpha R^{\mu\nu}R_{\mu\nu}(p)-\beta R^2(p))
\end{split}
\end{align}From the equivalence of the two-point functions obtained from the scalar-vector-tensor decomposition and the covariant two-point function, the pole prescription for the box operator $p^2$ must be identical, and given by~\eqref{eq:pole prescription}. That this is indeed the correct pole prescription may also be directly verified via~\eqref{eq:MJD}, without integrating out auxiliary fields.

Similarly, for the finite time case, one may treat the spin-0 propagator as the convolution of retarded and advanced Green's functions, $G(p)_{\rho\sigma}=\frac{\theta_{\rho\sigma}}{\sqrt{3(\alpha-3\beta)}p^2}$. The projector $\theta_{\rho\sigma}$ introduces no additional pole in the transverse sector and so, the retarded prescription for $1/p^2$ is the only one required.

We can also go back to the scalar-vector-tensor decomposition, and consider the action given by~\eqref{SDE:scalar} and \eqref{SDE:tensor}. Then, we simply assume that the two-point function is the convolution of two retarded Green's functions. If we do not wish to integrate out $\Phi$, we may instead take~\eqref{eq:twopointscalar} to be of the form $\mathcal{G}(p)\mathcal{G}(p')$ where $\mathcal{G}(p)$ is the matrix of retarded Green's functions for the stochastic differential equations giving rise to the action~\eqref{eq:scalar}. The exact form of $\mathcal{G}$ does not concern us apart from the fact that it is proportional to $\frac{1}{p^2}$, where we must use the retarded prescription.

Additionally, given that the matrix $\mathcal{G}(p)\mathcal{G}(p')$ must be symmetric in $p,p'$, we can rewrite the $\langle\Phi\Phi\rangle$ element in~\eqref{eq:twopointscalar} as
\begin{align}
    \langle\Phi(p)\Phi(p')\rangle=\frac{1}{8\alpha(\alpha-3\beta)} (\frac{8(\alpha-3\beta)}{k^2k'^2}+\frac{4\omega^2\omega'^2(\alpha-3\beta)}{k^2k'^2p^2p'^2}+\frac{(-2\alpha+8\beta)}{p^2p'^2})
\end{align}

\section{Calculational Details for the Spectral Acceleration}
\label{Details}
For the two-point function, we assume that it is the convolution of two retarded Green's functions (for an alternate derivation, see further below):
\begin{align}
\langle\Phi \Phi \rangle=\int ^{\infty}_{-\infty}d^3z\int^{t_f}_0dz^0G_{ret}(x-z)G_{ret}(y-z)=\int ^{\infty}_{-\infty}d^3z\int^{t_f}_0dz^0\int\frac{d^4p}{(2\pi)^4}\int\frac{d^4p'}{(2\pi)^4}G(p)G(p')
\end{align}
where
\begin{align}
    G(p)G(p')=\frac{1}{4\alpha(\alpha-3\beta)} (\frac{8(\alpha-3\beta)}{k^2k'^2}+\frac{4\omega^2\omega'^2(\alpha-3\beta)}{k^2k'^2p^2p'^2}+\frac{(-2\alpha+8\beta)}{p^2p'^2})
\end{align}
Using
\begin{align}
\label{eq:time}
    \int^\infty_0e^{-iz^0(\omega+\omega')}dz^0=\pi\delta(\omega+\omega')-iP.V(1/(\omega+\omega'))
\end{align}
we obtain,
\begin{align}
    \int ^{\infty}_{-\infty}d^3z\int^{t_f}_0dz^0\int\frac{d^4p}{(2\pi)^4}\int\frac{d^4p'}{(2\pi)^4}G(p)G(p')=\mathcal{G}_{\text{stationary}}+\mathcal{G}_{\text{non-stationary}}
\end{align}
where
\begin{align}
    \mathcal{G}_{\text{stationary}}=\int\frac{d^4p}{(2\pi)^4}\frac{1}{8\alpha(\alpha-3\beta)} (8(\alpha-3\beta)/k^4+4\omega^4(\alpha-3\beta)/k^4p^4+(-2\alpha+8\beta)/p^4)
\end{align}
where $p^4=(k^2-(\omega-i\epsilon)^2)(k^2-(\omega+i\epsilon)^2)$
Meanwhile, for $\mathcal{G}_{\text{non-stationary}}$ we need only consider the non-stationary parts of terms of the form $\omega^2\omega'^2/k^2k'^2p^2p'^2,1/p^2p'^2$. Let us start with $\omega^2\omega'^2/k^2k'^2p^2p'^2$: the Fourier transform of the non-stationary part is given by
 \begin{align}
    \begin{split}
    \mathcal{FT}(\omega^2\omega'^2/k^2k'^2p^2p'^2)_{\text{non-stationary}}&=-\int\frac{d^3k}{(2\pi)^3} (e^{-i\vec{k}.(\vec{x}-\vec{y})})e^{-\epsilon (x^0+y^0)}\\
    &\quad\frac{1}{4k^6}(4k^3\sin(k(x^0+y^0))+k^4\cos(k(x^0-y^0))/\epsilon)
    \end{split}
\end{align}
where we have only kept terms up to $e^0$ (for simplicity, we have kept the exponential as it is, but later on, when expanding, we will only keep terms up to an overall $e^0$). Integrating over angles:
\begin{align}
\mathcal{FT}(\omega^2\omega'^2/k^2k'^2p^2p'^2)_{\text{non-stationary}}= - \int\frac{dke^{-\epsilon (x^0+y^0)}}{8\pi^2|x-y|}\frac{1}{k^2}(4\sin(k(x^0+y^0))+k\cos(k(x^0-y^0))/\epsilon)\sin{(k(|x-y|))}
\end{align}
Similarly, for $1/p^4$,
\begin{align}
 \mathcal{FT}(1/p^2p'^2)_{\text{non-stationary}}= - \int\frac{dke^{-\epsilon (x^0+y^0)}}{8\pi^2|x-y|}\frac{1}{k^2}(\sin(k(x^0+y^0))+k\cos(k(x^0-y^0))/\epsilon)\sin{(k(|x-y|))}
\end{align}
Using the following integrals:
\begin{align}
\begin{split}
    \int^\infty_0 \frac{1}{k^2}\sin{ka}\sin{kb}&=ab\int^\infty_0 \text{sinc}ka \text{sinc}kb=\frac{ab\pi}{|a-b|+|a+b| }\\ \int^\infty_0 \frac{1}{k}\sin{ka}\cos{kb}&=a\int^\infty_0 \text{sinc} (ka)\cos{kb}\\&=\frac{\pi}{4}(\text{sgn}(a-b)+\text{sgn}(a+b)) \\\int^{\infty}_{-\infty}\frac{\pi}{4}(\text{sgn}(a-b)+\text{sgn}(a+b))e^{i\omega b}db&=\pi\sin{(\omega a)}/\omega.
   \end{split}
\end{align}

Thus, we obtain the following time-dependent parts, in frequency space, where by $\mathcal{FT}_{\omega}$ we refer to the full Fourier transform, followed by an inverse Fourier transform with respect to $(x^0-y^0)$:
\begin{align}
  \lim_{\epsilon\rightarrow0} \mathcal{FT}_{\omega}(1/p^4)=\frac{1}{8\pi|x-y|}(\frac{\sin{\omega|x-y|}}{\omega}(\frac{1}{\epsilon}-|x-y|)-\frac{|x-y|(x^0+y^0) }{|x^0+y^0-|x-y||+|x^0+y^0+|x-y||}\pi\delta(\omega))
\end{align}
where we have
\begin{align}
    \delta(\omega)=\frac{1}{2\pi}\int^\infty_{-\infty}d(x^0-y^0)e^{i\omega(x^0-y^0)}
\end{align}
and a similar expression from the time-dependent term of the form $\omega^4/k^4p^4$. Here $1/\epsilon$ is the regulated total elapsed time: the finite-time treatment of Appendix~\ref{Alternate}, in which the initial slice is held at $t=0$, replaces it by $x^0+y^0$, so that the combination $\tfrac{1}{\epsilon}-|x-y|$ becomes $x^0+y^0-|x-y|$ as in the results below.
For the time-independent part we have
\begin{align}
  \begin{split} \mathcal{FT}_{\omega}\frac{\omega^4}{k^4p^4}(x,\omega)=\frac{\omega^4}{8\pi x(\omega^2+\epsilon^2)^4}((-x^2\epsilon^4+4\omega^2-x^2\omega^4-2\epsilon^2(2+x^2\omega^2))+&\\\frac{e^{-\epsilon x}}{\epsilon\omega}(\sin{\omega x}(\epsilon^4-6\epsilon^2\omega^2+\omega^4)+\epsilon\cos(\omega x)(4\epsilon^2\omega-4\omega^3)))
    \end{split}
\end{align}
and
\begin{align}
\lim_{\epsilon\rightarrow0}\mathcal{FT}_{\omega}\frac{\omega^4}{k^4p^4}=&\frac{1}{8\pi x}(-x^2+\frac{4}{\omega^2}+(\frac{\sin{\omega x}}{\omega}(\frac{1}{\epsilon}-x)-\frac{4\cos(\omega x)}{\omega^2}))
\end{align}
\begin{align}
\lim_{\epsilon\rightarrow0}\mathcal{FT}_{\omega}\frac{1}{p^4}=\frac{\sin{\omega|x|}}{8\pi\omega|x|}(1/\epsilon-|x|)
    \end{align}
We thus obtain for the (wide-sense) stationary part
\begin{align}
  \int d^3k\frac{1}{(2\pi)^3(k^2-(\omega-i\epsilon)^2)(k^2-(\omega+i\epsilon)^2)}=\frac{1}{(2\pi)^2i|x|}\int ^{\infty}_{-\infty}dk(\frac{ke^{ik|x|}}{(k^2-(\omega-i\epsilon)^2)(k^2-(\omega+i\epsilon)^2)})&\\=\frac{\sin{\omega|x|e^{-\epsilon|x|}}}{8\pi\omega\epsilon|x|}
\end{align}And similarly,
\begin{align}
  \int d^3k\frac{k^2}{(2\pi)^3(k^2-(\omega-i\epsilon)^2)(k^2-(\omega+i\epsilon)^2)}=\frac{1}{(2\pi)^2i|x|}\int ^{\infty}_{-\infty}dk(\frac{k^3e^{ik|x|}}{(k^2-(\omega-i\epsilon)^2)(k^2-(\omega+i\epsilon)^2)})&\\=\frac{((\omega^2-\epsilon^2)\sin{\omega|x|+2\omega\epsilon\cos{\omega|x|})e^{-\epsilon|x|}}}{8\pi\omega\epsilon|x|} ,
\end{align}\begin{align}
  \int d^3k\frac{1}{(2\pi)^3k^2(k^2-(\omega-i\epsilon)^2)(k^2-(\omega+i\epsilon)^2)}=\frac{1}{(2\pi)^2i|x|}\int ^{\infty}_{-\infty}dk(\frac{e^{ik|x|}}{(k)(k^2-(\omega-i\epsilon)^2)(k^2-(\omega+i\epsilon)^2)})&\\=\frac{e^{-\epsilon|x|}}{8\pi\omega\epsilon|x|} (\frac{(\omega^2-\epsilon^2)\sin{\omega|x|}-2\omega\epsilon\cos{\omega|x|}+2e^{\epsilon|x|}\omega\epsilon}{(\omega^2+\epsilon^2)^2}))
\end{align}
where we have taken the principal value.
Finally,
\begin{align}
    \lim_ {\epsilon \rightarrow0} S(\vec{x},\omega)=\frac{1}{64\pi\alpha(\alpha-3\beta)|x-y|}((2\alpha-4\beta)\omega\sin(\omega|x-y|)(\frac{1}{\epsilon}-|x-y|)+2(20\beta-6\alpha)\cos{\omega|x-y|}+2(12\alpha-36\beta))
\end{align}One thus obtains
\begin{align}
\begin{split}
    \langle \Phi \Phi\rangle(|x-y|,x^0+y^0,\omega)&=\frac{1}{64(\alpha-3\beta)\alpha\pi}(-12(\alpha-3\beta)|x-y|+\frac{16(\alpha-3\beta)}{\omega^2|x-y|}\\&+\frac{(2\alpha- 4\beta)\sin{\omega |x-y|}}{\omega|x-y|}(x^0+y^0-|x-y|)\\&-\frac{\cos{\omega |x-y|}}{\omega^2|x-y|}(16(\alpha-3\beta))-\frac{2\pi\delta(\omega)(14\alpha-40\beta)(x^0+y^0)}{(|x^0+y^0-|x-y||+|x^0+y^0+|x-y||)})
    \end{split}
\end{align}
For the spectral density, we have
\begin{align}
\begin{split}
  \langle\partial_i \Phi \partial^i\Phi\rangle&=-\nabla^2\langle \Phi \Phi\rangle \\&=\frac{1}{64(\alpha-3\beta)\alpha\pi}(\frac{24(\alpha-3\beta)}{|x-y|}+(2\alpha-4\beta)(\frac{\omega\sin{\omega|x-y|}}{|x-y|}(x^0+y^0-|x-y|))+\frac{(40\beta-12\alpha)\cos{\omega |x-y|}}{|x-y|})
     \end{split}
\end{align}
where the second step follows by the linearity of the expectation value.

\subsubsection{An Alternative Derivation }
\label{Alternate}

Alternatively, one may start from $\mathcal{G}(t,s;k)$, i.e., the Green's function of the operator, $(\partial_t^2+k^2)^2$ subject to the boundary conditions, $\mathcal{G}(0)=\dot{\mathcal{G}}(0)=\hat{L}\mathcal{G}(T)=\dot{\hat{L}}\mathcal{G}(T)=0$ where $\hat{L}=(\partial_t^2+k^2)$. Note that $\mathcal{G}$ is in fact identical to the convolution of retarded and advanced Green's functions for the operator, $\hat{L}$ which we have discussed above. In momentum space, one thus finds, after making the change of variables, $t-s\rightarrow\tau,\frac{t+s}{2}\rightarrow T$
\begin{align}
    \mathcal{G}(\tau,T;k)=\frac{1}{4k^3}\left((2T-|\tau|)k\cos(k|\tau|)-\sin(2kT)+\sin(k|\tau|)\right)
\end{align}
where we have put absolute value signs on $\tau$ to allow for the possibility, $s>t$. Restoring factors of $2\pi$, which we have from Fourier transforming the $\Box^2$ operator in (spatial) momentum space, and then performing the inverse Fourier transform, we obtain:
\begin{align}
    \mathcal{G}(\tau,T;r)=\frac{\pi}{(2\pi)^3}\int_0^{\infty}dk\frac{\sin{(kr)}}{k^2r}\left(
    (2T-|\tau|)k\cos{(k|\tau|)-\sin{(2kT)}+\sin{(k|\tau|)}}\right)
\end{align}
Using the following integrals for $a>0$:
\begin{align}
\begin{split}
    \int^\infty_0 \frac{1}{k^2}\sin{ka}\sin{kb}
    &=\frac{ab\pi}{|a-b|+|a+b| }\\&=\frac{\pi}{2}\text{min}(a,b)\\ \int^\infty_0 \frac{1}{k}\sin{ka}\cos{kb}&=\frac{\pi}{4}(\text{sgn}(a-b)+\text{sgn}(a+b))\\&=\frac{\pi}{2}\Theta(a-|b|)
 \end{split}
\end{align}
we have
\begin{align}
\begin{split}
    \mathcal{G}(\tau,T;r)&=\frac{1}{16\pi r}((2T-|\tau|)\Theta(r-|\tau|)+(\text{min}(r,|\tau|)-\min(r,2T)))\\&=\frac{2T-r}{16\pi r}(\Theta(2T-r)\Theta(r-|\tau|)-\Theta(r-2T)\Theta(|\tau|-r))\\&=\frac{2T-r}{16\pi r}\Theta(2T-r)\Theta(r-|\tau|)
\end{split}
\end{align}
since $2T>|\tau|$.

The inverse Fourier transform with respect to $\tau$ gives us
\begin{align}
\begin{split}
 \mathcal{G}(\omega,T;r)&=\frac{2T-r}{16\pi r}(\Theta(2T-r)\int_{-r}^{r}\exp{-i\omega\tau}d\tau +(\Theta(r-2T)\int_{-r}^{r}\exp{-i\omega\tau}d\tau-\int_{-\infty}^{\infty}\exp{-i\omega\tau}d\tau))\\&=\frac{2T-r}{8\pi r}(\frac{\sin(\omega r)}{\omega}-\Theta(r-2T)\pi\delta(\omega))
\end{split}
\end{align}
where we use $\Theta(|\tau|-r)=1-\Theta(r-|\tau|)$.

Additionally, consider $\mathcal{S}$:
\begin{align}
\begin{split}
\mathcal{S}&=\frac{\sin(k|\tau|)}{k^3}\\
    \mathcal{S}(\omega,T;r)&=\frac{1}{4\pi r}(2\pi r\delta(\omega)-\frac{4\sin{^2(\frac{\omega r}{2})}}{\omega^2})\\\mathcal{S}(\tau,T;r)&=\frac{1}{4\pi r}(\min(r,\tau))
\end{split}
\end{align}

and one has
\begin{align}
\begin{split}
    \frac{\partial_t^2\partial_s^2}{k^4}\mathcal{G}(t,s;k)&=\mathcal{G}-\mathcal{S}+\frac{\delta(t-s)}{k^4}\\ \frac{\partial_t^2+\partial_s^2}{k^2}\mathcal{G}(t,s;k)&=-2\mathcal{G}+\mathcal{S}
\end{split}
\end{align}
on which one can perform the spatial Fourier transform.

Finally, the Laplacian:
\begin{align}
\begin{split}
  -\nabla^2\mathcal{G}(\omega,T;r)&=\frac{\omega(2T-r)\sin{(\omega r)}}{8\pi r}+\frac{\cos{(\omega r)}}{4\pi r}-\frac{\delta(\omega)\delta(r-2T)}{16T}\\-\nabla^2\mathcal{S}&=\frac{\cos{(\omega r) }}{2\pi r}
\end{split}
\end{align}

We can now begin to consider the two-point function of $\Phi$ and its spectral acceleration:

\begin{align}
    \langle \Phi \Phi\rangle(\omega,T,r)=\frac{(\alpha-2\beta)}{4\alpha(\alpha-3\beta)}\mathcal{G}(\omega,T,r)-\frac{1}{2\alpha}\mathcal{S}(\omega,r)-\frac{3}{2\alpha}\frac{r}{8\pi}
\end{align}

and the spectral acceleration:
\begin{align}
\begin{split}
  -\nabla^2 \langle \Phi \Phi\rangle(\omega,T;r)&=\frac{\alpha-2\beta}{4\alpha(\alpha-3\beta)}(\frac{\omega(2T-r)\sin{(\omega r)}}{8\pi r}+\frac{\cos{(\omega r)}}{4\pi r}-\frac{\delta(\omega)\delta(r-2T)}{16T})\\&-\frac{1}{2\alpha}\frac{\cos{(\omega r) }}{2\pi r}+\frac{3}{2\alpha}\frac{1}{4\pi r}
\end{split}
\end{align}

The two derivations agree for $r\neq2T$ (which we had assumed in the first derivation to simplify matters), as they must.

\subsection{Fourier transforms}

The two point function for $\langle \Phi\Phi\rangle$ has a $\frac{1}{k^4}$ as well as a $\frac{1}{p^4}$ piece which is common to the two point functions for $\psi, h_{ij}^{TT}$. One may obtain them directly via an (inverse) Fourier transform, as long as one identifies the IR-divergent terms as the total time elapsed, or the size of the box for $\frac{1}{p^4},\frac{1}{k^4}$ respectively. 

Let us quote the results directly. For $\frac{1}{k^4}$, one has
\begin{align}
    \lim_{m\rightarrow0}\int\frac{1}{(2\pi)^3}\frac{d^3\vec{k}e^{i\vec{k}\vec{x}}}{k^4+m^2}&=\frac{e^{-mr}}{8\pi m}\\&=\frac{1}{8\pi m}-\frac{r}{8\pi}+O(m).
\end{align}

The finite part may also be obtained directly via a Hadamard regularization. The key point however, is that the finite part is only conditionally PSD, of order one. That is, suppose one has the vector $\vec{v}$ with the i-th entry, $c_i$. Then the kernel $A$ ( $-\frac{r}{8\pi}$) is only PSD for $\sum_ic_ip(x)=0$ where $p(x)$ is a polynomial of degree strictly less than one. The obstruction to unconditional positive semi-definiteness of the finite part is the zero mode, which is removed by adding a constraint on $\vec{v}$. 

Similarly, one may also compute the Fourier transform of $\frac{1}{p^4}$:
\begin{align}
    \lim_{\epsilon\rightarrow0}\int\frac{1}{(2\pi)^4}\frac{d^4p e^{ipx}}{p^4+\epsilon^2}&=\frac{1}{4\pi^2 \epsilon s^2}-\frac{\text{sgn}(s^2)}{32\pi}+O(\epsilon)
\end{align}
(see \cite{grudka2024renormalisation} for calculational details)
which matches exactly with the results above under a particular identification of $\epsilon$ and $\frac{1}{T}$ .
In this case, the obstruction to positive semi-definiteness of the finite part is no longer a single mode, but rather the hypersurface, $k^2=\omega^2$ and hence it is not conditionally positive semi-definite to any finite order.

\section{Non-PSD Modes}
\label{VectorSection}

At this juncture, one has two alternatives to ensure a sensible  (i.e., normalizable and CP) CQ dynamics: a) to either dispense with the requirement that the noise have a covariant diffusion matrix or b) to restrict the kernel to time-like $p$.  The former corresponds to a preferred frame, while the latter would result in the noise kernel having correlations in space and time, which is more reminiscent of an effective field theory description. We do not speculate about the origins of either and suspend a detailed investigation to future work; however, we will briefly consider the implications of either possibility.

\subsubsection{A non-covariant diffusion matrix}

The local inverse diffusion matrix which gives a positive definite action that reproduces Einstein gravity in the quasi-static limit is given by 

  \begin{equation}
  \big(D_2^{-1}\big)_{\rm A}^{\mu\nu,\rho\sigma}
   =\alpha\,g^{\mu(\rho}g^{\sigma)\nu}-\beta\,g^{\mu\nu}g^{\rho\sigma}
   +2\alpha\big(1+c_1\big)\,n^{(\mu}h^{\nu)(\rho}n^{\sigma)},
  \qquad h^{\mu\nu}=g^{\mu\nu}+n^\mu n^\nu
  \end{equation}

where $n^\mu=(1,0,0,0)$. Introducing the matrix, $\mathcal{V}^{\mu\nu,\rho\sigma}\equiv n^{(\mu}h^{\nu)(\rho}n^{\sigma)}$, we have $X.\mathcal{V}.X=X_{0i}^2$ given that in flat space, $h=\text{diag}(0,1,1,1)$. Contracting $D_{2,A}$ with an arbitrary 2-tensor gives us
\begin{align}
   X.D_2^{-1}.X=\alpha\bar{X}_{ij}^2 +2\alpha c_1 X_{0i}^2+\begin{bmatrix}
       X_i^i&X_{00}
   \end{bmatrix}\begin{bmatrix}
       \frac{\alpha}{3}-\beta&\beta\\\beta&\alpha-\beta
   \end{bmatrix}\begin{bmatrix}
       X_i^i\\X_{00}
   \end{bmatrix}\qquad \bar{X}_{ij}=X_{ij}-\frac{1}{3}\delta_{ij}X_i^i.
\end{align}
This is positive definite for any symmetric 2 tensor for $\alpha>0,c_1>0,\alpha>4\beta$. In terms of the SVT decomposition we thus obtain the following vector and scalar actions:
  \begin{equation}
  \mathcal{L}^{(A)}_{\rm vector}=\frac{\alpha}{4}\Big[\big(\partial_j\dot V_i\big)^{2}
   +c_1\big(\triangle V_i\big)^{2}\Big]
   \;\simeq\;\frac{\alpha}{4}\Big(\partial_j\Big[\dot V_i
   +\sqrt{c_1}\,\sqrt{-\triangle}\;V_i\Big]\Big)^{2}
  \end{equation}
and 
\begin{equation}
  \mathcal{L}^{\rm (A)}_{\rm scalar}
   =\frac{\alpha}{3}\Big[\triangle\big(\Phi+\psi\big)\Big]^{2}
   +\frac{2\big(\alpha-3\beta\big)}{3}\Big[\triangle\Phi-2\triangle\psi+3\ddot\psi\Big]^{2}+4\alpha\big(1+c_1\big)\big(\partial_i\dot\psi\big)^{2}
  \end{equation}
with the $\Phi$-integrated out action given by
\begin{equation}
\mathcal{L}^{(A)}_{\psi}=\frac{2\alpha\big(\alpha-3\beta\big)}{\alpha-2\beta}
   (\Box\psi)^{2}
+4\alpha\big(1+c_1\big)\big(\partial_i\dot\psi\big)^{2}.
  \end{equation}

The scalar two point function is then given by
  \begin{equation}
  \begin{pmatrix}
  \big\langle\Phi\Phi\big\rangle & \big\langle\Phi\psi\big\rangle\\[4pt]
  \big\langle\psi\Phi\big\rangle & \big\langle\psi\psi\big\rangle
  \end{pmatrix}
  =\frac{1}{\Lambda}
  \begin{pmatrix}\hat T^{2}&\hat T\\[2pt]\hat T&1\end{pmatrix}
  +\frac{1}{2\big(\alpha-2\beta\big)k^{4}}
  \begin{pmatrix}1&0\\[2pt]0&0\end{pmatrix} 
  \end{equation}
where
 \begin{equation}
  \hat T=\frac{-2(\alpha-3\beta)\omega^{2}+(\alpha-4\beta)k^{2}}{(\alpha-2\beta)k^{2}},
  \qquad
  \Lambda
   =\frac{4\alpha\big(\alpha-3\beta\big)}{\alpha-2\beta}p^4
   +8\alpha\big(1+c_1\big)\,\omega^{2}k^{2}
  \end{equation}
$\langle \psi \psi\rangle$ now has a decaying two point function, while the positivity condition is also altered from $\alpha>3\beta\rightarrow\alpha> 4\beta$.

This is in contrast to a non-local inverse diffusion matrix which still leads to a local action: 
 \begin{equation}
  \big(D_2^{-1}\big)_{\rm C}=\alpha\,P^{(2)}+\big(\alpha-3\beta\big)\,P^{(0-s)}+2\alpha\big(1+c_1\big)\,\mathcal{N},
  \qquad \mathcal{N}^{\mu\nu,\rho\sigma}=N^{(\mu}\!\perp^{\nu)(\rho}\!N^{\sigma)}
  \end{equation}

where 
 \begin{equation}
   N^{\mu}=\theta^{\mu}{}_{\nu}n^{\nu}
  =n^{\mu}+\frac{\omega}{p^{2}}p^{\mu}
  =n^{\mu}-\frac{(n\cdot\partial)\,\partial^{\mu}}{\Box},
  \qquad
  \perp^{\mu\nu}=\theta^{\mu\nu}-\frac{N^{\mu}N^{\nu}}{N\cdot N}
  =\theta^{\mu\nu}+\frac{p^{2}}{k^{2}}N^{\mu}N^{\nu}
  \end{equation}
  \begin{equation}
  \begin{gathered}
  N\cdot p=0,\qquad N\cdot N=-\frac{k^{2}}{p^{2}},\qquad
  \theta^{\mu\nu}p_\nu=\perp^{\mu\nu}p_\nu=0,\qquad \perp^{\mu\nu}N_\nu=0\\ N^{\alpha}X_{\alpha\beta}=X_{0\beta},\qquad
  \perp^{\nu\rho}X_{0\nu}X_{0\rho}=\big|X_{0a}\big|^{2},\quad a\in\{x,y\}
  \end{gathered}
  \end{equation}
  \begin{equation}
  n^{\mu}=(1,0,0,0),\qquad
  N^{\mu}=\frac{k}{p^{2}}\big(k,0,0,\omega\big),\qquad
  \perp^{\mu\nu}=\mathrm{diag}(0,1,1,0)
  \end{equation}

This inverse diffusion matrix is equivalent to the following matrix on the transverse subspace:
\begin{equation}
  \big(D_2^{-1}\big)_{\rm C'}^{\mu\nu,\rho\sigma}
     =\alpha\,g^{\mu(\rho}g^{\sigma)\nu}-\beta\,g^{\mu\nu}g^{\rho\sigma}
     +2\alpha\big(1+c_1\big)\,n^{(\mu}h^{\nu)(\rho}n^{\sigma)}
     -2\alpha\big(1+c_1\big)\frac{\omega^{2}}{k^{2}}\,n^{\mu}n^{\nu}n^{\rho}n^{\sigma}
  \end{equation}

This is the minimal modification case (of the action): only the vector action is modified, while the scalar and tensor actions, along with the positivity condition, remain unchanged. In either case, one has the same low frequency behavior, as described in \ref{Exp}.

Preferred frame effects could therefore theoretically provide another handle on this theory although given that the tensor sector is unchanged, it is unlikely that cosmological anisotropy hunts could detect them. On the other hand, sidereal modulation of the scalar sector is also much suppressed, and hence such effects are unlikely to be seen at the table-top level. Thus, one must turn towards the scalar and the tensor sector to constrain the parameters of this theory.

\section{Spin Projectors in Linearized Gravity}

The Barnes--Rivers spin projection operators provide a standard decomposition of symmetric rank-2 tensor fields in momentum space~\cite{barnes1963thesis,rivers1964lagrangian}, and have been widely used in higher-derivative gravity theories~\cite{van1973ghost,julve1978quantum}.

  \subsubsection{Transverse and Longitudinal Projectors}

  Given a momentum $k^\mu$ with $k^2 = k^\mu k_\mu$, define the
  transverse and longitudinal rank-2 projectors:
  \begin{align}
      \theta_{\mu\nu} &= \eta_{\mu\nu} - \frac{k_\mu k_\nu}{k^2}, \\
      \omega_{\mu\nu} &= \frac{k_\mu k_\nu}{k^2},
  \end{align}
  satisfying $\theta_{\mu\nu} + \omega_{\mu\nu} = \eta_{\mu\nu}$,\;
  $\theta_{\mu}{}^{\alpha}\theta_{\alpha\nu} = \theta_{\mu\nu}$,\;
  $\omega_{\mu}{}^{\alpha}\omega_{\alpha\nu} = \omega_{\mu\nu}$,\;
  $\theta_{\mu}{}^{\alpha}\omega_{\alpha\nu} = 0$.

\subsubsection{Space-time Representation}
We may write the following in space-time as
  \begin{align}
      \theta_{\mu\nu} &= \eta_{\mu\nu} - \frac{\partial_{\mu}\partial_{\nu}}{\Box}, \\
      \omega_{\mu\nu} &= \frac{\partial_{\mu}\partial_{\nu}}{\Box},
  \end{align}
where the expressions above are to be understood more properly as the Fourier transform of the momentum-space projectors.

  \subsubsection{Barnes--Rivers Spin Projection Operators}
  \label{Barnes-Rivers}
The identity operator on the space of symmetric rank-2 tensors is
  \begin{equation}
      I_{\mu\nu,\rho\sigma} = \frac{1}{2}\bigl(\eta_{\mu\rho}\eta_{\nu\sigma}
      + \eta_{\mu\sigma}\eta_{\nu\rho}\bigr).
  \end{equation}

  It decomposes into four orthogonal spin projectors:
  \paragraph{Spin-2 (transverse traceless):}
  \begin{equation}
  \label{eq:Spin-2}
      P^{(2)}_{\mu\nu,\rho\sigma} = \frac{1}{2}
      \bigl(\theta_{\mu\rho}\theta_{\nu\sigma}
          + \theta_{\mu\sigma}\theta_{\nu\rho}\bigr)
      - \frac{1}{3}\,\theta_{\mu\nu}\,\theta_{\rho\sigma}.
  \end{equation}

  \paragraph{Spin-1 (transverse vector):}
  \begin{equation}
      P^{(1)}_{\mu\nu,\rho\sigma} = \frac{1}{2}
      \bigl(\theta_{\mu\rho}\omega_{\nu\sigma}
          + \theta_{\mu\sigma}\omega_{\nu\rho}
          + \theta_{\nu\rho}\omega_{\mu\sigma}
          + \theta_{\nu\sigma}\omega_{\mu\rho}\bigr).
  \end{equation}

  \paragraph{Spin-0, scalar ($s$):}
  \begin{equation}
  \label{eq:spin-0}
      P^{(0-s)}_{\mu\nu,\rho\sigma} = \frac{1}{3}\,\theta_{\mu\nu}\,\theta_{\rho\sigma}.
  \end{equation}

  \paragraph{Spin-0, scalar ($w$):}
  \begin{equation}
      \label{eq:spin0}P^{(0-w)}_{\mu\nu|\rho\sigma} = \omega_{\mu\nu}\,\omega_{\rho\sigma}.
  \end{equation}

  \subsubsection{Completeness and Orthogonality}

  \begin{align}
      I_{\mu\nu|\rho\sigma}
      &= P^{(2)}_{\mu\nu|\rho\sigma}
       + P^{(1)}_{\mu\nu|\rho\sigma}
       + P^{(0-s)}_{\mu\nu|\rho\sigma}
       + P^{(0-w)}_{\mu\nu|\rho\sigma}, \\[4pt]
      P^{(i)}_{\mu\nu|}{}^{\rho\sigma}\,P^{(j)}_{\rho\sigma|\kappa\lambda}
      &= \delta^{ij}\,P^{(i)}_{\mu\nu|\kappa\lambda}.
  \end{align}

  The degrees of freedom count ($d=4$) is $5 + 3 + 1 + 1 = 10$, consistent with a symmetric rank-2 tensor.

  \subsubsection{Off-Diagonal Scalar Mixer}

  When diagonalizing the scalar sector it is useful to define the
  transition operator
  \begin{equation}
      P^{(0-sw)}_{\mu\nu|\rho\sigma} =
      \frac{1}{\sqrt{3}}\,\theta_{\mu\nu}\,\omega_{\rho\sigma},\qquad
      P^{(0-ws)}_{\mu\nu|\rho\sigma} =
      \frac{1}{\sqrt{3}}\,\omega_{\mu\nu}\,\theta_{\rho\sigma},
  \end{equation}
  satisfying $P^{(0-sw)} = \bigl(P^{(0-ws)}\bigr)^\dagger$ and
  $P^{(0-s)}\,P^{(0-sw)} = P^{(0-sw)}$,\;
  $P^{(0-sw)}\,P^{(0-w)} = P^{(0-sw)}$.

\section{Boundary Terms}
Given that the Einstein-Hilbert (EH) action has higher derivatives, in the presence of a boundary the Gibbons-Hawking-York (GHY) term is required to make the variational principle well-defined, or alternatively to make the composition law well-defined
\begin{align}
    \langle g_{ij}(t_3)|g_{ij}(t_1)\rangle=\sum_{g_{ij}(t_2)}\langle g_{ij}(t_3)|g_{ij}(t_2)\rangle\langle g_{ij}(t_2)|g_{ij}(t_1)\rangle
\end{align}
where $g_{ij}(t)$ is the metric on a spacelike surface for a particular foliation.

Given that in the stochastic gravity case an initial time boundary is necessary to ensure that the correlation functions are finite, the following question naturally arises: must we similarly add any additional boundary terms? The answer is, in fact, no. One can see this both from the action itself, as well as by writing down the action to quadratic order in terms of the linearized metric.

To see the latter, note that the EH action has terms of the form $\ddot{q}q$ in addition to terms of the form $(\dot{q})^2$. The variation of the $\ddot{q}q$ terms introduces, at the boundary, terms of the form $\delta(\dot{q})q$. One cannot set $\delta\dot{q}$ to be zero at the boundary without over-determining the system, and thus, at the boundary, one needs to add counter-terms that cancel with these. Alternatively, integration by parts sends these terms to the boundary, but one cannot set their value without over-determining the system.

On the other hand, the stochastic gravity (SG) action, which may be obtained from~\eqref{eq:Mink}, only has terms of the form $\ddot{q}^2$. Consequently, as long as we can set $\delta q$ as well as $\delta\dot{q}$ to zero, there is no need for additional counter-terms. Since for unconditioned stochastic path integrals we have a single time boundary, we can now assign \textit{two} sets of initial conditions instead of a single one.

To see this from the action
\begin{align}
    \mathcal{S}_{G}=\int\ d^4x\sqrt{-g}(\alpha R_{\mu\nu}R^{\mu\nu}-\beta R^2),
\end{align}
we vary this with respect to the inverse metric
\begin{align}
\begin{split}
    \delta\mathcal{S}_{G}&=\int\ d^4x\sqrt{-g}\delta g^{\mu\nu}((\alpha(\Box R_{\mu\nu}+\frac{1}{2}g_{\mu\nu}\Box R-\nabla_{\mu}\nabla_{\nu}R+2R_{\mu\rho\nu\sigma}R^{\rho\sigma}-\frac{1}{2}g_{\mu\nu}R_{\rho\sigma}R^{\rho\sigma})\\&-\beta(2RR_{\mu\nu}-\frac{1}{2}g_{\mu\nu}R^2+2g_{\mu\nu}\Box R-2\nabla_{\mu}\nabla_{\nu}R)))\\&+\int_{\partial M} d^3x \sqrt {|h|}(\delta h^{ij}(f_1(h_{ij}, K_{ij}))+\delta K(f_2(h_{ij}))+\delta K^{ij}(f_3(h_{ij})))
\end{split}
\end{align}
where $f_1,f_2,f_3$ are some known functions.

Thus, as long as we can fix $g_{\mu\nu}, K_{ij}$ at the boundary, the path integral is well-defined. That is, for a conditioned pdf, one must specify $24-8=16$ pieces of initial and final data, i.e., 12 at each boundary less 8 due to diffeomorphism invariance. Then, for the unconditioned pdf, we must specify 8 pieces of initial data.

This matches nicely with what we have in terms of the SVT decomposition: we must specify 8 pieces of initial data for the unconditional pdf: 4 for the tensor, 2 for the scalar, and 2 for the vector modes. That we only have a single piece for each vector mode follows from the fact that it is not a true degree of freedom.

\section{The gravitational action with matter in the SVT decomposition}

We perform a similar SVT decomposition on $T_{\mu\nu}$, matching each gravitational mode to a corresponding matter one:
\begin{equation}
  T_{00}=\density,\qquad T_{0i}=q_{,i}+q_i,\qquad
  T_{ij}=p\,\delta_{ij}+2\Pi_{,ij}+\Pi_{i,j}+\Pi_{j,i}+\Pi^{TT}_{ij}
  \end{equation}
  \begin{equation}
  S_{i,i}=F_{i,i}=q_{i,i}=\Pi_{i,i}=0,\qquad
  h^{TT}_{ii}=h^{TT}_{ij,i}=\Pi^{TT}_{ii}=\Pi^{TT}_{ij,i}=0
  \end{equation}
  \begin{equation}
  \partial^\mu T_{\mu\nu}=0:\qquad
  \dot\density=\triangle q,\qquad \dot q=p+2\triangle\Pi,\qquad \dot q_i=\triangle\Pi_i
  \end{equation}
  \begin{equation}
  q=\triangle^{-1}\dot\density,\qquad p=\triangle^{-1}\ddot\density-2\triangle\Pi,\qquad
  \Pi_i=\triangle^{-1}\dot q_i,\qquad
  T^{\mu}{}_{\mu}=-\density+3\triangle^{-1}\ddot\density-4\triangle\Pi
  \end{equation}
  \begin{equation}
  R^{(1)}=2\big(\triangle\Phi-2\triangle\psi+3\ddot\psi\big)
  \end{equation}

Saturating the action, we obtain
  \begin{equation}
  \mathcal{I}^{\rm sat}_{CQ}=\int d^4x\Big[-i\big(\mathcal{L}_Q[T^{+}]-\mathcal{L}_Q[T^{-}]\big)
  +\mathcal{L}_{\rm tensor}+\mathcal{L}_{\rm vector}+\mathcal{L}_{\rm scalar}\Big],
  \qquad \bar X=\tfrac12(X^++X^-),\ \ \Delta X=X^+-X^-
  \end{equation}
  \begin{equation}
  \mathcal{L}_{\rm tensor}=\frac{\alpha}{8}\Big(\Box h^{TT}_{ij}+16\pi G_N\bar\Pi^{TT}_{ij}\Big)^{2}
  +\frac{\alpha}{8}\Big(8\pi G_N\Delta\Pi^{TT}_{ij}\Big)^{2}
  \end{equation}
  \begin{equation}
  \mathcal{L}_{\rm vector}=\frac{\alpha}{4}\bigg[\Big(\dot V_{i,j}
  +16\pi G_N\triangle^{-1}\dot{\bar q}_{i,j}\Big)^{2}
  -\Big(\triangle V_i+16\pi G_N\bar q_i\Big)^{2}\bigg]
  +\frac{\alpha}{4}\bigg[\Big(8\pi G_N\triangle^{-1}\Delta\dot q_{i,j}\Big)^{2}
  -\Big(8\pi G_N\Delta q_i\Big)^{2}\bigg]
  \end{equation}
  \begin{equation}
  \begin{split}
  \mathcal{L}_{\rm scalar}=&-(-\frac{\alpha}{3}\Big[\triangle(\Phi+\psi)
   +8\pi G_N\big(\bar\density-2\triangle\bar\Pi\big)\Big]^{2}
   \\[4pt]&-\frac{\alpha}{3}\Big[4\pi G_N\big(\Delta\density-2\triangle\Delta\Pi\big)\Big]^{2}\\[4pt]
  &-\frac{2(\alpha-3\beta)}{3}\Big[\triangle\Phi-2\triangle\psi+3\ddot\psi
   -4\pi G_N\big(\bar\density-3\triangle^{-1}\ddot{\bar\density}+4\triangle\bar\Pi\big)\Big]^{2}
   \\[4pt]&-\frac{2(\alpha-3\beta)}{3}\Big[2\pi G_N\big(\Delta\density-3\triangle^{-1}\Delta\ddot\density
   +4\triangle\Delta\Pi\big)\Big]^{2})
  \end{split}
  \end{equation}  

\end{document}

%% file: definitions.tex
\newcommand{\q}{q}

\def\<{\langle}
\def\>{\rangle}

\newcommand{\be}{\begin{eqnarray} \begin{aligned}}
\newcommand{\ee}{\end{aligned} \end{eqnarray} }
\newcommand{\benn}{\begin{eqnarray*} \begin{aligned}}
\newcommand{\eenn}{\end{aligned} \end{eqnarray*} }

\newcommand{\ben}{\begin{eqnarray} \begin{aligned}}
\newcommand{\een}{\end{aligned} \end{eqnarray} }

\newcommand{\bc}{\begin{center}}
\newcommand{\ec}{\end{center}}

\newcommand{\beq}{\begin{eqnarray} \begin{aligned}}
\newcommand{\eeq}{\end{aligned} \end{eqnarray} }
\newcommand{\bea}{\begin{array}}
\newcommand{\eea}{\end{array}}

\newcommand{\bee}{\begin{enumerate}}
\newcommand{\eee}{\end{enumerate}}
\newcommand{\bei}{\begin{itemize}}
\newcommand{\eei}{\end{itemize}}

\usepackage{amsfonts}

\def\01{\{0,1\}}

\newcommand{\cE}{\mathcal{E}}

\newcommand{\str}{{\vec{x}}}

\def\<{\langle}
\def\>{\rangle}

\def\s{\,\,\,\,}

\newtheorem*{rep@theorem}{\rep@title}
\newcommand{\newreptheorem}[2]{%
\newenvironment{rep#1}[1]{%
 \def\rep@title{#2 \ref{##1} (restatement)}%
 \begin{rep@theorem}}%
 {\end{rep@theorem}}}
\makeatother

\newreptheorem{thm}{Theorem}
\newreptheorem{lem}{Lemma}